\documentclass[aps,prb,twocolumn,superscriptaddress,floatfix,longbibliography]{revtex4-2}
\usepackage[english]{babel}
\usepackage[utf8]{inputenc}
\usepackage{mathtools}
\usepackage{tikz} \usetikzlibrary{decorations.pathreplacing, arrows.meta, calc}
\usepackage{physics}
\usepackage{float}
\usepackage{braket}
\usepackage{bbm}
\usepackage{bm}
\usepackage{amsbsy}
\usepackage{amsthm}
\usepackage{amssymb}
\usepackage{amsfonts}
\usepackage{amsmath}
\usepackage{soul}
\setstcolor{red}
\usepackage{dsfont} % for symbols like the identity: \mathds{1}
\usepackage{graphicx} % for graphics
\usepackage{hhline}
\usepackage{epsfig}
\usepackage{epstopdf}
\usepackage{dsfont}
\usepackage{xcolor}
\usepackage{color}
\usepackage{verbatim}
\usepackage{nomencl}
\usepackage[colorlinks]{hyperref}
\usepackage{float}
\usepackage[normalem]{ulem}
\usepackage{multirow}
\usepackage{makecell}
\usepackage{url}
\usepackage{tikz}
\usepackage{subcaption}
\usetikzlibrary{arrows.meta, positioning, decorations.pathmorphing}

\DeclareUnicodeCharacter{0301}{\'{e}}

\begin{document}
\title{Collective dynamics versus entanglement in quantum battery performance}
%\title{Role of  Entanglement in the Charging Dynamics of Many-Body Quantum Batteries}
\author{Rohit Kumar Shukla}
\email[]{rohitkrshukla.rs.phy17@itbhu.ac.in}
\affiliation{Department of Chemistry; Institute of Nanotechnology and Advanced Materials; Center for
Quantum Entanglement Science and Technology, Bar-Ilan University, Ramat-Gan, 5290002 Israel}
\author{Sunil K. Mishra}
\email{sunilkm.app@iitbhu.ac.in}
\affiliation{Department of Physics, Indian Institute of Technology (Banaras Hindu University) Varanasi - 221005, India}
\author{Ujjwal Sen}
\email{ujjwal@hri.res.in}
\affiliation{Harish-Chandra Research Institute, A CI of Homi Bhabha National Institute, Chhatnag Road, Jhunsi, Prayagraj 211019, India}

\date{\today}

\begin{abstract}

We investigate charging dynamics in many-body quantum batteries by examining the relationship between instantaneous charging power and the emergence of bipartite, tripartite, and multipartite quantum correlations across different battery--charger configurations. We find that the charging power reaches its maximum before the correlation measures peak, revealing a temporal separation between rapid energy transfer and the buildup of quantum correlations. We further study the role of interaction structure using local and many-body charging Hamiltonians under both unconstrained and constrained charging Hamiltonians, in which the total available charging resources are fixed to scale proportionally with the number of spins, thereby isolating the influence of interaction geometry and particle participation on the charging dynamics. In the unconstrained case, the enhanced charging energy observed for higher-order interactions originates primarily from the larger energy scale of the charging Hamiltonian. Under constrained conditions, however, increasing the interaction order alone provides no advantage over parallel charging when interaction and local spin-flip contributions are balanced, whereas interaction-dominated protocols significantly enhance charging performance, highlighting the importance of collective many-body dynamics. Fully collective interactions are found to yield the largest enhancement in charging power together with stronger multipartite correlations, while partially collective interactions provide only limited improvements. Finally, extending the interaction range through next-nearest-neighbor couplings suppresses charging power, demonstrating that, under constrained energetic resources, charging performance is governed primarily by interaction structure, particle participation, and collective many-body dynamics rather than by interaction order alone.
\end{abstract}

\maketitle
\section{Introduction}
\label{Introduction}

Quantum batteries (QBs) are devices designed to store and deliver energy through controlled quantum dynamics and have emerged as a central platform in quantum thermodynamics and quantum technologies. Noting by the limitations of classical chemical batteries and the rapid miniaturization of quantum devices, it is natural to look towards QBs, which exploit genuinely quantum features such as quantum coherence, entanglement, and collective many-body dynamics to enhance performance beyond classical bounds~\cite{alicki2013entanglement,campaioli2017enhancing,seah2021quantum,Zhu2023Charging,downing2024hyperbolic,friis2018precision,Lu2025topological,rossini2020quantum,gyhm2022quantum,gyhm2024beneficial,rosa2020ultra,rodriguez2021collective,konar2024quantum,mazzoncini2023optimal,zhang2019powerful,yang2023battery,ferraro2018high,andolina2019quantum}. A wide range of theoretical strategies has been proposed to optimize charging protocols and performance of QBs~\cite{ferraro2018high,andolina2019quantum,PhysRevE.100.032107,rossini2020quantum,julia2020bounds,le2018spin,ghosh2020enhancement,PhysRevA.104.032207,PhysRevA.105.022628,PhysRevA.104.L030402,PhysRevB.102.245407,PhysRevResearch.4.033216,PhysRevA.106.022618,PhysRevA.110.022226,konar2024quantum,chaki2024positive,PhysRevA.106.062609,PhysRevA.107.032203,cyrc-ms34,chaki2024universal,PhysRevA.108.042618,PhysRevA.110.012227,PhysRevApplied.19.064069,PhysRevLett.132.090401,PhysRevLett.133.180401,mitra2025bound,perciavalle2025extractable,PhysRevLett.131.240401,PhysRevB.100.115142,PhysRevE.108.064106,gyhm2024beneficial,romero2025kicked,sahoo2025power,bukov2015universal,eckardt2017colloquium,Puri2024Floquet,Lu2025topological,ferraro2018high,andolina2019extractable,GarciaPintos2020}, many of which have been realized experimentally in platforms such as quantum dots \cite{PhysRevLett.131.260401}, superconducting transmons \cite{PhysRevA.107.023725,hu2022optimal,gemme2022ibm}, organic semiconductors \cite{quach2022superabsorption}, superconducting circuits~\cite{hu2026quantum}, and nuclear magnetic resonance systems \cite{PhysRevA.106.042601}. In addition, QBs have been extensively studied in open quantum system settings, where environmental effects such as dissipation, decoherence, and engineered reservoirs strongly influence charging efficiency, stability, and extractable work \cite{Quach2020,Tabesh2020EnvMediated,Morrone2022NonMarkov,Yao2025Reservoir,QuantumErgotropy2025,NonMarkovNSpin2025,Xu2023CompositeEnv,Zhang2024OpenInteractive,Zhang2024Open,EntropicUncertaintyQB2022,liu2025open}.

Interacting many-body systems provide a natural platform for quantum batteries, where collective dynamics and correlations can significantly enhance charging power, accelerate energy storage, and improve work extraction compared to independent cells \cite{binder2015quantacell,campaioli2017enhancing,ferraro2018high,andolina2019quantum,andolina2019extractable,le2018spin,rodriguez2021collective}. Such systems, including interacting spin chains, not only exhibit nontrivial scaling with system size but also serve as versatile platforms for probing nonequilibrium quantum thermodynamics and many-body dynamics \cite{alicki2013entanglement,Puri2024Floquet,romero2025kicked}. Quantum battery architectures consist  broadly of interacting batteries driven by noninteracting systems or noninteracting batteries charged by interacting many-body chargers, with performance strongly dependent on the Hamiltonian structure, interaction range, and 
%normalization 
relative strength of the charger and the battery~\cite{le2018spin,ghosh2020enhancement,zhao2021quantum,rossini2020quantum,ferraro2018high,julia2020bounds}. Motivated by this strong sensitivity to microscopic details, periodically driven and time-dependent protocols have attracted growing attention, as they enable the engineering of effective interactions, enhance robustness, and further improve charging performance \cite{mondal2022periodically,romero2025kicked,bai2020floquet,Puri2024Floquet,downing2023quantum,crescente2020charging,shukla2026many}.  In most existing studies, the charger and the battery are treated as two distinct subsystems, with energy transferred between them through an interaction term. In contrast, our framework considers a single many-body spin system, and the notions of battery and charger are introduced through a decomposition of the total Hamiltonian rather than through physically separate subsystems. 
Depending on the physical realization, the charging protocol can be implemented in two complementary ways. In the first realization, the system is initially governed by a non-interacting Hamiltonian, and charging is achieved by switching on spin-spin interactions during the evolution. In the second realization, the system is initially interacting, and charging is induced by activating an external driving field. Thus, in both cases, the battery and the charger are not distinct physical entities but are instead defined by the role played by different Hamiltonian components during the charging protocol. Accordingly, we refer to the Hamiltonian generating the dynamics during the charging process as the \textit{charging Hamiltonian} (or ``chargoid''), while the Hamiltonian describing the reference (unperturbed) configuration is identified as the \textit{battery Hamiltonian}.
\par
Entanglement has long been considered a potential resource for enhancing quantum battery performance \cite{ferraro2018high,PhysRevResearch.2.023113}, as global collective driving protocols can generate entanglement among battery units and potentially increase instantaneous charging power compared to local parallel schemes \cite{binder2015quantacell,campaioli2017enhancing,andolina2019quantum}. This led to the view that multipartite entanglement could be a key ingredient for fast charging. However, subsequent research has clarified that entanglement alone does not guarantee improved charging or work extraction; instead, performance is primarily determined by the structure of interactions and collective effects, rather than the presence of entanglement itself \cite{PhysRevE.102.052109,le2018spin,gyhm2024beneficial}.

%Motivated 
Intrigued by the unresolved interplay between charging power and the buildup of quantum correlations, we investigate the temporal relationship between energy flow and entanglement generation during the charging process. To this end, we compare the instantaneous charging power with several entanglement-based measures, including bipartite, tripartite, and multipartite correlations. This analysis allows us to explore the temporal ordering between energy transfer and the buildup of quantum correlations, and to determine whether the emergence of entanglement precedes or follows the attainment of maximum instantaneous charging power.

 To further understand the origin of charging enhancement, we next investigate charging protocols in which a non-interacting battery is driven by a chargoid composed of $\kappa$-local interaction terms, where $\kappa$ denotes the number of spins participating in each interaction term. By varying $\kappa$, the interaction architecture can be continuously tuned from local interactions ($\kappa=2$) to fully collective interactions ($\kappa=N$, where $N$ is the total number of spins), providing a unified framework for exploring how the range and order of interactions influence charging performance.

To distinguish genuine many-body advantages from trivial increases in the energy scale of the charging Hamiltonian, we examine both unconstrained and norm-constrained charging protocols. We also analyze the interaction-only limit to isolate the role of cooperative many-body dynamics. This approach enables us to identify the conditions under which higher-order interactions lead to enhanced charging performance and to clarify the interplay between interaction structure, collective dynamics, and quantum correlations. Our analysis shows that the interaction order alone is not sufficient to guarantee improved charging, but that the organization of the interaction network and the resulting cooperative dynamics play the central role in determining the charging advantage.

The paper is organized as follows. Section~\ref{Set_up} introduces the quantum battery and chargoid and defines measures for bipartite, tripartite, and multipartite entanglement. Section~\ref{result} presents the charging dynamics for both configurations: interacting battery with noninteracting chargoid, and noninteracting battery with interacting chargoid and examines the effects of $\kappa$-local interactions on energy transfer and correlations of particles. Section~\ref{conclusion} summarizes the main findings of this work, discusses their broader implications, and provides an outlook on possible future directions.

\section{set up}
\label{Set_up}

\begin{figure}[t]
\centering
\resizebox{\columnwidth}{!}{%
\begin{tikzpicture}[x=1cm,y=1cm,line cap=round,line join=round]

% ===================== COLORS =====================
\definecolor{batterycol}{RGB}{0,85,100}
\definecolor{chargoidcol}{RGB}{85,0,160}

% ===================== PANEL (a) =====================
\node[font=\bfseries\fontsize{12}{14}\selectfont] at (0,8.1)
{(a) Protocol I: Non-interacting Battery, Interacting Chargoid};

\node[font=\bfseries\fontsize{11}{12}\selectfont,text=batterycol] at (0,7.45)
{Battery (Energy Storage)};

\node[font=\fontsize{12}{14}\selectfont,text=batterycol] at (0,6.95)
{$\hat{H}_B^{({\rm I})} = h_z \sum_{j=1}^{N} \hat{\sigma}_j^z$};

% top brace pointing upward
\draw[batterycol,line width=1pt,decorate,decoration={brace,amplitude=8pt}]
(-4.8,5.88) -- (4.8,5.88);

% spins
\node[circle,draw=black,line width=0.9pt,fill=white,minimum size=0.58cm] (a1) at (-3.8,4.45) {$\hat{\sigma}_1$};
\node[circle,draw=black,line width=0.9pt,fill=white,minimum size=0.58cm] (a2) at (-1.25,4.45) {$\hat{\sigma}_2$};
\node[font=\fontsize{18}{18}\selectfont] at (0.55,4.45) {$\cdots$};
\node[circle,draw=black,line width=0.9pt,fill=white,minimum size=0.58cm] (aN) at (3.85,4.45) {$\hat{\sigma}_N$};

% hz arrows upward
\draw[->,batterycol,line width=1pt] (-3.8,4.73)--(-3.8,5.5);
\node[font=\bfseries\fontsize{10}{10}\selectfont,text=batterycol] at (-3.5,5.0) {$h_z$};

\draw[->,batterycol,line width=1pt] (-1.25,4.73)--(-1.25,5.5);
\node[font=\bfseries\fontsize{10}{10}\selectfont,text=batterycol] at (-0.95,5.0) {$h_z$};

\draw[->,batterycol,line width=1pt] (3.85,4.73)--(3.85,5.5);
\node[font=\bfseries\fontsize{10}{10}\selectfont,text=batterycol] at (4.15,5.0) {$h_z$};

% dotted vertical lines downward
\draw[chargoidcol,dotted,line width=1pt] (-3.8,4.18)--(-3.8,3.55);
\draw[chargoidcol,dotted,line width=1pt] (-1.25,4.18)--(-1.25,3.55);
\draw[chargoidcol,dotted,line width=1pt] (3.85,4.18)--(3.85,3.55);

% hx arrows diagonal upward
\draw[->,chargoidcol,dashed,line width=1pt] (-4.55,3.6)--(-4.1,4.02);
\node[font=\bfseries\fontsize{11}{11}\selectfont,text=chargoidcol] at (-4.62,4.08) {$h_x$};

\draw[->,chargoidcol,dashed,line width=1pt] (-2.3,3.6)--(-1.85,4.02);
\node[font=\bfseries\fontsize{11}{11}\selectfont,text=chargoidcol] at (-2.47,4.05) {$h_x$};

\draw[->,chargoidcol,dashed,line width=1pt] (3.05,3.6)--(3.5,4.02);
\node[font=\bfseries\fontsize{11}{11}\selectfont,text=chargoidcol] at (3.0,3.98) {$h_x$};

% wavy line
\draw[chargoidcol,line width=1pt,decorate,decoration={snake,amplitude=1.2pt,segment length=8pt}]
(-3.8,3.45)--(3.85,3.45);

% bottom brace
\draw[chargoidcol,line width=1pt]
(-4.8,2.95)--(-4.8,2.82)
.. controls (-4.8,2.6) and (-4.65,2.48) .. (-4.42,2.48)
-- (-0.22,2.48)
.. controls (-0.08,2.48) and (-0.06,2.32) .. (0,2.32)
.. controls (0.06,2.32) and (0.08,2.48) .. (0.22,2.48)
-- (4.42,2.48)
.. controls (4.65,2.48) and (4.8,2.6) .. (4.8,2.82)
-- (4.8,2.95);

\node[font=\bfseries\fontsize{11}{12}\selectfont,text=chargoidcol] at (0,1.95)
{Chargoid (Driving Field)};

\node[font=\fontsize{11}{12}\selectfont,text=chargoidcol,align=center] at (0,1.2)
{$\hat{H}_C^{({\rm I})} =
\sum_{\alpha=x,y,z}
\left(
J_{\alpha}\sum_{j=1}^{N-\kappa+1}\prod_{r=1}^{\kappa}\hat{\sigma}^{\alpha}_{j+r-1}
\right)
+ h_x \sum_{j=1}^{N}\hat{\sigma}^{x}_{j}$};

% ===================== PANEL (b) =====================
\node[font=\bfseries\fontsize{12}{14}\selectfont] at (0,-0.2)
{(b) Protocol II: Interacting Battery, Non-interacting Chargoid};

\node[font=\bfseries\fontsize{11}{12}\selectfont,text=batterycol] at (0,-0.65)
{Battery (Energy Storage)};

\node[font=\fontsize{11}{12}\selectfont,text=batterycol,align=center] at (0,-1.25)
{$\hat{H}_B^{({\rm II})} =
\sum_{\alpha=x,y,z}
\left(
J_{\alpha}\sum_{j=1}^{N-\kappa+1}\prod_{r=1}^{\kappa}\hat{\sigma}^{\alpha}_{j+r-1}
\right)
+ h_x \sum_{j=1}^{N}\hat{\sigma}^{x}_{j}$};

% upper curly brace in panel b
\draw[batterycol,line width=1pt,decorate,decoration={brace,amplitude=6pt}]
(-4.75,-1.85) -- (4.75,-1.85);

% upper wavy line in panel b
\draw[batterycol,line width=1pt,decorate,decoration={snake,amplitude=1.2pt,segment length=8pt}]
(-4.37,-1.95)--(4.37,-1.95);

% spins
\node[circle,draw=black,line width=0.9pt,fill=white,minimum size=0.58cm] (b1) at (-3.45,-3.0) {$\hat{\sigma}_1$};
\node[circle,draw=black,line width=0.9pt,fill=white,minimum size=0.58cm] (b2) at (-1.3,-3.0) {$\hat{\sigma}_2$};
\node[font=\fontsize{18}{18}\selectfont] at (0.55,-3.0) {$\cdots$};
\node[circle,draw=black,line width=0.9pt,fill=white,minimum size=0.58cm] (bN) at (3.85,-3.0) {$\hat{\sigma}_N$};

% dotted lines upward
\draw[batterycol,dotted,line width=1pt] (-3.45,-2.72)--(-3.45,-1.95);
\draw[batterycol,dotted,line width=1pt] (-1.3,-2.72)--(-1.3,-1.95);
\draw[batterycol,dotted,line width=1pt] (3.85,-2.72)--(3.85,-1.95);

% hx dashed diagonal upward
\draw[->,batterycol,dashed,line width=1pt] (-4.35,-3.85)--(-3.88,-3.42);
\node[font=\bfseries\fontsize{11}{11}\selectfont,text=batterycol] at (-4.38,-3.48) {$h_x$};

\draw[->,batterycol,dashed,line width=1pt] (-2.2,-3.85)--(-1.73,-3.42);
\node[font=\bfseries\fontsize{11}{11}\selectfont,text=batterycol] at (-2.23,-3.38) {$h_x$};

\draw[->,batterycol,dashed,line width=1pt] (3.0,-3.85)--(3.47,-3.42);
\node[font=\bfseries\fontsize{11}{11}\selectfont,text=batterycol] at (2.97,-3.45) {$h_x$};

% hz arrows upward from below
\draw[->,chargoidcol,line width=1pt] (-3.45,-4.65)--(-3.45,-3.72);
\node[font=\bfseries\fontsize{10}{10}\selectfont,text=chargoidcol] at (-3.15,-4.22) {$h_z$};

\draw[->,chargoidcol,line width=1pt] (-1.3,-4.65)--(-1.3,-3.72);
\node[font=\bfseries\fontsize{10}{10}\selectfont,text=chargoidcol] at (-1.0,-4.22) {$h_z$};

\draw[->,chargoidcol,line width=1pt] (3.85,-4.65)--(3.85,-3.72);
\node[font=\bfseries\fontsize{10}{10}\selectfont,text=chargoidcol] at (4.15,-4.22) {$h_z$};

% bottom brace panel b
\draw[chargoidcol,line width=1pt]
(-4.75,-5.02)--(-4.75,-5.15)
.. controls (-4.75,-5.37) and (-4.6,-5.5) .. (-4.37,-5.5)
-- (-0.22,-5.5)
.. controls (-0.08,-5.5) and (-0.06,-5.66) .. (0,-5.66)
.. controls (0.06,-5.66) and (0.08,-5.5) .. (0.22,-5.5)
-- (4.37,-5.5)
.. controls (4.6,-5.5) and (4.75,-5.37) .. (4.75,-5.15)
-- (4.75,-5.02);

\node[font=\bfseries\fontsize{11}{12}\selectfont,text=chargoidcol] at (0,-5.95)
{Chargoid (Driving Field)};

\node[font=\fontsize{12}{14}\selectfont,text=chargoidcol] at (0,-6.55)
{$\hat{H}_C^{({\rm II})} = h_z \sum_{j=1}^{N}\hat{\sigma}_j^z$};

\end{tikzpicture}%
}
\caption{Schematic illustration of the two complementary charging protocols considered in this work, both defined on the same chain of $N$ spin-$\frac{1}{2}$ particles. (a) In Protocol I, the battery Hamiltonian is non-interacting, while the chargoid Hamiltonian introduces the $\kappa$-body interaction terms together with a local magnetic field $h_x$ applied along the $x$-direction. (b) In Protocol II, the interacting and non-interacting contributions are assigned oppositely: the battery Hamiltonian contains the $\kappa$-body interaction terms and the local magnetic field $h_x$, while the chargoid Hamiltonian consists of a non-interacting  local field $h_z$ acting along the $z$-direction.} 
\label{fig:charging_protocols}
\end{figure}

\subsection*{Quantum Battery and Chargoid Hamiltonian}

To investigate the role of interactions in quantum battery charging, we consider two complementary charging protocols applied to a single system of $N$ spin-$\frac{1}{2}$ particles, as illustrated in Fig.~\ref{fig:charging_protocols}. It is important to emphasize that the ``battery" and ``chargoid" in our framework do not represent separate physical hardware; instead, they correspond to different components of the total Hamiltonian acting on the same set of spins, $\hat{H}=\hat{H}_{B}+\hat{H}_{C}$. The battery Hamiltonian $\hat{H}_{B}$ defines the observable associated with the stored energy, whereas the chargoid Hamiltonian $\hat{H}_{C}$ governs the driving dynamics responsible for energy transfer.

We consider two complementary realizations in which the interacting and non-interacting parts of the Hamiltonian are assigned differently between the battery and chargoid. This allows us to distinguish whether charging enhancement arises from interactions embedded in the battery Hamiltonian itself or from interactions introduced through the charging dynamics.

\subsubsection{Protocol I: Non-interacting battery and interacting chargoid}

In Protocol I [see Fig.~\ref{fig:charging_protocols}(a)], we consider a quantum battery consisting of $N$ noninteracting spin-$\frac{1}{2}$ particles. The corresponding battery Hamiltonian is defined as

\begin{equation}
\hat{H}_{B}^{({\rm I})} = h_{z} \sum_{j=1}^{N} \hat{\sigma}_{j}^{z},
\label{Battery_Ham}
\end{equation}
where $h_z$ sets the energy splitting between the spin-up ($\ket{\uparrow}$) and spin-down ($\ket{\downarrow}$) states of each qubit. The operator $\hat{\sigma}_{j}^{z}$ acts on the $j$-th spin. This Hamiltonian defines the intrinsic energy structure of the battery, where the ground state corresponds to the uncharged configuration with all spins aligned in the down state, while excited states correspond to configurations with stored energy.

At time $t=0$, the charging process is initiated by switching on interactions among the battery spins together with an external transverse field in the $x$ direction. The charging dynamics is therefore governed by the interacting Hamiltonian

\begin{equation}
\hat H_C^{({\rm I})} =  \sum_{\alpha=x,y,z} \left( J_{\alpha} \sum_{j=1}^{N-\kappa+1} \left[\prod_{r=1}^{\kappa} \hat \sigma_{j+r-1}^{\alpha}\right] \right) + h_x\sum_{j=1}^{N}\hat \sigma_j^x .
\label{Charger_Ham}
\end{equation}
Here, $r=1,2,\dots,\kappa$ denotes the site index within each interacting block of $\kappa$ consecutive spins. The parameter $\kappa$ characterizes the locality of the interaction, with $\kappa=2$ corresponding to pairwise interactions and $\kappa>2$ describing higher-order many-body terms. By varying $\kappa$, we investigate how the interaction order influences charging performance and the development of multipartite correlations. Since the battery Hamiltonian in Protocol I is purely local, all correlations and entanglement generated during the charging process originate from the chargoid Hamiltonian. The total evolution is governed by $\hat H^{({\rm I})}=\hat H_B^{({\rm I})}+\hat H_C^{({\rm I})}$.

\subsubsection{Protocol II: Interacting battery and noninteracting chargoid}

In Protocol II [Fig.~\ref{fig:charging_protocols}(b)], we reverse the assignment of the interacting and non-interacting components of the total Hamiltonian. Here, the battery Hamiltonian itself contains the many-body interaction terms,

\begin{equation}
\hat H_B^{({\rm II})} =
\sum_{\alpha=x,y,z}
\left(
J_{\alpha}
\sum_{j=1}^{N-\kappa+1}
\prod_{r=1}^{\kappa}
\hat \sigma_{j+r-1}^{\alpha}
\right)
+h_x\sum_{j=1}^{N}\hat \sigma_j^x .
\label{Ham_B_II}
\end{equation}
The parameter $\kappa$ again determines the locality of the interaction, where $\kappa=2$ corresponds to two-body interactions and larger values of $\kappa$ introduce higher-order many-body terms. Unlike Protocol I, where the interactions are activated only during the charging process, the many-body correlations are already encoded in the energy structure of the battery Hamiltonian.

The charging process is generated by a non-interacting chargoid Hamiltonian,

\begin{equation}
\hat{H}_{C}^{({\rm II})}=h_z\sum_{j=1}^{N}\hat{\sigma}_{j}^{z}.
\label{Ham_C_II}
\end{equation}
Thus, in this protocol, the energy increase is induced through a local driving field acting on a battery whose internal structure already contains many-body interactions. The total Hamiltonian governing the evolution is given by

\begin{equation}
\hat H^{({\rm II})}=\hat H_B^{({\rm II})}+\hat H_C^{({\rm II})}.
\end{equation}
By comparing the charging dynamics of Protocol I and Protocol II, we can distinguish the contribution of interactions arising from the charging drive from those intrinsically present in the battery energy landscape.

\subsection*{Energy Storage and Power Dynamics}
To characterize the performance of a quantum battery, we focus on key quantities related to energy storage and its rate of transfer. These quantities allow us to analyze how efficiently the battery is charged over time.  

\paragraph*{Stored Energy:} 
The average energy stored in the battery at a given time $t$ is defined as
\begin{equation}
W(t) \equiv \Delta E(t) = \mathrm{Tr}\big[\hat H_B \hat \rho(t)\big] - \mathrm{Tr}\big[\hat H_B \hat \rho(0)\big],
\end{equation}
where $\hat \rho(t) = \hat U(t) \hat \rho(0) \hat U^\dagger(t)$ is the time-evolved state of the battery under the Hamiltonian $\hat H$. Here, $\hat \rho(0)$ is the initial state of the battery, typically chosen as the ground state of $\hat H_B$, and $\hat U(t) = e^{-i \hat H t}$ is the unitary evolution operator. The quantity $W(t)$ measures the net increase in the battery's energy due to the charging process, providing a direct quantification of energy storage.

\paragraph*{Instantaneous Power:} 
The instantaneous power captures the rate at which energy is being transferred to the battery at a specific moment in time. It is defined as the time derivative of the stored energy:
\begin{equation}
P_i(t) = \frac{d W(t)}{dt}.
\end{equation}  
\par
In the study of quantum battery dynamics, the instantaneous power provides a time-resolved characterization of the energy-transfer process, in contrast to the average power, which describes the energy change averaged over a finite charging interval. The analysis of $P_i(t)$ allows us to identify the temporal profile of energy injection and to compare it with the evolution of quantum correlations generated during the charging process. To investigate the possible relationship between energy transfer and correlation development, we consider several entanglement-based measures, including bipartite measures such as concurrence and bipartite entanglement entropy, tripartite correlations quantified by the tripartite mutual information, and multipartite measures such as the quantum Fisher information and average bipartite entanglement entropy. In the following, we introduce the definitions of these correlation measures used throughout our analysis.

\noindent
\subsection*{Entanglement Measures}

\paragraph*{Concurrence:}
Concurrence is one of the most widely used measures of bipartite entanglement for two-qubit systems. It provides a quantitative value ranging from $0$ for separable (uncorrelated) states to $1$ for maximally entangled states. In the context of quantum batteries, concurrence is useful for quantifying the entanglement between specific pairs of spins, for example, the first and last spins of a chain, which can reveal the formation of long-range correlations during the charging process.

For a two-qubit density matrix $\rho$, \emph{spin-flipped} matrix can br defined as \cite{wootters1998entanglement,subrahmanyam2004quantum,osborne2002entanglement,osterloh2002scaling,arnesen2001natural}
$\tilde{\rho} = (\sigma_y \otimes \sigma_y) \rho^* (\sigma_y \otimes \sigma_y)$, where $\rho^*$ is the complex conjugate in the computational basis and $\sigma_y$ is the Pauli-$y$ matrix. Using this, we construct $
R = \rho \tilde{\rho}$. Let $\{\mu_i\}_{i=1}^4$ be the eigenvalues of $R$ (which are always real and nonnegative), and let their square roots be arranged in decreasing order: 
$\lambda_1 \ge \lambda_2 \ge \lambda_3 \ge \lambda_4 \ge 0$.
The concurrence is then given by
\[
C(\rho) = \max\Big\{0, \lambda_1 - \lambda_2 - \lambda_3 - \lambda_4 \Big\}.
\]

Concurrence provides insight into how entanglement between two specific subsystems evolves in time. 

\paragraph*{Bipartite Entanglement Entropy:}

Bipartite entanglement entropy (BEE) quantifies the degree of entanglement between a subsystem $X$ and its complement $X^c$ in a many-body system. It is computed from the entropy of the reduced density matrix associated with the subsystem \cite{hamma2005bipartite,PhysRevLett.96.181602,calabrese2004entanglement,page1993average,jiang2025entanglement,kadar2010entanglement},
\begin{equation}
S_X = - \mathrm{Tr}_X \left[ \hat{\rho}_X \log \hat{\rho}_X \right],
\end{equation}
where $\hat{\rho}_X = \mathrm{Tr}_{X^c} \left[ \hat{\rho} \right]$ is obtained by tracing out all degrees of freedom outside $X$. Here, $\hat{\rho}$ denotes the density matrix of the full system, which we consider to be a pure state, $\hat{\rho} = |\psi_0\rangle \langle \psi_0|$, with $|\psi_0\rangle$ representing the ground state of the battery Hamiltonian $\hat{H}_B$.

The BEE captures the total amount of quantum correlations shared between a chosen subsystem and the rest of the system. It reveals how local parts of the system become entangled with the rest during the charging process.

\paragraph*{Tripartite Mutual Information:}

To study correlations among three subsystems simultaneously, we use the tripartite mutual information (TMI), defined as \cite{hosur2016chaos,ding2016conditional,iyoda2018scrambling,seshadri2018tripartite,shukla2025scrambling,PhysRevB.100.224302}
\begin{equation}
I_3(X:Y:Z) = I_2(X:Y) + I_2(X:Z) - I_2(X:YZ),
\end{equation}
where $I_2(X:Y) = S_X + S_Y - S_{XY}$ is the bipartite mutual information. Here, $S_X$ is the von Neumann entropy of subsystem $X$, and $S_{XY}$ corresponds to the joint entropy of subsystems $X$ and $Y$. \par
TMI characterizes the distribution of correlations among three different subsystems by quantifying whether the correlations are shared redundantly, synergistically, or exclusively among them. In the context of quantum batteries, TMI provides a measure of the development and redistribution of multipartite correlations during the charging dynamics. By comparing its temporal behavior with the instantaneous charging power, we examine how the emergence of higher-order correlations is related to the energy-transfer process beyond pairwise correlations.

\paragraph*{Quantum Fisher Information:}

Quantum Fisher information (QFI) provides a \textit{direct and quantitative witness of multipartite entanglement} in quantum batteries \cite{hauke2016measuring,Naik2019Controlled,pappalardi2017multipartite,PhysRevA.85.022322,PhysRevA.85.022321}. 
For a pure state $\ket{\psi_0}$ with Hamiltonian $\hat H_B$, it is defined as
\begin{equation}
F_Q[\hat H_B] = 4 \, (\Delta \hat H_B)^2 = 4 \Big( \langle \psi_0 | \hat H_B^2 | \psi_0 \rangle - \langle \psi_0 | \hat H_B | \psi_0 \rangle^2 \Big),
\end{equation}
and its magnitude directly reflects the number of spins that are genuinely entangled. Specifically, for an $N$-spin battery, exceeding the threshold
\begin{equation}
F_Q[\hat H_B] > \Big\lfloor \frac{N}{\kappa} \Big\rfloor \kappa^2 + \Big(N - \Big\lfloor \frac{N}{\kappa} \Big\rfloor \kappa \Big)^2
\end{equation}
guarantees at least $(\kappa+1)$-partite entanglement, providing a rigorous lower bound on the depth of correlations \cite{Naik2019Controlled}.

Multipartite entanglement captures the collective quantum correlations that can emerge among multiple spins during the charging dynamics. The QFI provides a quantitative measure of these collective correlations and the degree of coherent participation of the system's constituents. Therefore, a larger QFI indicates the presence of stronger multipartite quantum correlations, allowing us to investigate how the development of collective quantum features is temporally related to the energy-transfer process in quantum batteries.

\paragraph*{Average Bipartite Entanglement Entropy:}
While BEE measures correlations for a single partition, the \emph{average bipartite entanglement entropy} (ABEE) provides a global view of entanglement across the system \cite{igloi2008finite}. It is computed by averaging the BEE over all contiguous bipartitions:
\begin{equation}
\overline{S} = \frac{1}{\mathcal{N}} \sum_{X=1}^{\mathcal{N}} \Big[ - \mathrm{Tr}_{\{1,\dots,X\}} \left( \hat{\rho}_{\{1,\dots,X\}} \log \hat{\rho}_{\{1,\dots,X\}} \right) \Big],
\end{equation}
where $\hat{\rho}_{\{1,\dots,X\}} = \mathrm{Tr}_{\{X+1,\dots,N\}} \left[ \hat{\rho} \right]$ and $\mathcal{N}=N/2$ for even $N$ and $(N-1)/2$ for odd $N$, so that equivalent bipartitions are not double counted. ABEE captures the typical entanglement shared across multiple subsystems, providing a quantitative measure of the system’s multipartite entanglement. High ABEE indicates that entanglement is distributed extensively, while lower ABEE reflects more localized correlations. By comparing the temporal evolution of ABEE with energetic quantities such as the instantaneous charging power, we investigate the relationship between the spreading of multipartite correlations and the energy-transfer dynamics in many-body quantum batteries.

\begin{figure*}
\includegraphics[width=.32\linewidth, height=.23\linewidth]{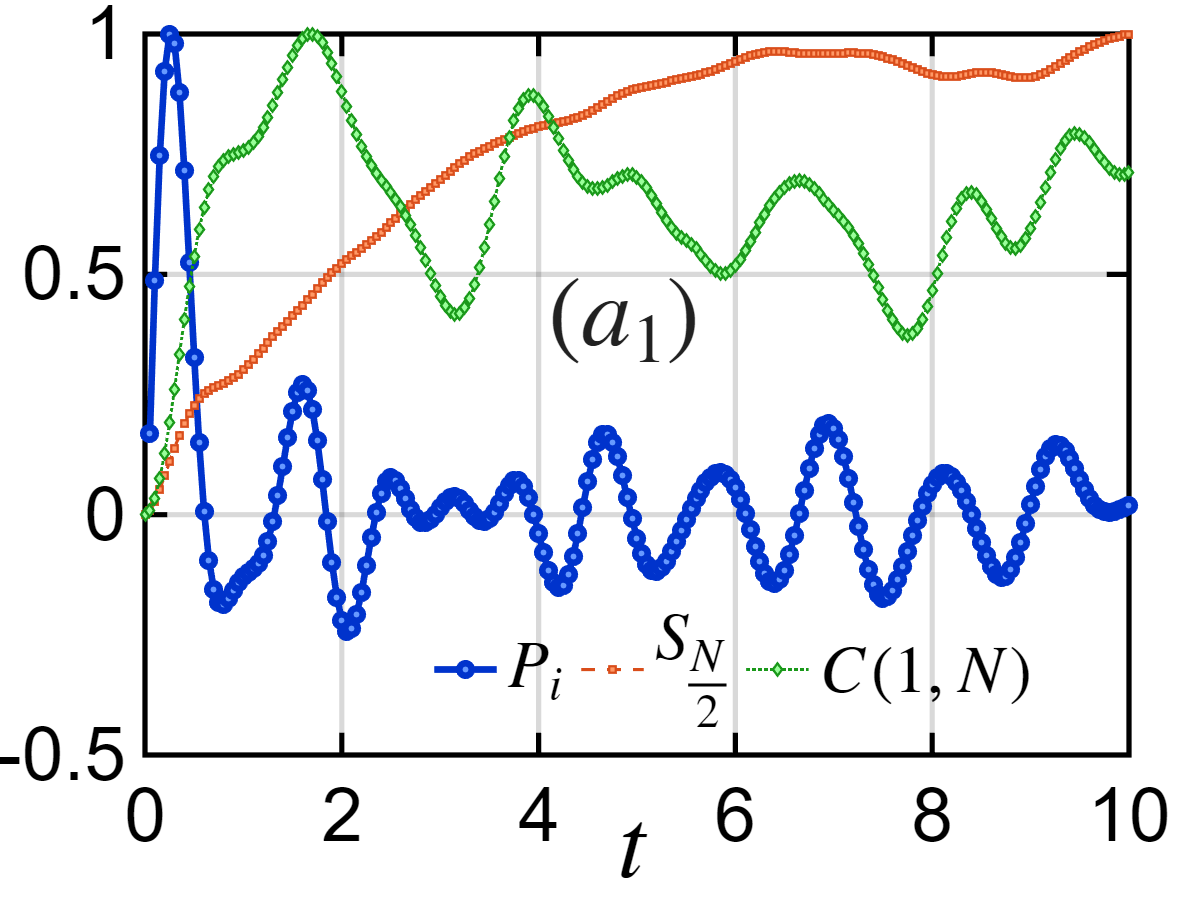}
\includegraphics[width=.32\linewidth, height=.23\linewidth]{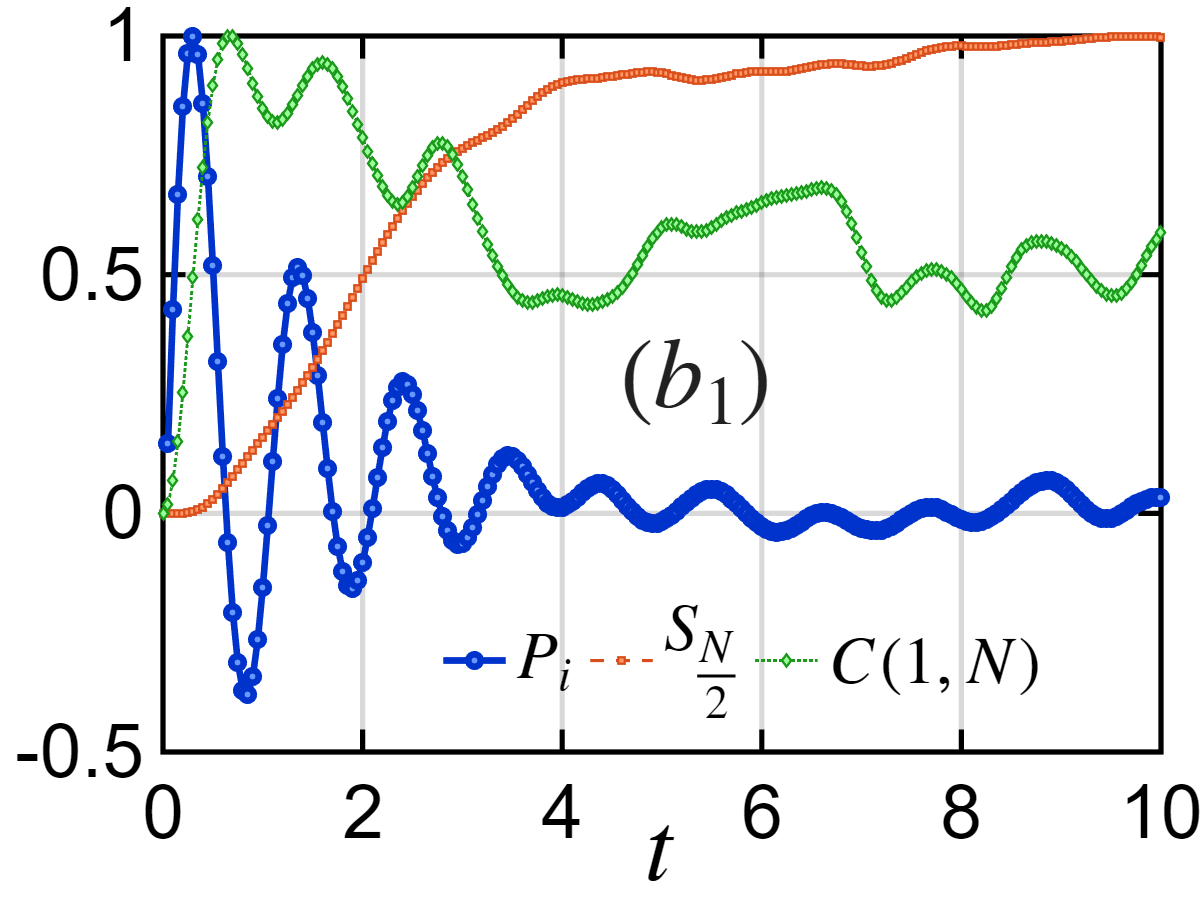}
\includegraphics[width=.32\linewidth, height=.23\linewidth]{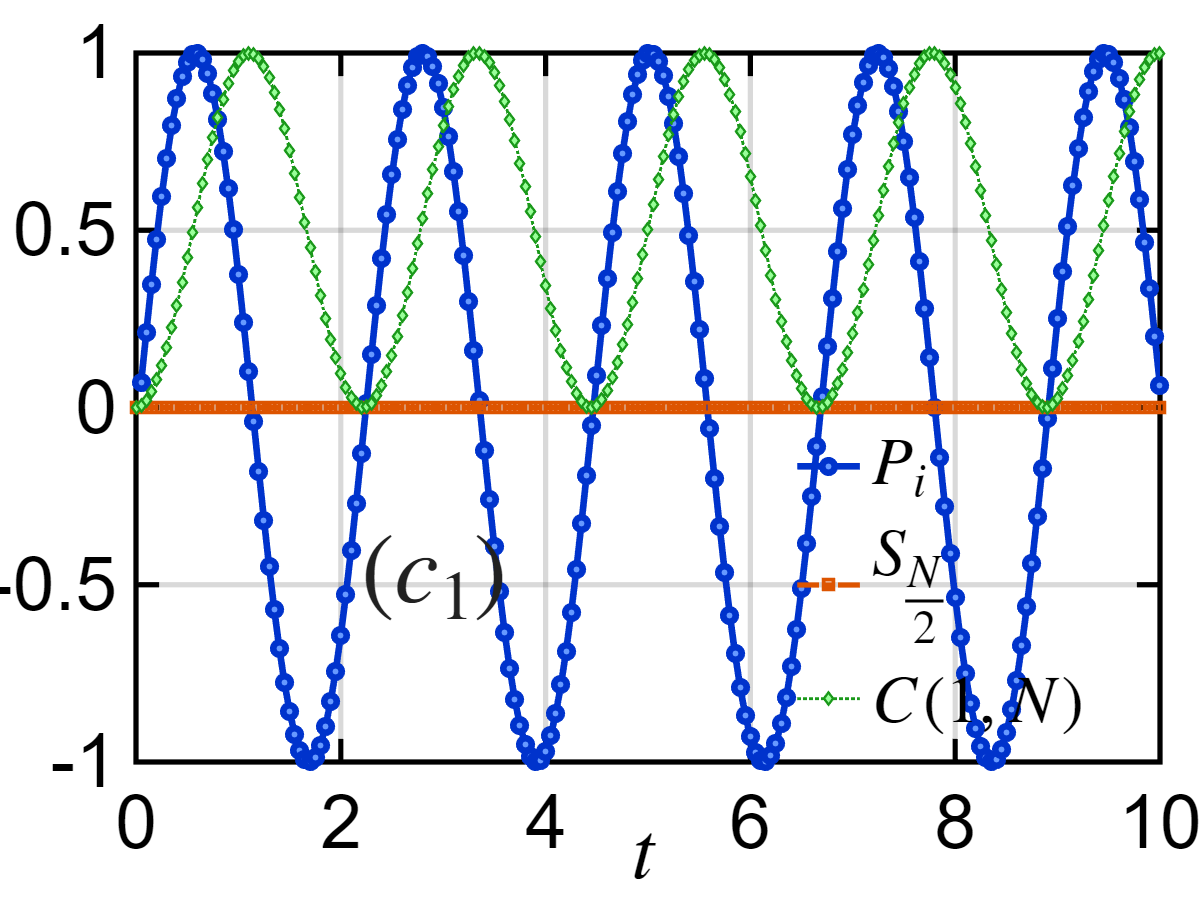}
\includegraphics[width=.32\linewidth, height=.23\linewidth]{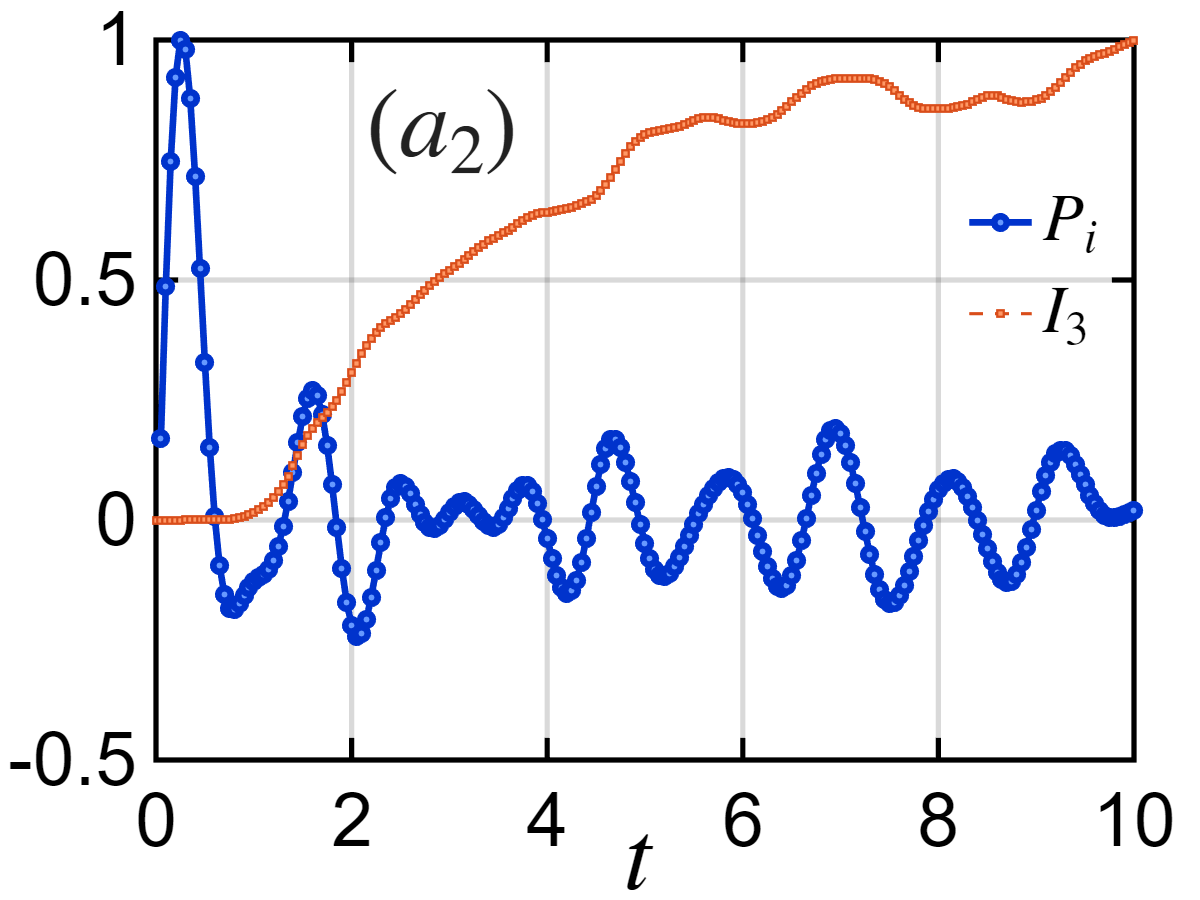}
\includegraphics[width=.32\linewidth, height=.23\linewidth]{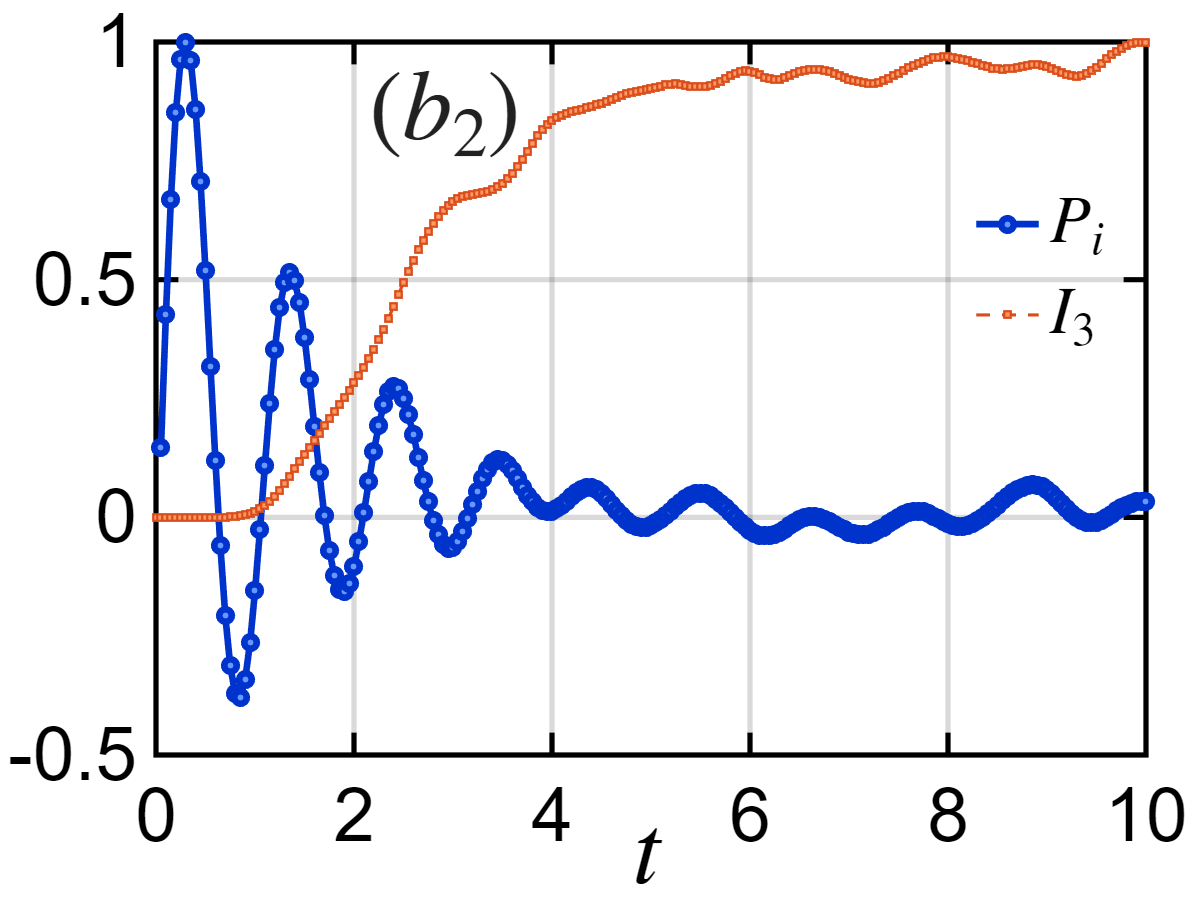}
\includegraphics[width=.32\linewidth, height=.23\linewidth]{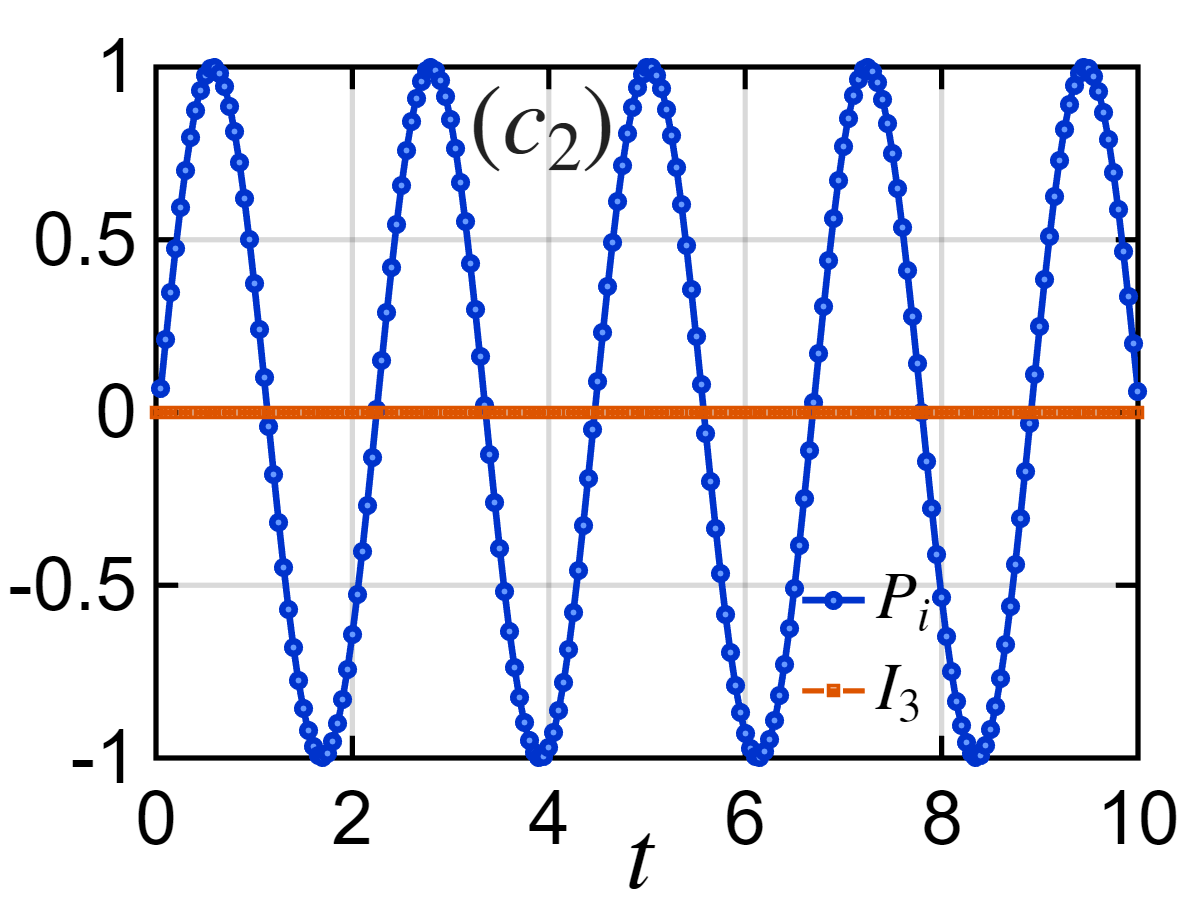}
\includegraphics[width=.32\linewidth, height=.23\linewidth]{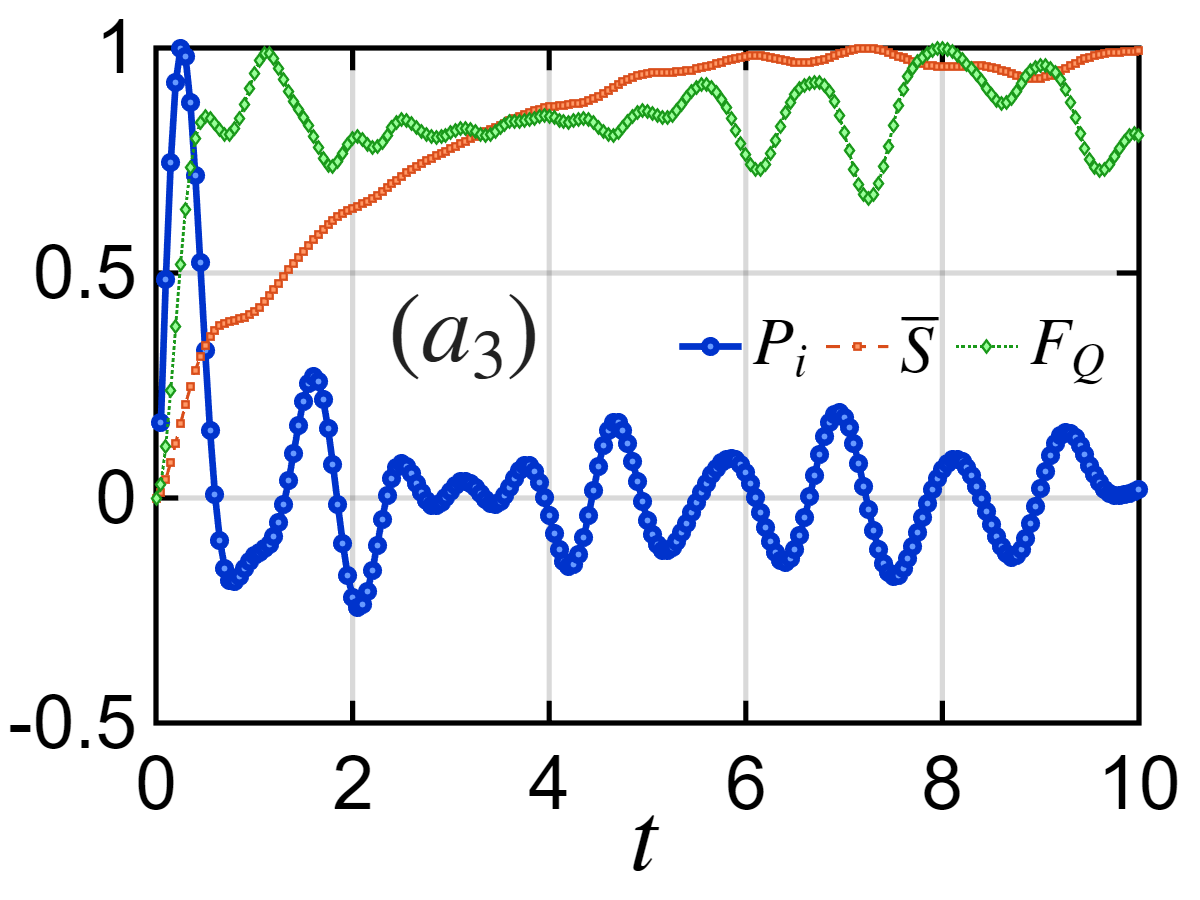}
\includegraphics[width=.32\linewidth, height=.23\linewidth]{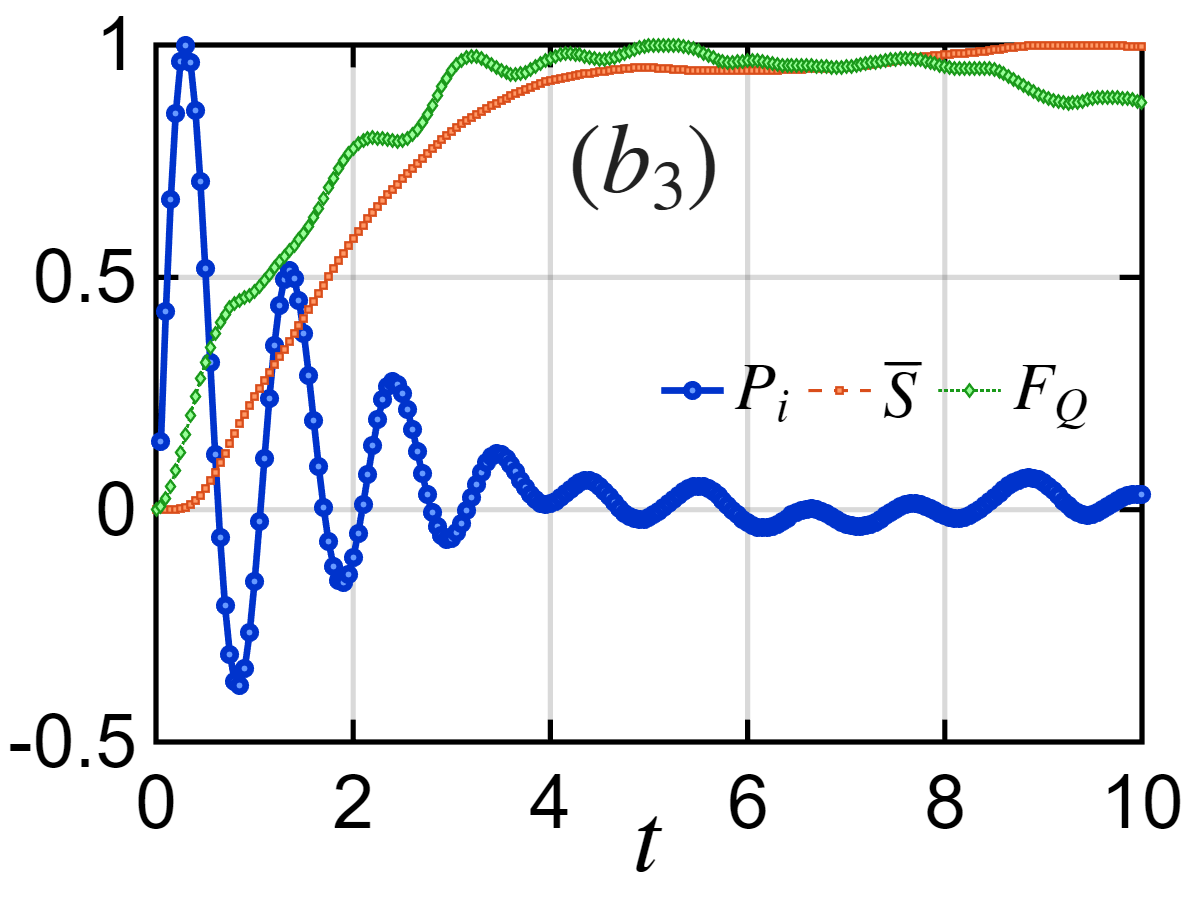}
\includegraphics[width=.32\linewidth, height=.23\linewidth]{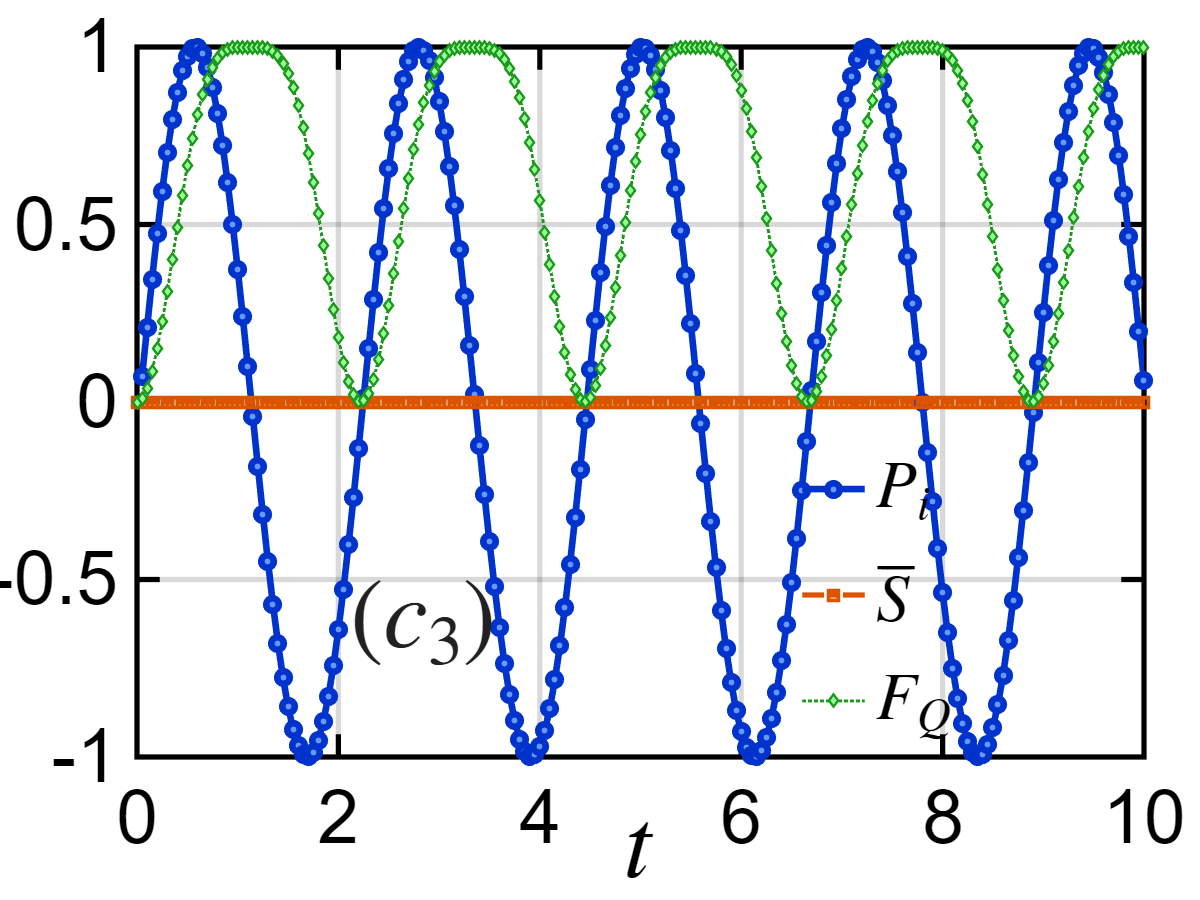}
\caption{\textbf{Noninteracting battery charged by an interaction-induced chargoid.} 
\textit{Bipartite entanglement $(a_1,b_1,c_1)$:} Time evolution of the instantaneous power $P_i$, concurrence between the first and last spins $C(1,N)$, and BEE $S_{N/2}$.  
\textit{Tripartite entanglement $(a_2,b_2,c_2)$:} Time evolution of $P_i$ and TMI $I_3$.  
\textit{Multipartite entanglement $(a_3,b_3,c_3)$:} Time evolution of $P_i$, QFI $F_Q$, and ABEE $\overline{S}_{N/2}$.  
Panels correspond to: $(a_1, a_2, a_3)$ $J_x = 1$, $J_y = J_z = 0$; $(b_1, b_2, b_3)$ $J_x = J_y = 1 $, $J_z = 0$; $(c_1, c_2, c_3)$ $J_x = J_y = J_z = 1$.  
Number of spins: $N = 12$ and $h_x=h_z=1$.}

\label{Conc_Fig}
\end{figure*}

\section{Results}
\label{result}
we investigate the relationship between energy transfer and the development of quantum correlations in many-body quantum batteries. To this end, we compare the temporal evolution of energetic quantities such as instantaneous charging power, with a hierarchy of information-theoretic measures that probe correlations at different levels: bipartite entanglement measures (concurrence and BEE), tripartite correlations (TMI), and multipartite correlation measures (QFI and ABEE). By analyzing these temporal profiles together, we examine the interplay between energy transfer and correlation buildup during the charging dynamics, providing insight into how different classes of quantum correlations evolve in many-body quantum battery systems.

\subsection*{Charging Rate and Entanglement with $2$-local Interactions}
\subsubsection*{Protocol I: Noninteracting Battery, Interacting chargoid}
We first consider Protocol ${\rm I}$ in the battery–chargoid configuration, where the battery is initially a noninteracting spin system, described by the Hamiltonian \(H_B^{\rm (I)}\) [Eq.~\ref{Battery_Ham}]. Charging is implemented by turning on spin-spin interactions among the same spins, described by the Hamiltonian \(H_C^{(\rm I)}\)[Eq.~\ref{Charger_Ham}], which act as the chargoid. These interactions are active only during the charging interval \([0, \tau]\), where $\tau$ denotes the total charging time, driving energy flow into the battery while simultaneously generating quantum correlations (a similar setup was discussed in Ref.~\cite{alicki2013entanglement}). In this setup, the noninteracting battery serves as a passive energy receiver, allowing us to isolate and study the effects of chargoid-induced correlations on the charging dynamics.
\par
We investigate the dynamical relationship between the battery charging rate and bipartite correlations. To quantify bipartite correlations, we evaluate two complementary measures: the concurrence and the BEE. These quantities are computed for three representative chargoid interaction configurations, namely $J_x = 1$, $J_y = 0$, $J_z = 0$; $J_x = 1$, $J_y = 1$, $J_z = 0$; and $J_x = 1$, $J_y = 1$, $J_z = 1$. In all cases, a local field is applied along the $x$-direction with a fixed strength $h_x=1$. Concurrence quantifies pairwise entanglement between two qubits; here, we focus on the first and last spins to probe long-range correlations. In contrast, the  BEE measures the entanglement between a chosen subsystem and its complement through the BEE of the reduced density matrix. In our calculations, we consider equal bipartitions of the spin chain. For a meaningful comparison, all three quantities are normalized by their respective maximum values, allowing us to identify the time regimes in which each quantity attains its peak.
\par
For the case of $J_x = 1$, $J_y = 0$, $J_z = 0$ [Fig.~\ref{Conc_Fig}$(a_1)$] and $J_x = 1$, $J_y = 1$, $J_z = 0$ [Fig.~\ref{Conc_Fig}$(b_1)$],  during the initial charging stage the instantaneous power $P_i(t)$ rises rapidly as the chargoid injects excitations into the battery; this rise is driven by coherent excitation transfer enabled by the chargoid’s internal couplings. Concurrence and BEE also increase during this phase. Crucially, however, the peak of instantaneous power is reached \emph{before} concurrence and BEE attain their maxima (see Fig.~\ref{Conc_Fig} $(a_1,b_1)$). Concurrence typically grows faster than BEE and peaks earlier than BEE, but both lag the power peak. 
In contrast, for the isotropic Heisenberg interaction ($J_x=J_y=J_z=1$) [Fig.~\ref{Conc_Fig}$(c_1)$], the concurrence exhibits oscillatory dynamics throughout the evolution. Although the concurrence reaches its maximum after the instantaneous power attains its peak value, the separation between these two timescales is considerably smaller than in the anisotropic cases. Interestingly, in this parameter regime, the BEE remains identically zero during the entire evolution [Fig.~\ref{Conc_Fig}$(c_1)$].

\par
To probe correlations beyond pairwise contributions, we evaluate the TMI, which quantifies how information and correlations are distributed among three subsystems. For this calculation, the spin chain of length $N$ is partitioned into four contiguous segments of size $N/4$ each, which serves as a symmetric and balanced choice for defining the subsystems involved in the TMI. The initial state is taken to be the ground state of the battery Hamiltonian, providing a well-defined reference for both the injected energy and the resulting correlation dynamics during the charging process. For the parameter regime $J_x=1, J_y=J_z=0$ [Fig.~\ref{Conc_Fig} $(a_2)$] and $J_x=1, J_y=1, J_z=0$ [Fig.~\ref{Conc_Fig} $(b_2)$], the TMI increases substantially only after the early-time rise of power, and its maximum occurs at later times compared to energetic quantities [Fig.~\ref{Conc_Fig} $(a_2,b_2)$]. This delayed growth of TMI reveals that three-way and higher-order correlations require additional time to form: pairwise coherences must first be established and propagated before genuine three-body information sharing emerges. From a physical standpoint, this hierarchical buildup reflects the propagation and interference of excitations through the interacting chargoid, leading ultimately to delocalized correlation structures (scrambling) that are slower to develop than local energy transfer. In contrast, for the isotropic Heisenberg interaction $J_x=J_y=J_z=1$, the TMI remains zero throughout the entire evolution [Fig.~\ref{Conc_Fig} $(c_2)$].
\par

To characterize genuinely global correlations we evaluate the QFI and the ABEE. The QFI serves as a witness of multipartite entanglement and quantifies the state’s collective sensitivity to global parameter shifts; ABEE captures the typical entanglement across all contiguous bipartitions of the chain. For the case of  $J_x=1, J_y=J_z=0$ [Fig.~\ref{Conc_Fig} $(a_3)$], and $J_x=1, J_y=1, J_z=0$ [Fig.~\ref{Conc_Fig} $(b_3)$], QFI and ABEE grow more slowly than the instantaneous power and reach their maxima at later times [Fig.~\ref{Conc_Fig} $(a_3,b_3)$]. The delayed peak of multipartite correlation measures suggests that many-body entanglement is not created instantaneously. Instead, energy transfer and local correlations develop first, and only later do these processes give rise to genuine multipartite entanglement. High late-time ABEE values signify that entanglement becomes extensively distributed across the system only after most of the energy transfer has already occurred. 
For the isotropic case $J_x=J_y=J_z=1$ [Fig.~\ref{Conc_Fig} $(c_3)$], the QFI exhibits oscillatory behavior. However, the time at which QFI reaches its maximum remains larger than the corresponding time for the instantaneous power. Notably, the time difference between these two maxima is smaller compared to the cases with anisotropic couplings (all the J's are not equal). In the same regime, the ABEE remains zero throughout the entire evolution [Fig.~\ref{Conc_Fig} $(c_3)$].

 Energy injection from an interacting chargoid leads to a rapid increase in the instantaneous charging power during the early stages of the dynamics. In contrast, the correlation measures considered in this work (concurrence, BEE, TMI, QFI, and ABEE) typically reach their extrema at later times compared to the peak of the instantaneous power. This temporal ordering reflects a dynamical separation between the charging power and the evolution of the correlation measures considered here. As excitations propagate through the system, increasingly complex correlation structures emerge and continue to evolve even after the charging power has attained its maximum value. Consequently, the peak of the instantaneous charging power is not synchronized with the extrema of the correlation measures examined here, while leaving open the broader question of how different forms of quantum correlations contribute to charging performance.

To verify that our conclusions are not artifacts of small system sizes, we extend our calculations to larger system sizes up to $N=20$ (depending on computational cost). As shown in Appendix~\ref{Finit_size_ana}, the normalized instantaneous charging power and the various correlation measures exhibit qualitatively similar behavior across different system sizes. While system-size effects become more visible at longer times for the Ising and XY interactions, the short-time dynamics including the time at which the charging power reaches its maximum remain largely unaffected. Moreover, the key observation that the peak charging power is attained before the maxima of concurrence, BEE, TMI, ABEE, and QFI persists for all system sizes considered. These results indicate that our main conclusions remain consistent for larger system sizes within the range studied.

\subsubsection*{Protocol II: Interacting Battery, Noninteracting Chargoid}
We next consider the complementary charging scenario, referred to as Protocol II, in which the roles of the battery and chargoid are interchanged compared with Protocol I. Here, the battery itself consists of interacting spins, described by Eq.~(\ref{Ham_B_II}), while the charging process is driven by an external field acting independently on each spin, as defined in Eq.~(\ref{Ham_C_II}) (similar scenarios are considered in Refs.~\cite{le2018spin,ghosh2020enhancement}). This configuration enables us to examine how interactions embedded within the battery Hamiltonian affect the energy-transfer dynamics and the development of quantum correlations, while the charging drive itself remains free of interaction-induced correlation generation.
\par
\begin{figure*}
\includegraphics[width=.32\linewidth, height=.23\linewidth]{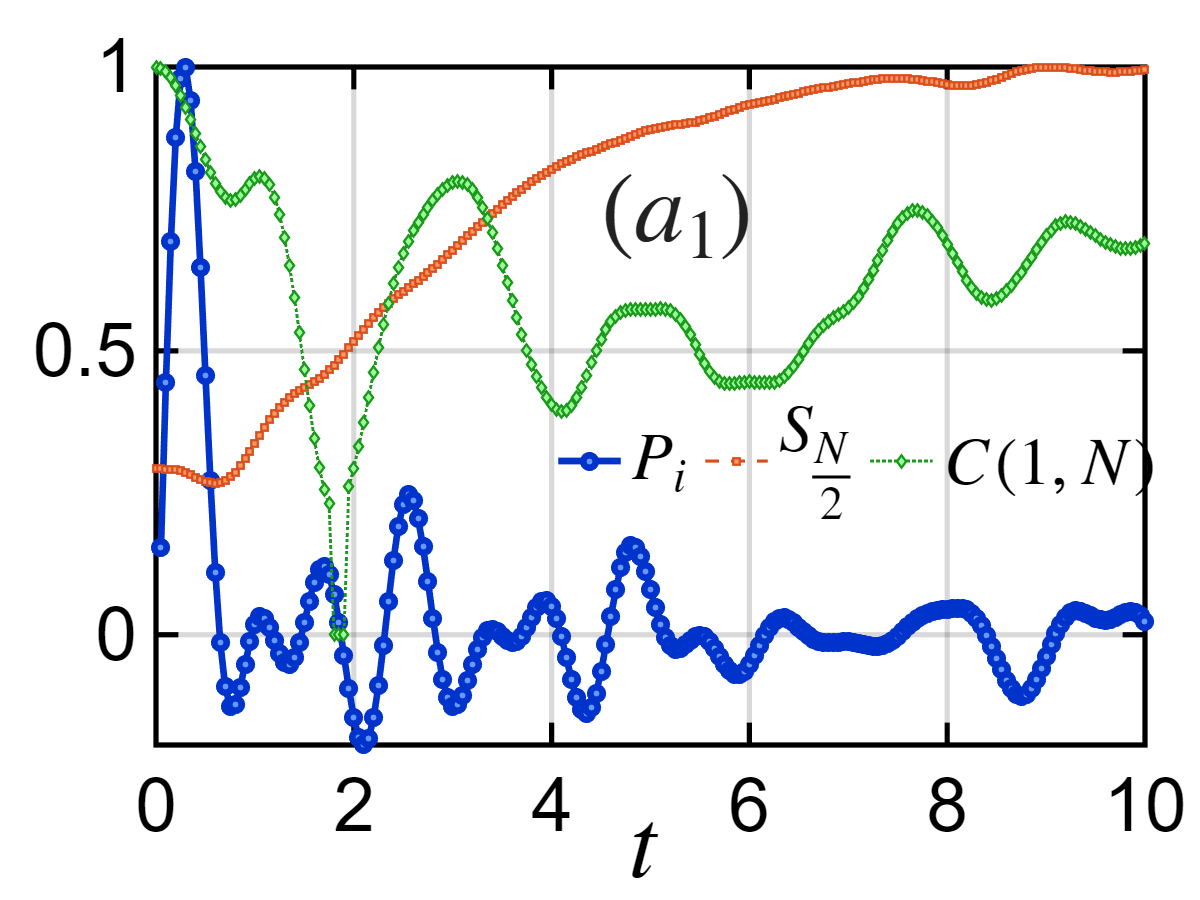}
\includegraphics[width=.32\linewidth, height=.23\linewidth]{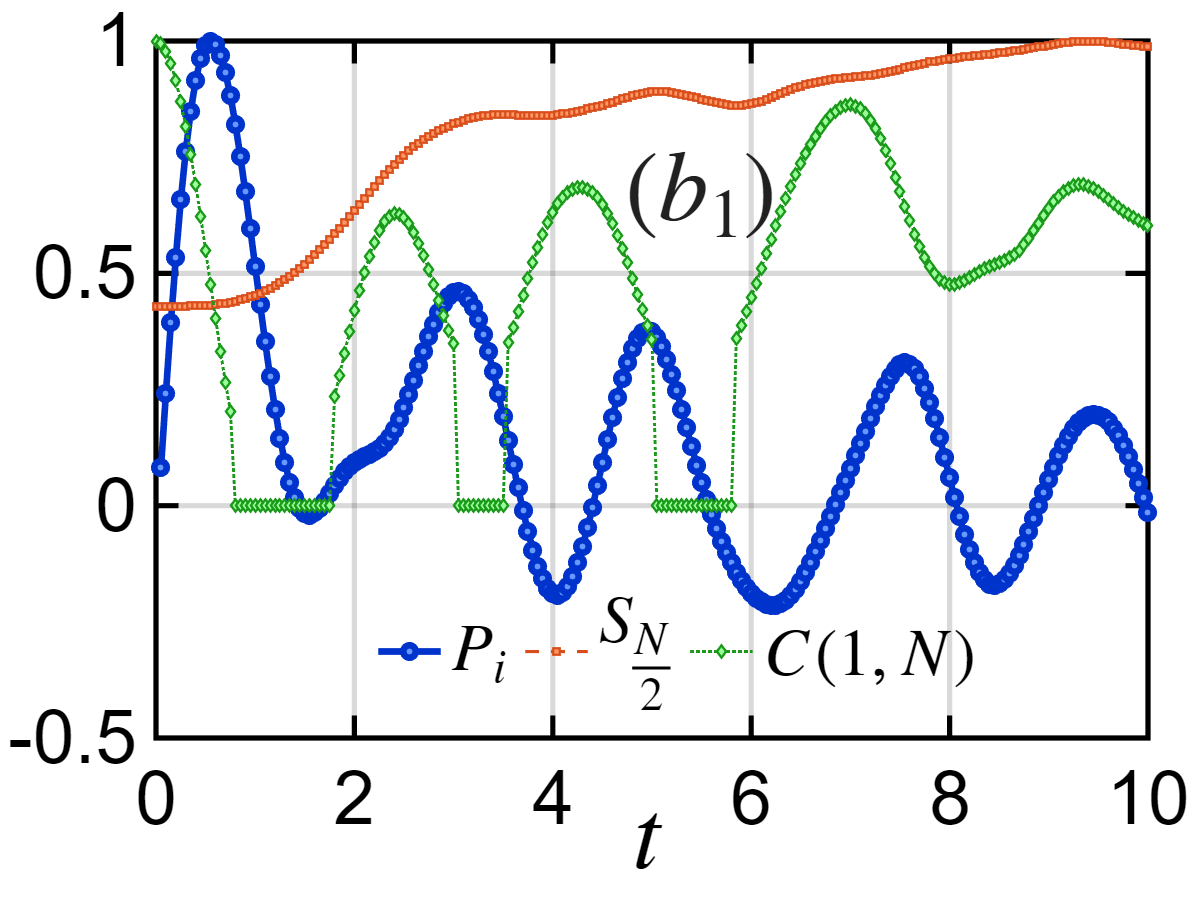}
\includegraphics[width=.32\linewidth, height=.23\linewidth]{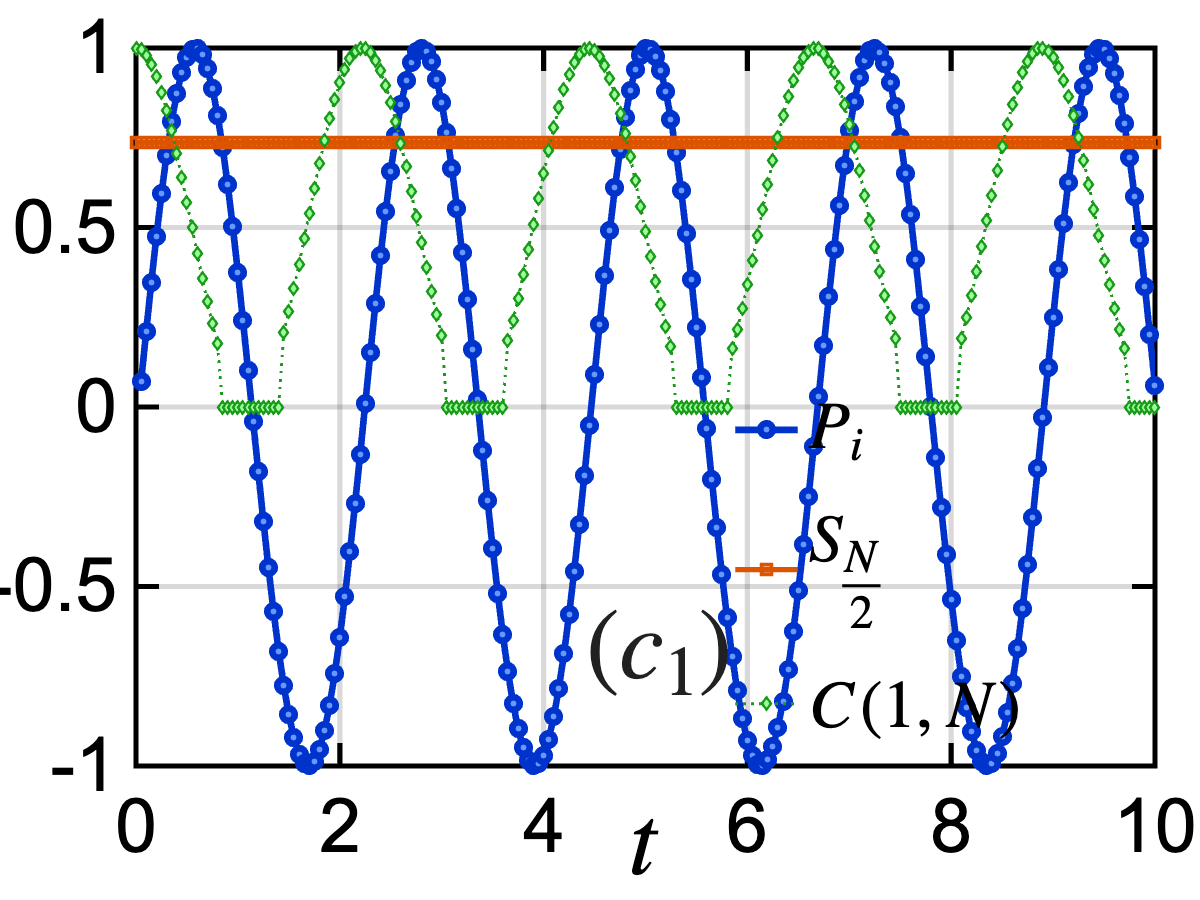}
 \includegraphics[width=.32\linewidth, height=.23\linewidth]{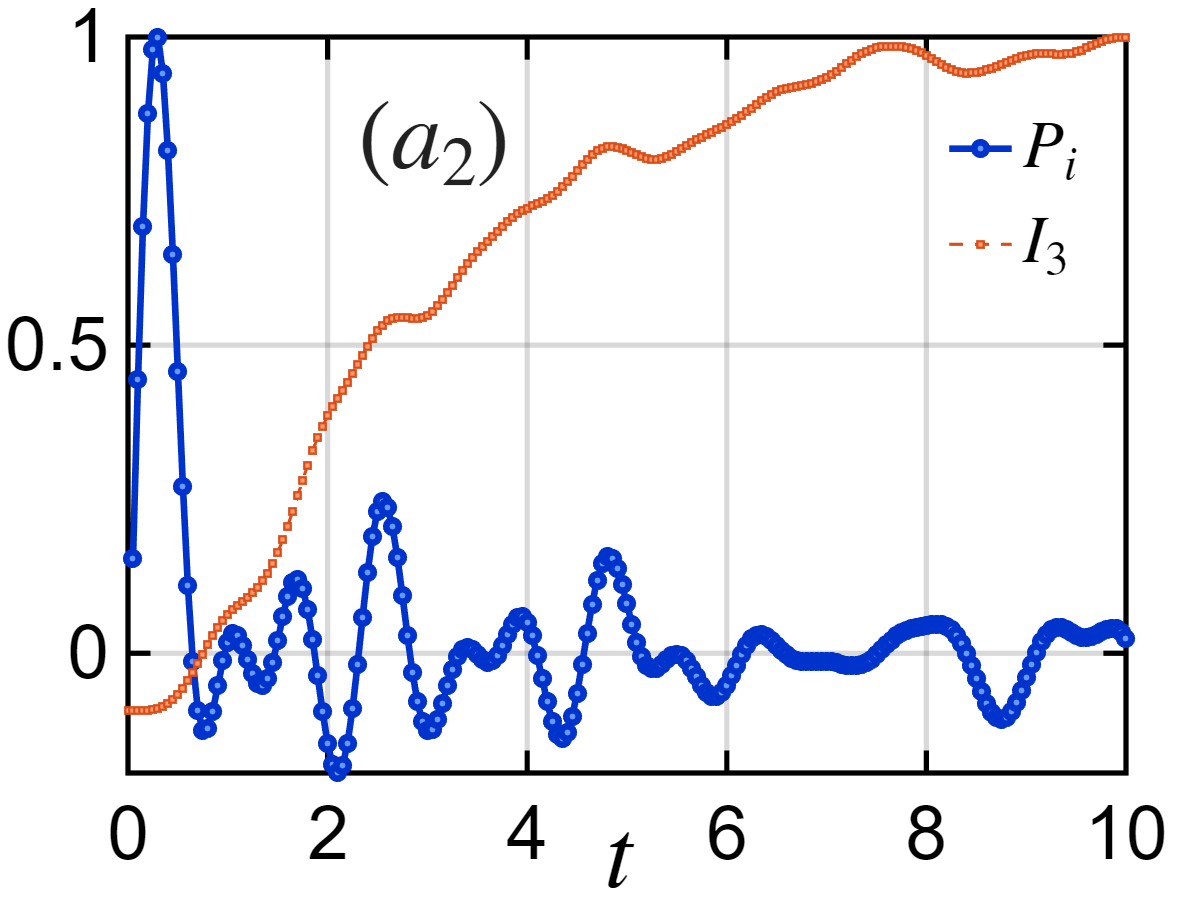}
\includegraphics[width=.32\linewidth, height=.23\linewidth]{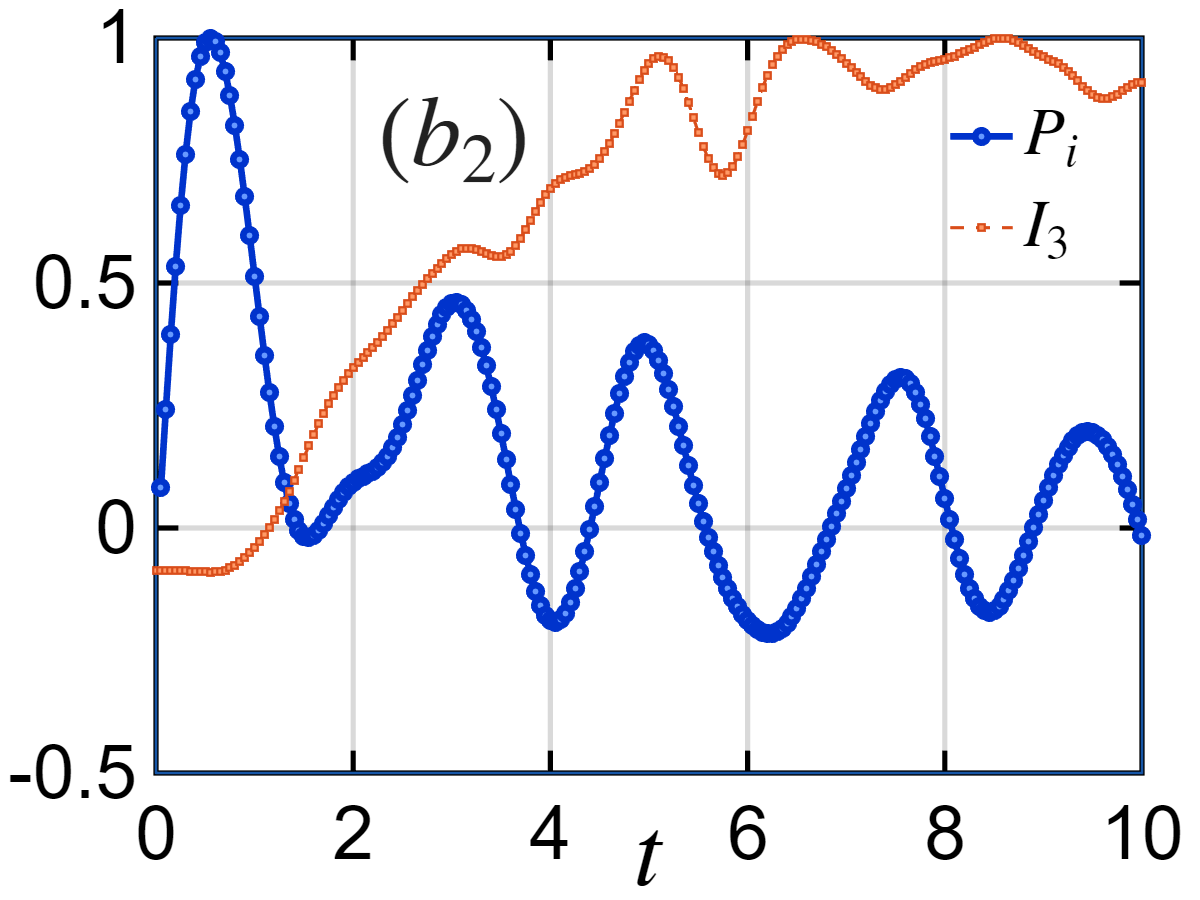}
\includegraphics[width=.32\linewidth, height=.23\linewidth]{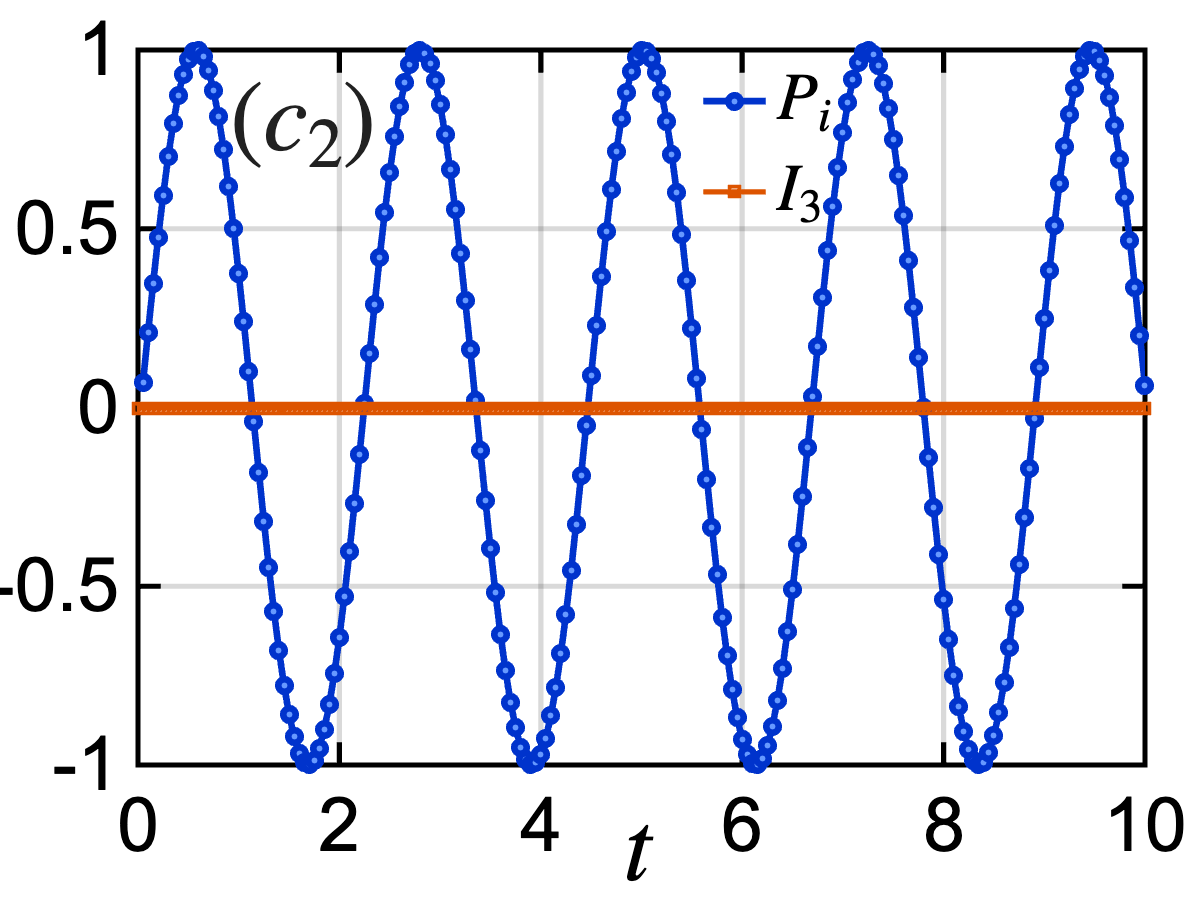}
\includegraphics[width=.32\linewidth, height=.23\linewidth]{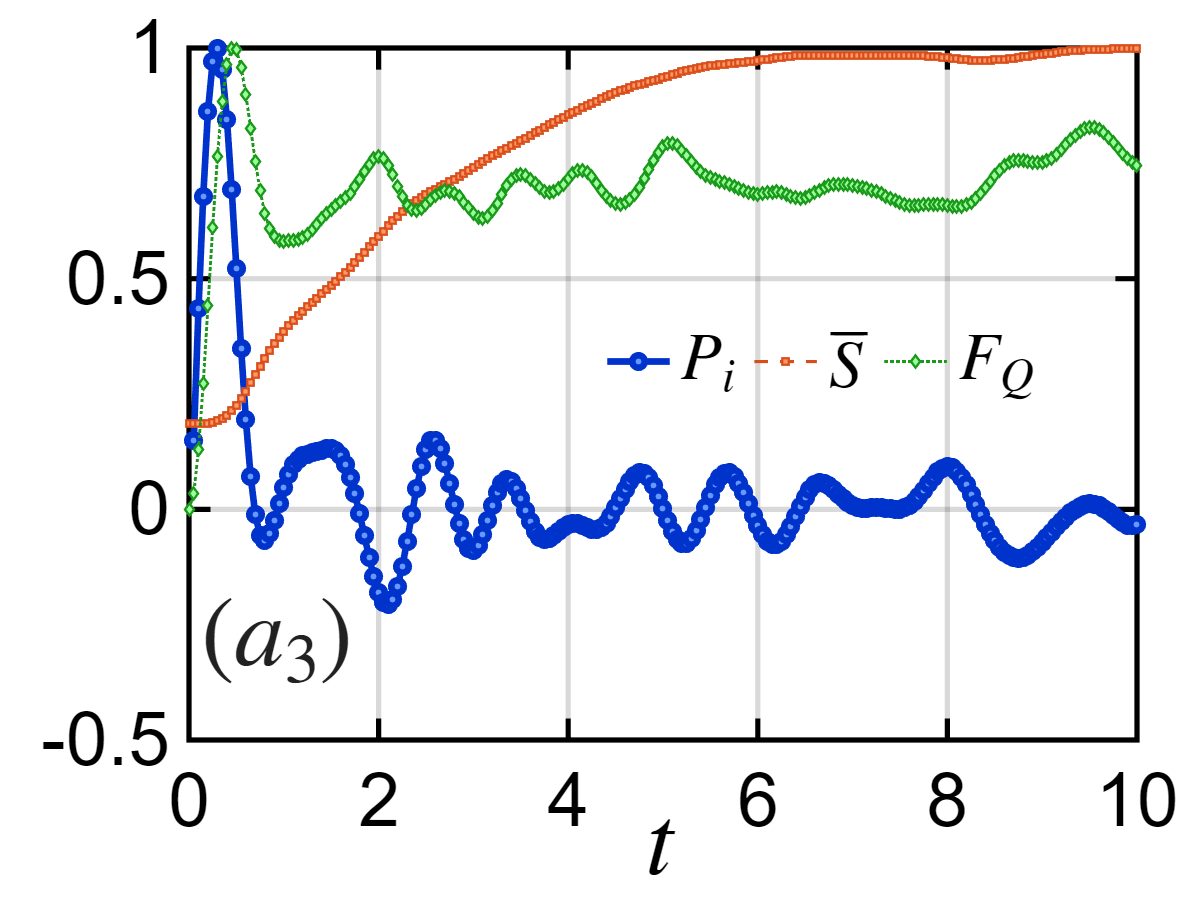}
\includegraphics[width=.32\linewidth, height=.23\linewidth]{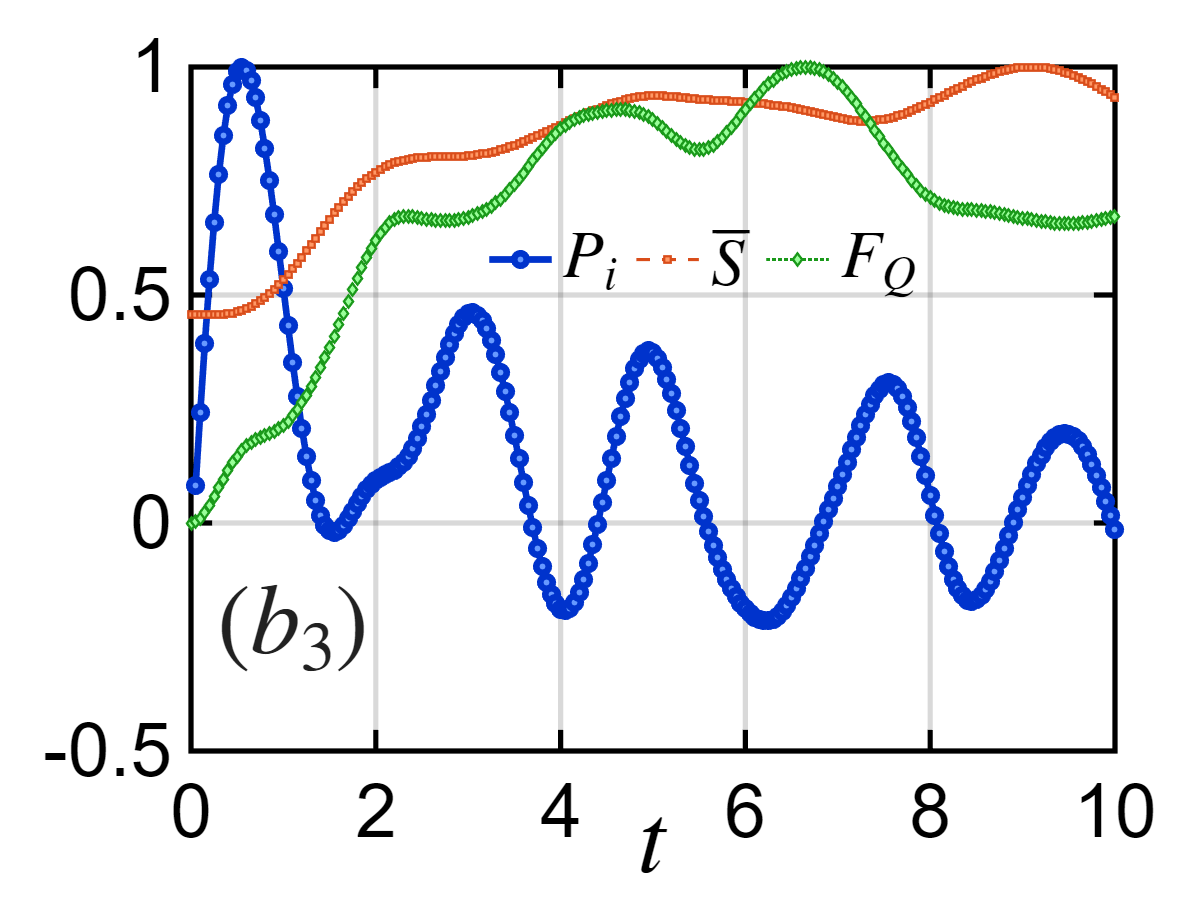}
\includegraphics[width=.32\linewidth, height=.23\linewidth]{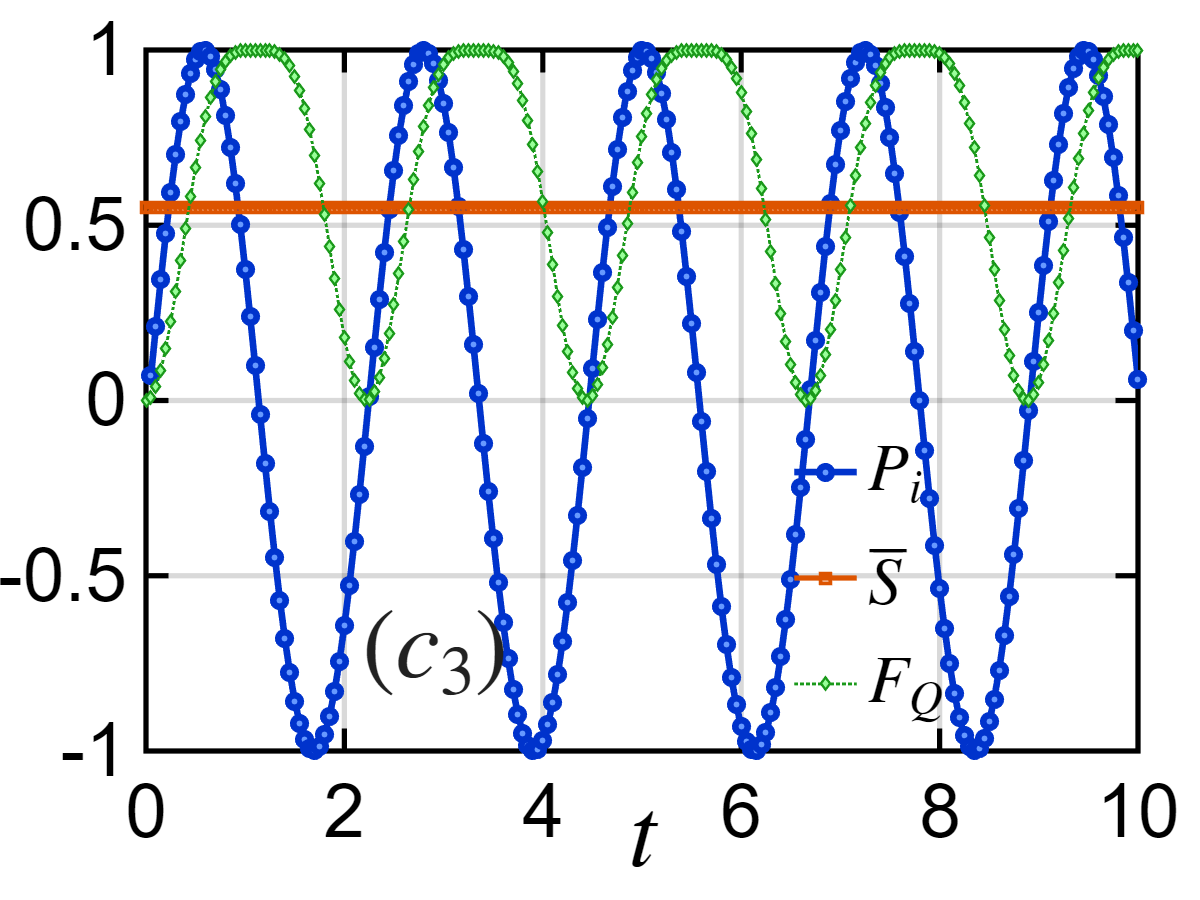}
\caption{\textbf{Interacting battery charged by noninteracting chargoid.} 
\textit{Bipartite entanglement $(a_1,b_1,c_1)$:} Time evolution of the instantaneous power $P_i$, concurrence between the first and last spins $C(1,N)$, and BEE $S_{N/2}$.  
\textit{Tripartite entanglement $(a_2,b_2,c_2)$:} Time evolution of $P_i$ and TMI $I_3$.  
\textit{Multipartite entanglement $(a_3,b_3,c_3)$:} Time evolution of $P_i$, QFI $F_Q$, and ABEE $\overline{S}_{N/2}$.  
Panels correspond to: $(a_1, a_2, a_3)$ $J_x = 1$, $J_y = J_z = 0$; $(b_1, b_2, b_3)$ $J_x = J_y = 1$, $J_z = 0$; $(c_1, c_2, c_3)$ $J_x = J_y = J_z = 1$. Number of spins: $N = 12$} and $h_x=h_z=1$.
\label{Conc_HB_int}
\end{figure*}

For bipartite correlations, we again compute the concurrence and the BEE. Across all interaction configurations, the concurrence exhibits a qualitatively different temporal behavior compared to the previous configuration: it starts with a finite value at $t=0$ and subsequently decreases as the instantaneous power increases [Fig.~\ref{Conc_HB_int} ($a_1,b_1,c_1$)]. For the cases $J_x = 1$,  $J_y = J_z = 0$ [Fig.~\ref{Conc_HB_int} ($a_1$)] and $J_x = J_y = 1, J_z = 0$ [Fig.~\ref{Conc_HB_int} ($b_1$)], the BEE grows more gradually than the instantaneous power and reaches its maximum at later times, consistent with the general trend observed in the interacting-chargoid setup [Fig.~\ref{Conc_HB_int} ($a_1,b_1$)]. However, when all interaction terms are present ($J_x = J_y = J_z = 1$) [Fig.~\ref{Conc_HB_int}$(c_1)$], the BEE remains constant throughout the evolution (the data are shown without normalization by the maximum value, since normalization would yield a trivial constant value of one), indicating a suppression of entanglement dynamics despite strong coupling among the spins. Notably, the BEE maximum, when present, still appears \emph{after} the peak in instantaneous power [Fig.~\ref{Conc_HB_int} $(a_1, b_1, c_1)$], maintaining the same temporal hierarchy as in the Protocol ${\rm I}$. 
\par
We next examine higher-order correlations using the TMI. For the cases $J_x = 1$,  $J_y = J_z = 0$ [Fig.~\ref{Conc_HB_int} $(a_2)$] and $J_x = J_y = 1, J_z = 0$ [Fig.~\ref{Conc_HB_int} $(b_2)$], TMI exhibits a slower growth rate compared to energetic quantities, reaching its maximum at later times [Fig.~\ref{Conc_HB_int} $(a_2, b_2)$].  This delay indicates that the emergence of genuine multipartite correlations is not instantaneous, even when the battery possesses intrinsic interactions. Instead, these correlations develop progressively as excitations propagate and interfere across different regions of the interacting battery. At early times, the charging process is dominated by localized excitations and short-range coherence, while nonlocal correlation sharing becomes prominent only after the system approaches energetic saturation. However, when all interaction terms are present, $J_x = J_y = J_z = 1$ [Fig.~\ref{Conc_HB_int} $(c_2)$], similar to the BEE, the TMI also remains constant throughout the evolution (the data are shown without normalization by the maximum value, since normalization would yield a trivial constant value of one)[Fig.~\ref{Conc_HB_int}($c_2$)]. This behavior can be attributed to the high degree of symmetry in the fully isotropic spin interaction. The isotropic Heisenberg dynamics preserves the global correlation structure of the system, restricting the redistribution of information among different subsystems. As a result, multipartite correlations do not grow or decay during the evolution, leading to a time-invariant TMI despite the ongoing unitary dynamics.
\par
To quantify global quantum correlations, we evaluate the QFI and ABEE. For the interaction configurations $J_x = 1, J_y = J_z = 0$ [Fig.~\ref{Conc_HB_int} $(a_3)$] and $J_x = J_y = 1, J_z = 0$ [Fig.~\ref{Conc_HB_int} $(b_3)$], both measures increase more gradually than the instantaneous power and reach their maxima at later times [Fig.~\ref{Conc_HB_int} $(a_3, b_3)$], consistent with the qualitative behavior observed in the Protocol ${\rm I}$. However, when all interaction terms are present ($J_x = J_y = J_z = 1$) [Fig.~\ref{Conc_HB_int}$(c_3)$], the ABEE remains nearly constant throughout the evolution (shown without normalization by its maximum value), indicating a suppression of global entanglement dynamics despite the presence of strong couplings. In contrast, the QFI continues to exhibit the characteristic delayed rise, achieving its maximum well after the power peak [Fig.~\ref{Conc_HB_int}$(c_3)$]. This confirms that multipartite entanglement develops over longer timescales than energy transfer, as the interactions must coherently couple multiple subsystems before extensive global correlations are established. Thus, although strong internal interactions facilitate the redistribution of correlations within the battery, they do not alter the fundamental temporal hierarchy: energy transfer and power amplification occur first, followed by the progressive emergence of bipartite, tripartite, and finally multipartite entanglement.
\par
Interestingly, when all interaction terms are present ($J_x = J_y = J_z = 1$), we also observe that the instantaneous power $P_i(t)$ oscillates roughly with  same frequency of the QFI [Fig.~\ref{Conc_HB_int}$(c_3)$]. This correlation arises from the fact that both quantities are determined by the time-dependent state of the system and the fluctuations of the battery Hamiltonian. While the instantaneous power characterizes the rate of energy transfer, the QFI probes the fluctuations of the battery Hamiltonian through its variance. Consequently, the QFI and instantaneous power display oscillations with a similar characteristic frequency, despite probing different aspects of the charging dynamics.
\par
We observe a strikingly similar behaviour of the instantaneous power in both protocols:  a noninteracting battery charged with an interacting chargoid, and an interacting battery charged with a noninteracting chargoid (compare Fig.~\ref{Conc_Fig} and Fig.~\ref{Conc_HB_int}). The instantaneous power of the quantum battery is defined as $P_i(t)=\frac{d}{dt}\langle \hat{H}_B\rangle = i\langle[\hat{H},\hat{H}_B]\rangle$,  (Here $H_B/H_B/H \equiv H_B^{\rm (I)}/H_C^{\rm (I)}/H^{\rm (I)}$ or $H_B^{\rm (II)}/H_C^{\rm (II)}/H^{{\rm (II)}}$) where the total Hamiltonian is taken as $\hat{H}=\hat{H}_B+\hat{H}_C$. Since $[\hat{H}_B,\hat{H}_B]=0$, the power is governed entirely by the commutator $P_i(t)=i\langle[\hat{H}_C,\hat{H}_B]\rangle$. Thus, the instantaneous power originates purely from the noncommutativity between the battery Hamiltonian and the chargoid Hamiltonian. In the case of a noninteracting battery with an interacting chargoid, the interacting nature of $\hat{H}_C$ ensures that $[\hat{H}_C,\hat{H}_B]\neq 0$, leading to a nontrivial power profile. Conversely, when the battery is interacting and the chargoid is noninteracting, the structure of $\hat{H}_B$ now plays the dominant role in producing a similar nonvanishing commutator. In both situations, the source of power is effectively the same mathematical object, the commutator between the two subsystems, resulting in qualitatively similar early-time behaviour of the instantaneous power. The differences between the two scenarios are expected to manifest only at later times, where the interacting Hamiltonian induces many-body correlations and enhanced entanglement dynamics in the system.

\begin{figure*}
 \includegraphics[width=.45\linewidth, height=.30\linewidth]{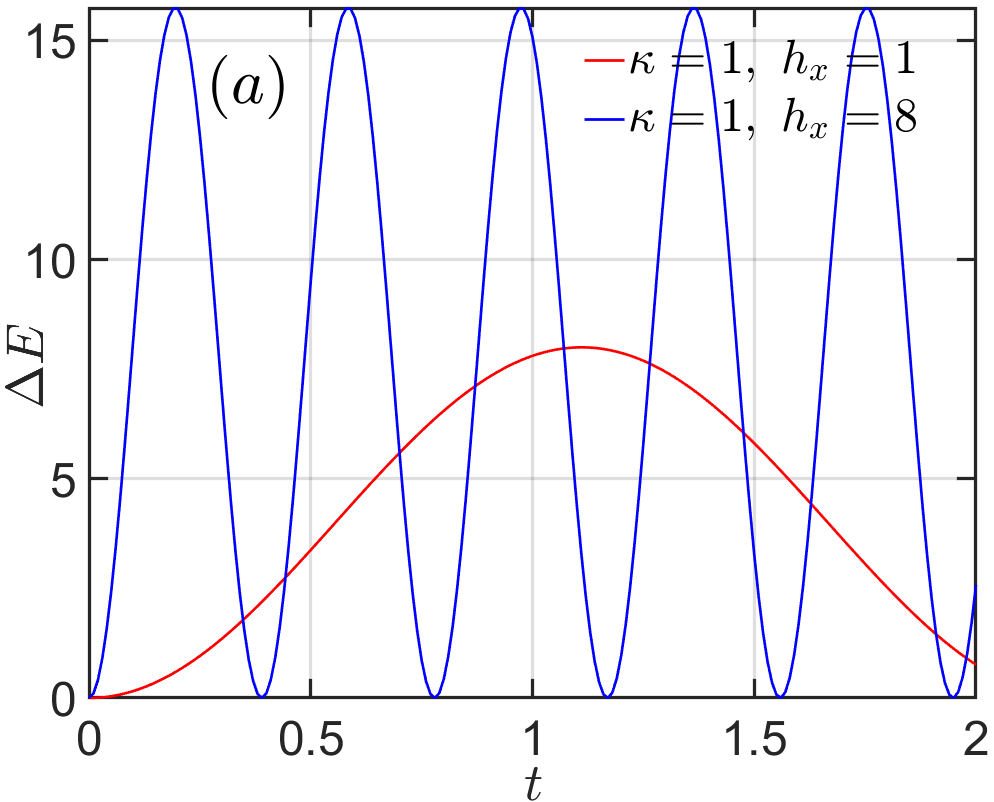}
 \includegraphics[width=.45\linewidth, height=.30\linewidth]{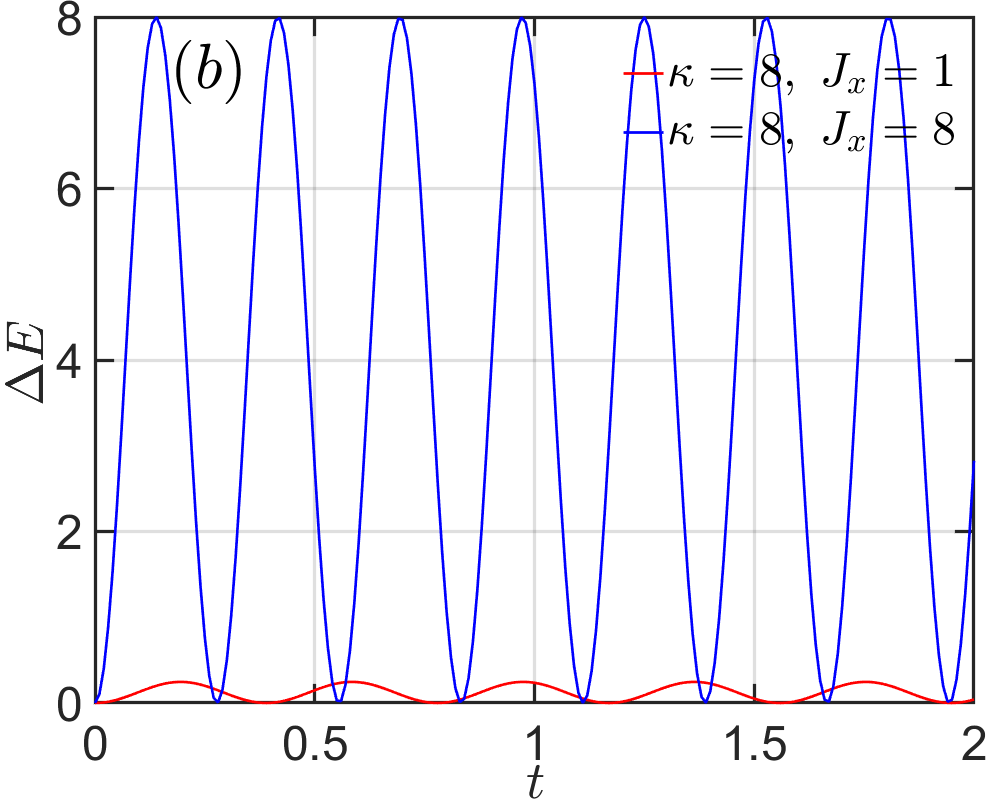}
\caption{
\textbf{Unfair charging.}
(\textit{a}) \textit{Parallel charging:} Time evolution of the stored energy $\Delta E$
for a noninteracting battery, where each spin
is charged independently by a local transverse field of strength
$h_x=1$ and $h_x=N$ (with all spin-spin interactions set to zero). (\textit{b}) \textit{Global charging:} Time evolution of the stored energy $\Delta E$
for the same noninteracting battery, charged by an all-to-all interacting spin
chargoid with interaction strengths $J_x=1$ and $J_x=N$, in the absence of any external
transverse field. $N=8$ (open boundary conditions), with $h_z=1$ fixing the battery energy scale.
}
\label{Par_E}
  \end{figure*}

\subsection*{Charging Dynamics under $\kappa$-Local Interactions: Role of Interaction Structure}
\label{k_locality}

Understanding the role of interaction structure in quantum battery charging is central to identifying genuine quantum advantages. Early theoretical studies have demonstrated that collective interactions can enhance charging power compared to parallel protocols \cite{campaioli2017enhancing}. However, it was later clarified that such advantages may arise trivially from increasing the overall energy scale of the chargoid Hamiltonian rather than from intrinsically quantum effects \cite{gyhm2022quantum}. Therefore, a meaningful comparison between different charging protocols requires fixing the operator norm of the chargoid Hamiltonian, allowing the effects of interaction-induced many-body dynamics to be isolated.

Motivated by this consideration, we investigate the role of the interaction order $\kappa$, defined as the number of spins simultaneously coupled in the chargoid Hamiltonian, on the energy storage, instantaneous charging power, and quantum correlations under both unconstrained and norm-constrained charging conditions. Throughout this analysis, we consider Protocol ${\rm I}$, in which the battery remains noninteracting [Eq.~\ref{Battery_Ham}], while the chargoid consists of $\kappa$-local interacting terms [Eq.~\ref{Charger_Ham}] with $J_y=J_z=0$. We fix the system size to $N=8$ and examine how varying the interaction order modifies the charging performance and the underlying quantum properties.

\subsubsection*{Unconstrained Charging: Classical Scaling Effects}

We first analyze the charging dynamics without imposing any constraint on the operator norm of the chargoid Hamiltonian $\|\hat H_C^{\rm (I)}\|$. This setting isolates classical contributions to energy storage arising purely from an increased energy scale of the chargoid, rather than from inherently quantum effects such as entanglement or multipartite correlations.

  \begin{figure}
  \includegraphics[width=.90\linewidth, height=.60\linewidth]{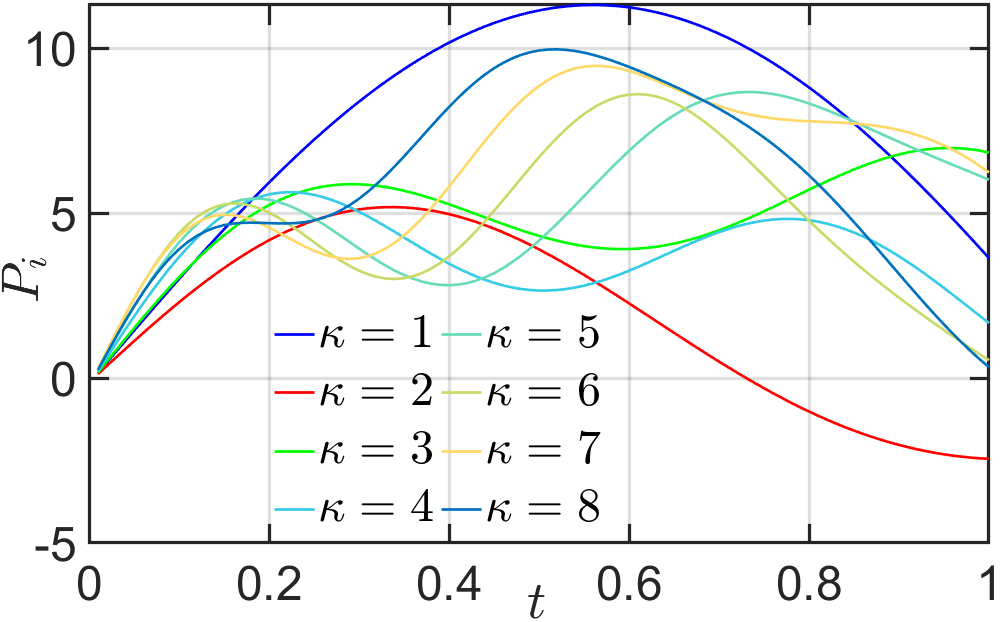}
\caption{
\textbf{Fair charging with $\boldsymbol{h_x \neq 0}$ and $\boldsymbol{J_x = h_x}$.}
Instantaneous power $P_i$ as a function of time $t$ for a noninteracting battery with $h_z = 1$, charged by a $\kappa$-local interacting chargoid in the presence of an additional transverse field of strength $h_x$. The chargoid Hamiltonian is normalized such that $\lVert \hat{H}_C^{\rm (I)} \rVert = N (=8)$ by tuning the parameters $J_x$ and $h_x$ under the constraint $J_x = h_x$. The system size is $N = 8$.
}
\label{Pi_k1_N}
  \end{figure}
 \begin{figure*}
 \includegraphics[width=.45\linewidth, height=.30\linewidth]{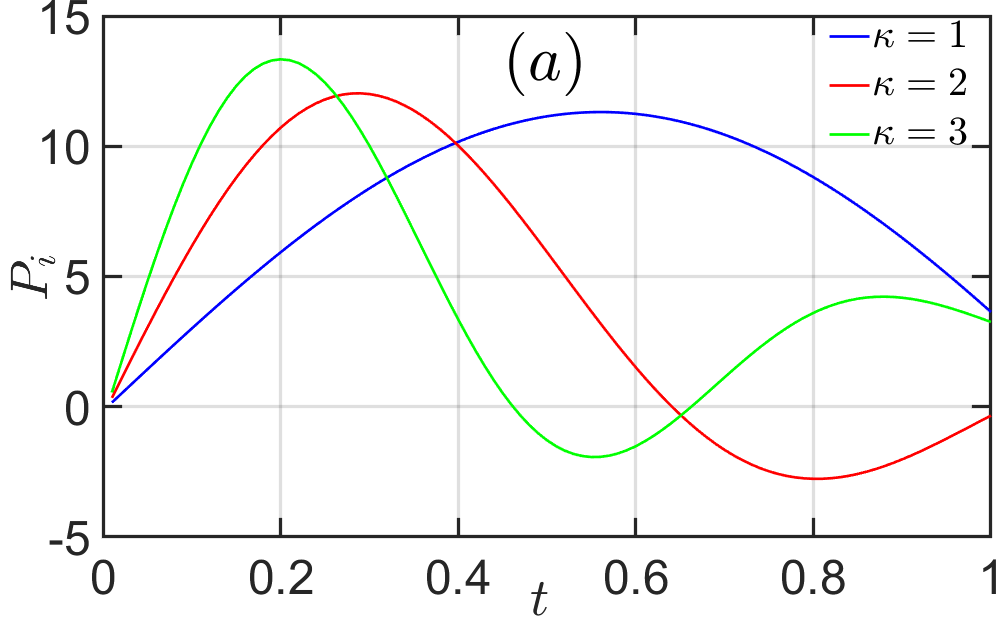}
  \includegraphics[width=.45\linewidth, height=.30\linewidth]{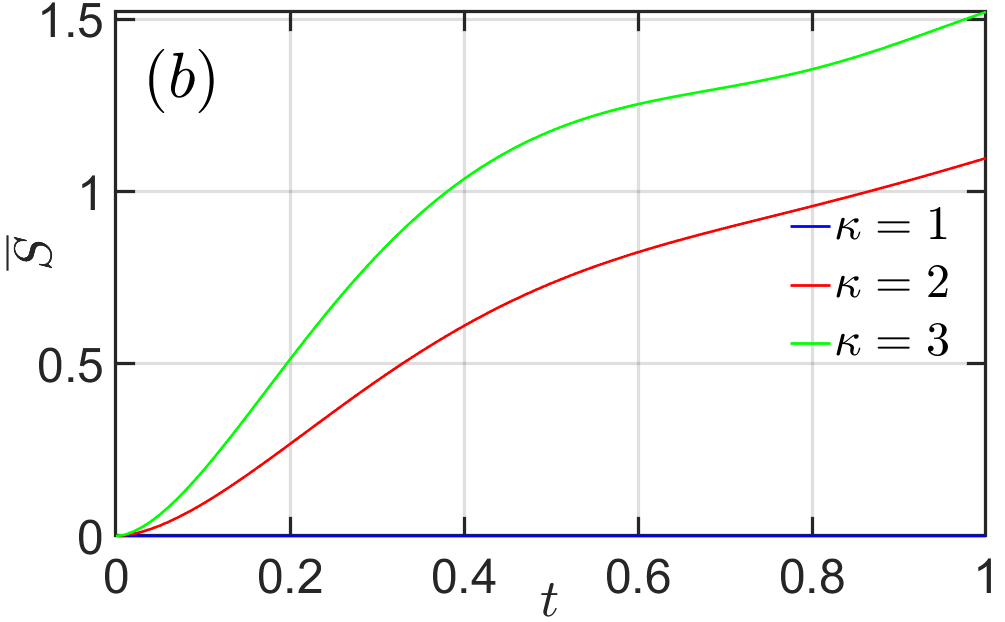}
\caption{
\textbf{Fair charging with $\boldsymbol{h_x \neq 0}$ and $\boldsymbol{J_x > h_x}$.}
A noninteracting battery with $h_z = 1$ is charged by a $\kappa = 1,2,3$-local interacting chargoid in the presence of an additional transverse field of strength $h_x$. The chargoid Hamiltonian is normalized such that $\lVert \hat{H}_C^{\rm (I)} \rVert = N (=8)$ by tuning the parameters under the constraint $J_x > h_x$. Panels (a) and (b) show, respectively, the instantaneous power $P_i$ and the ABEE $\overline{S}$ as functions of time $t$.
}
\label{Pi_k1_k2_k3}
  \end{figure*}
%\paragraph*{Parallel Charging}  
To demonstrate this classical advantage, we first examine the simplest parallel charging scenario, where both the battery and the chargoid are composed of noninteracting spins. In this case, the battery Hamiltonian is given by $H_B^{\rm (I)}=h_z\sum_{j=1}^N\sigma_j^z$, with $h_z=1$, while the chargoid Hamiltonian is described by $H_C^{\rm (I)}=h_x\sum_{j=1}^N\sigma_j^x$, where $h_x$ denotes the transverse field strength. This corresponds to Protocol ${\rm I}$ in the absence of interactions, i.e., $J_x=J_y=J_z=0$. The operator norm of the chargoid scales with the applied field: for $h_x = h_z$, we have $\lVert \hat H_C^{\rm (I)} \rVert = Nh_z$, while for $h_x = Nh_z$, it increases as $\lVert \hat H_C^{\rm (I)} \rVert = N^2h_z^2$. As a consequence, the charging strength grows with $h_x$. Under $h_x = h_z$, the average stored energy reaches only half of the battery's maximum capacity, $\Delta E_\mathrm{max} = N h_z$, because the battery Hamiltonian partially counteracts the spin flips induced by the chargoid \cite{shukla2025optimizing}. Increasing $h_x$ further allows the chargoid to dominate the dynamics, coherently flipping all spins and achieving maximal energy storage, $\Delta E_\mathrm{max} = 2N h_z$ at $h_x = N h_z$ [Fig.~\ref{Par_E}($a$)]. Next, we consider the opposite limit of a fully interacting chargoid, corresponding to $\kappa = N$, where all spins are coupled simultaneously through an all-to-all interaction in the $x$ direction ($J_y=J_z=0$) i.e. 
$H_C^{\rm (I)} = J_x \prod_{j=1}^{N} \sigma_j^x$. We vary the interaction strength $J_x$.  For $J_x = h_z$, the operator norm is small, $\lVert \hat H_C^{\rm (I)} \rVert =h_z(= 1)$, yielding modest maximum energy storage, $\Delta E_\mathrm{max} \approx 0.625$. By increasing the interaction to $J_x = N h_z$, the operator norm rises to $\lVert \hat H_C^{\rm (I)} \rVert = Nh_z$, resulting in $\Delta E_\mathrm{max} = N h_z $, which matches the maximal stored energy achieved in the parallel scenario with $h_x = h_z$ [Fig.~\ref{Par_E}($b$)]. This comparison demonstrates that augmenting the chargoid strength either via the transverse field in parallel charging or via interaction strength in global charging produces increases in stored energy through classical effects alone.  Therefore, the enhanced charging performance observed in both cases should not be interpreted as a genuine cost--benefit advantage. Instead, it arises solely from increasing the overall energy scale of the driving (chargoid) Hamiltonian, rather than from a more efficient utilization of the available resources. In other words, the improvement in stored energy is achieved by supplying a larger energetic input, not by exploiting interaction-induced many-body effects. Consequently, such an enhancement does not represent a genuinely improved charging protocol and offers limited practical significance when the available charging resources are fixed.

The equivalence of these two protocols under unconstrained norms emphasizes that the observed enhancements in the \emph{unfair charging scenario} are predominantly classical. Consequently, to assess genuine quantum advantages in charging performance, it is necessary to normalize the operator norm of the chargoid across different configurations. This \emph{fair charging condition} ensures that any observed improvement in instantaneous power or energy transfer can be attributed to collective quantum effects, rather than trivial scaling of the chargoid Hamiltonian.

\subsubsection*{Fair Charging with Flipping Term}
To isolate genuine quantum effects in charging dynamics, we impose the \textit{fair charging condition} by fixing the norm of the chargoid Hamiltonian, $\lVert \hat H_C^{\rm (I)} \rVert = N$. This removes any classical advantage arising purely from differences in energy scale. Under this constraint, we study the role of the \textit{interaction order}, $\kappa$, which defines the number of spins simultaneously coupled in the chargoid Hamiltonian, on the instantaneous power $P_i$. We emphasize that fixing the operator norm provides a practical and operational criterion for eliminating trivial enhancements arising solely from changes in the overall energy scale, thereby enabling a meaningful comparison between different charging protocols. While this is not the only possible notion of fairness, alternative criteria based on locality, connectivity, or interaction complexity depend on the specific physical setting and the resources under consideration. In this work, we therefore adopt the fixed-norm condition as a natural benchmark to isolate the impact of the interaction order on the charging dynamics.

We impose the fair charging condition by fixing the chargoid Hamiltonian norm through the choice $J_x=h_x$. Starting from $\kappa=1$, which corresponds to parallel charging with noninteracting spins, we systematically increase the interaction order to include multi-spin couplings ($\kappa=2,3,\ldots,N$). We observe that [see Fig.~\ref{Pi_k1_N}] the instantaneous power $P_i$ initially increases with $\kappa$, reaches a maximum, and subsequently decreases. Importantly, under the fixed-norm constraint, increasing the interaction order alone does not enhance the maximum instantaneous power beyond the parallel charging case [Fig.~\ref{Pi_k1_N}]. This demonstrates that the presence of multi-spin interactions or correlations by themselves is insufficient to generate a power advantage. This result is consistent with the unconstrained case, where the observed enhancement primarily originates from an increase in the overall energy scale of the chargoid Hamiltonian rather than from the interaction structure itself.

A different behavior emerges when the interaction strength $J_x$ becomes larger than the flipping term $h_x$. In this regime, $P_i$ reaches its peak more quickly, especially for larger $\kappa$ values [Fig.~\ref{Pi_k1_k2_k3}($a$)], demonstrating that \textit{cooperative multi-spin interactions} can enhance charging power.  To further characterize the dynamics, we compare the instantaneous power $P_i$ with the ABEE. Interestingly, ABEE remains small at the time when $P_i$ reaches its maximum and increases only at later times [Fig.~\ref{Pi_k1_k2_k3}($b$)]. This indicates a temporal separation between the peak of energy transfer and the growth of ABEE.

Under fair charging conditions, simply adding an external flipping term $h_x$ does not increase instantaneous power. Genuine enhancement occurs only when the interaction strength $J_x$ dominates, allowing cooperative dynamics among multiple spins to produce a measurable advantage. This highlights the critical role of strong interactions and system complexity in achieving {\it quantum advantages} in energy transfer.

\begin{figure*}
 \includegraphics[width=.32\linewidth, height=.23\linewidth]{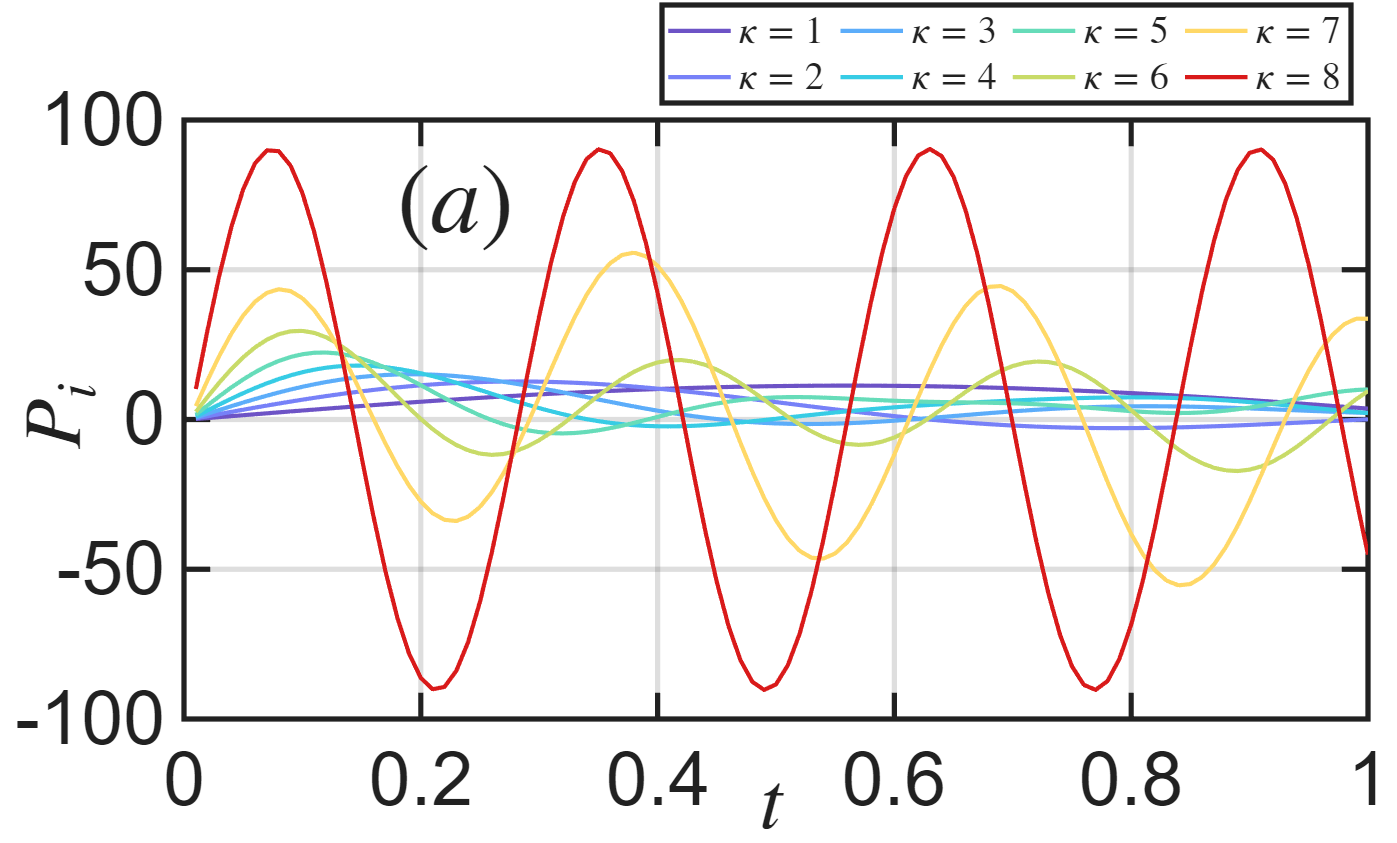}
\includegraphics[width=.32\linewidth, height=.23\linewidth]{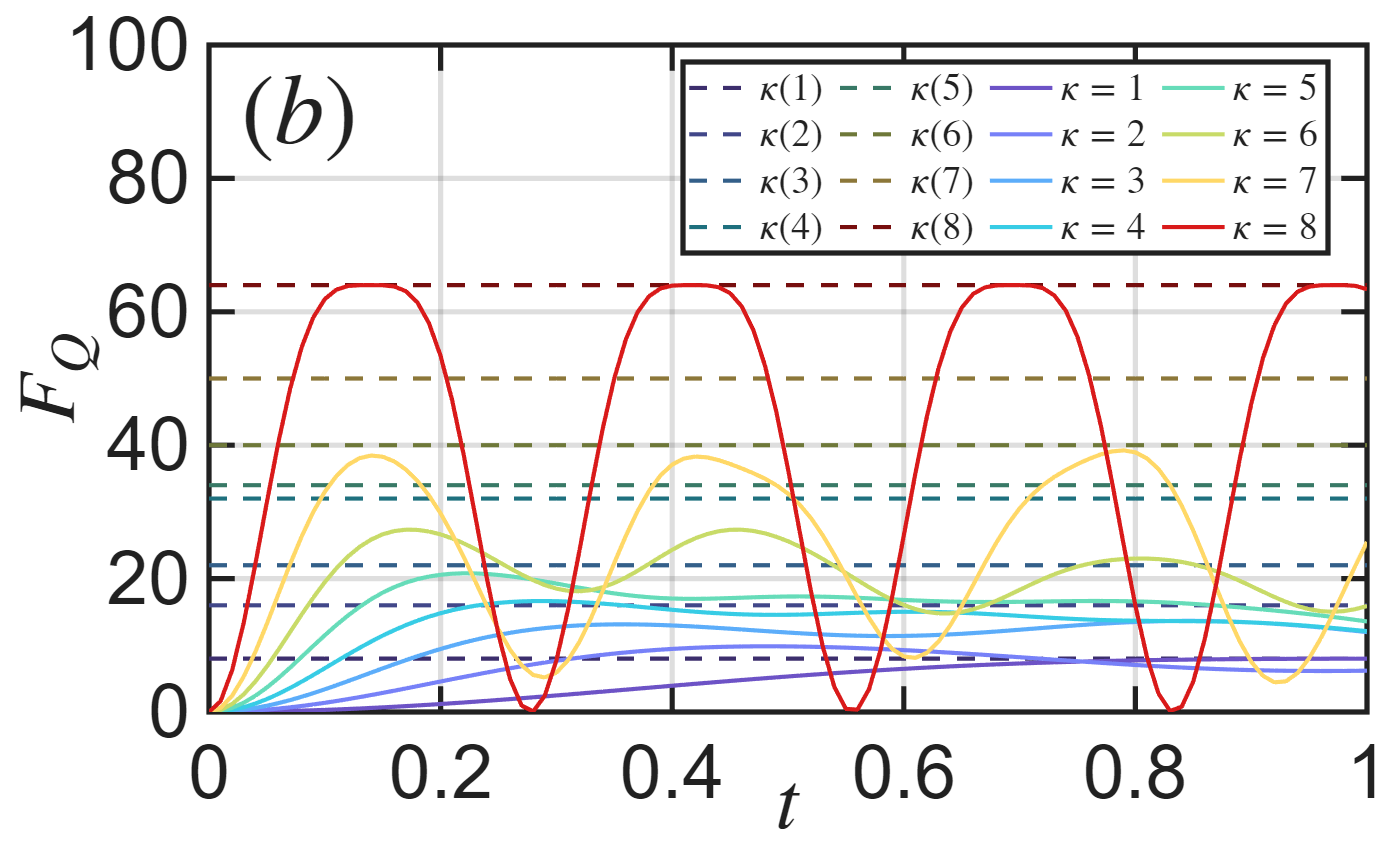}
\includegraphics[width=.32\linewidth, height=.23\linewidth]{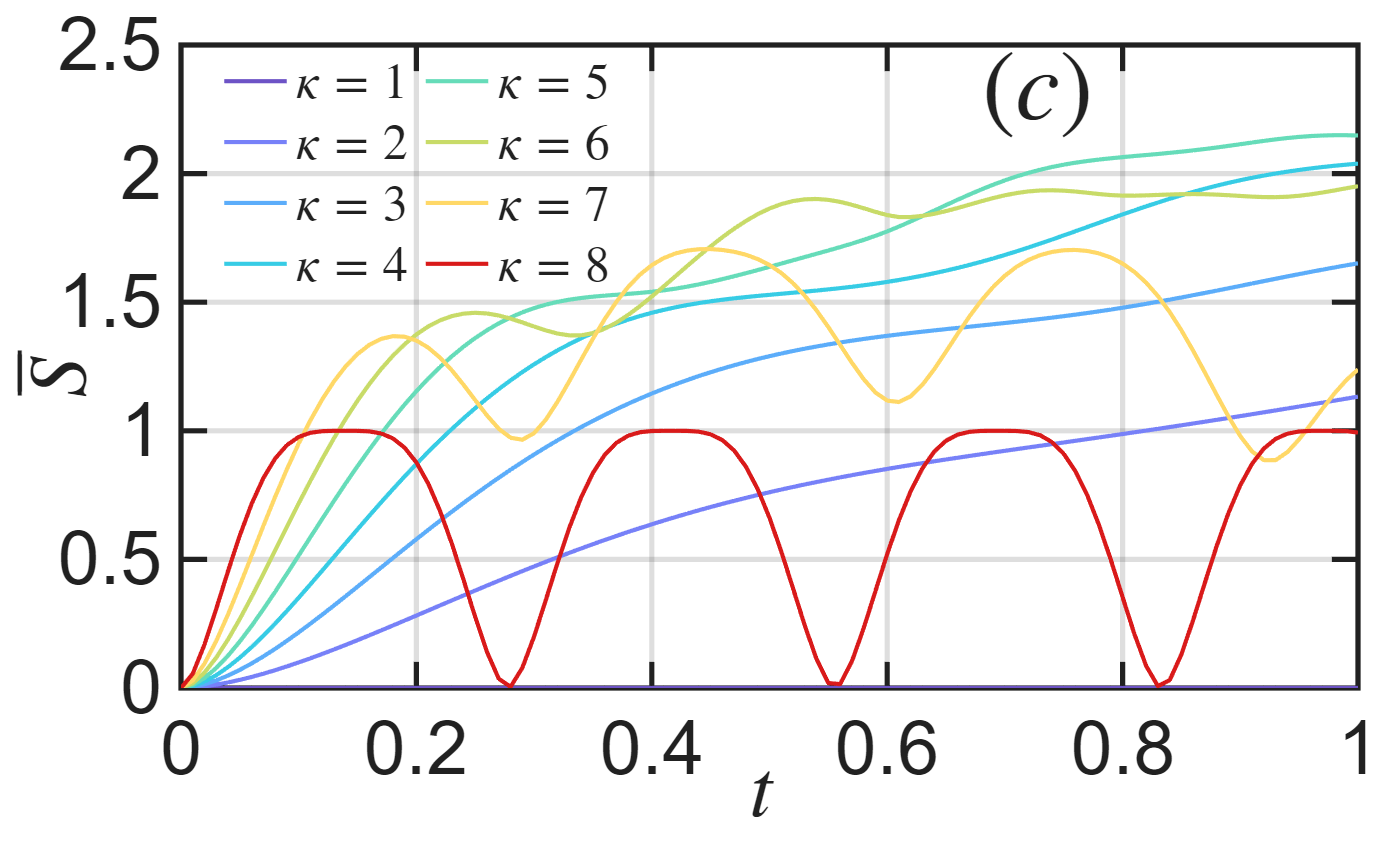}
\caption{
\textbf{Fair charging with $\boldsymbol{h_x = 0}$.} 
Noninteracting battery with $h_z=1$, charged by a $\kappa$-local interacting chargoid ($\kappa=1$ to $N$) in the absence of a transverse field ($h_x=0$). 
The chargoid Hamiltonian is normalized as $\lVert \hat H_C^{\rm (I)} \rVert = N (=8)$ by tuning the interaction strength $J_x$. 
(a) Instantaneous power $P_i$, (b) QFI, and (c) ABEE $\overline{S}$ as functions of time $t$, for different participation numbers $\kappa$, as indicated in legend. In panel (b), the horizontal dashed lines correspond to the entanglement bounds $\kappa(x)$, with $\kappa(1), \kappa(2), \dots$ indicated in the legend.}

 \label{Pi_QFI_SA}
  \end{figure*}

\subsubsection*{Fair Charging without the Flipping Term}
To isolate the role of spin-spin interactions in the charging process, we eliminate single-spin rotations by setting the external flipping field to zero, $h_x=0$, in the charging Hamiltonian. This allows us to focus exclusively on interaction-driven dynamics and assess whether collective many-body effects alone can enhance charging performance. The Hamiltonian is normalized such that $\lVert \hat{H}_C^{\rm (I)} \rVert = N$ for all interaction ranges, ensuring fair charging conditions.

In this interaction-only regime, the charging dynamics differ qualitatively from cases where local fields are present. The instantaneous power $P_i$ increases systematically with the interaction order $\kappa$, indicating that extending the number of spins participating simultaneously in the interaction promotes cooperative energy transfer. The enhancement becomes most pronounced in the global interaction limit $\kappa=N$, where all spins are collectively coupled.  In this case, the maximum instantaneous power $P_i^{\mathrm{max}}$ reaches nearly $N$ times the value obtained under parallel charging and exhibits a periodic behavior [Fig.~\ref{Pi_QFI_SA}$(a)$] . This demonstrates that highly efficient charging can be achieved solely through collective interactions, even in the absence of local spin-flip processes.

To gain insight into the underlying quantum correlations, we analyze the QFI [Fig.~\ref{Pi_QFI_SA}(b)] and the ABEE [Fig.~\ref{Pi_QFI_SA}(c)] for all values of $\kappa$ while keeping the chargoid norm fixed, $\lVert \hat{H}_C^{\rm (I)} \rVert = N$. Both quantities increase with the interaction range and exhibit similar qualitative behavior [Figs.~\ref{Pi_QFI_SA}(b) and \ref{Pi_QFI_SA}(c)], indicating the development of stronger multipartite quantum correlations as the interactions become more global. For all interaction orders, the maxima of multipartite entanglement are reached after the instantaneous power reaches its peak, showing that the generation of strong quantum correlations is delayed relative to the optimal charging rate. In the fully interacting case, $\kappa=N$, similar to instantaneous power, both the QFI and ABEE display periodic behavior, and the time separation between the maximum instantaneous power and the maximum multipartite entanglement becomes smaller compared with lower interaction orders [Fig.~\ref{Pi_QFI_SA}$(b,c)$]. Nevertheless, even in this case, the peaks of the QFI and ABEE do not coincide with the maximum charging power, demonstrating that the strongest quantum correlations are not directly associated with the optimal charging speed.

\par
The QFI provides a measure of the number of particles effectively participating in entanglement. In Fig.~\ref{Pi_QFI_SA}(b), the horizontal lines $\kappa(x)$ indicate specific participation numbers, and each time the QFI curve (solid line) crosses a line $\kappa(x)$, it signifies that $x+1$ particles are entangled. As the interaction order $\kappa$ increases, more particles participate in entanglement. For $\kappa < N$, only subsets of spins contribute, and the number of entangled particles does not grow monotonically with $\kappa$. Consequently, the instantaneous power increases with interaction order but not proportionally. In the fully connected regime ($\kappa = N$), all spins participate, resulting in a significantly larger instantaneous power compared to the $\kappa < N$ case. 
\par

\begin{figure*}
 \includegraphics[width=.32\linewidth, height=.22\linewidth]{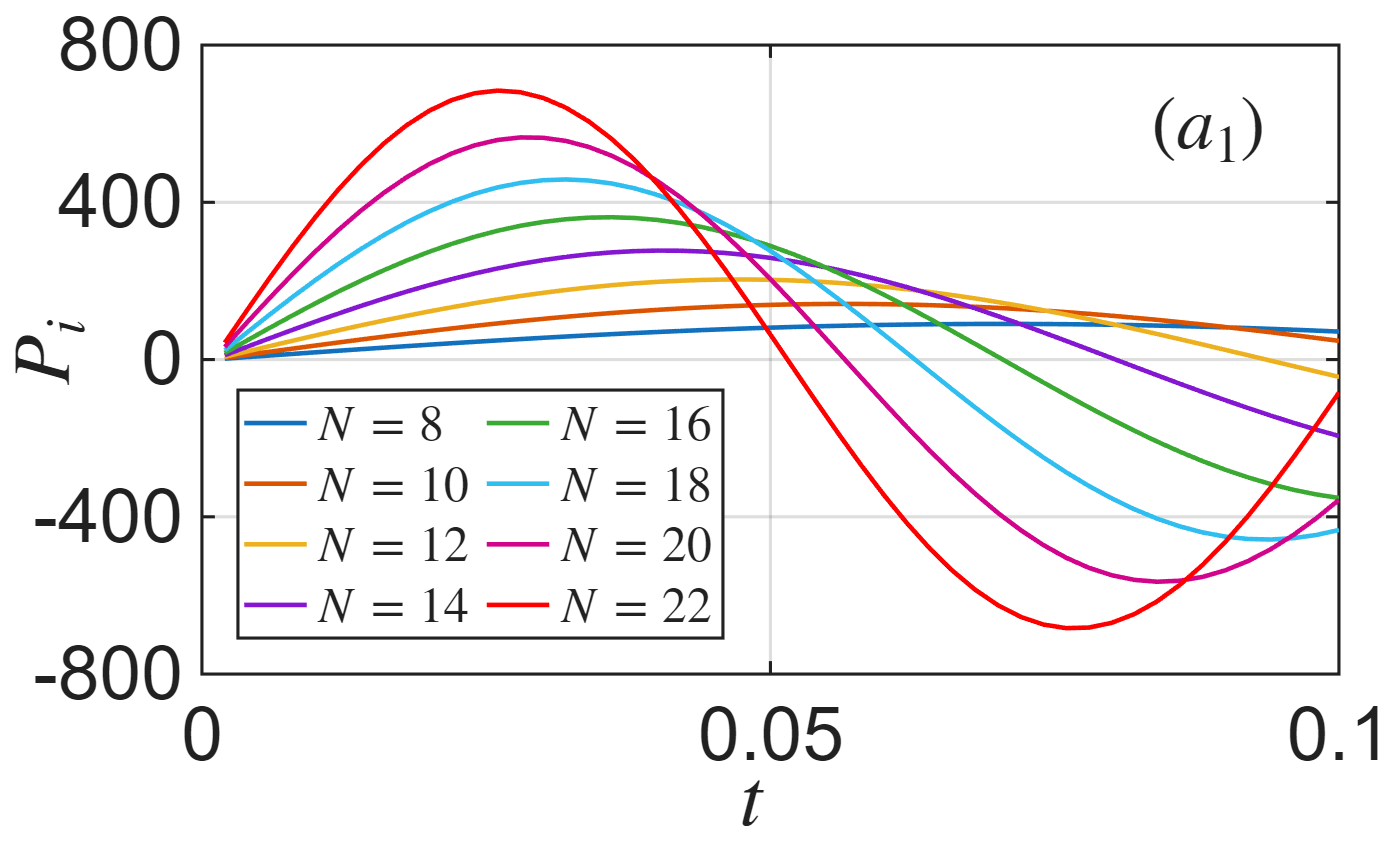}
 \includegraphics[width=.32\linewidth, height=.22\linewidth]{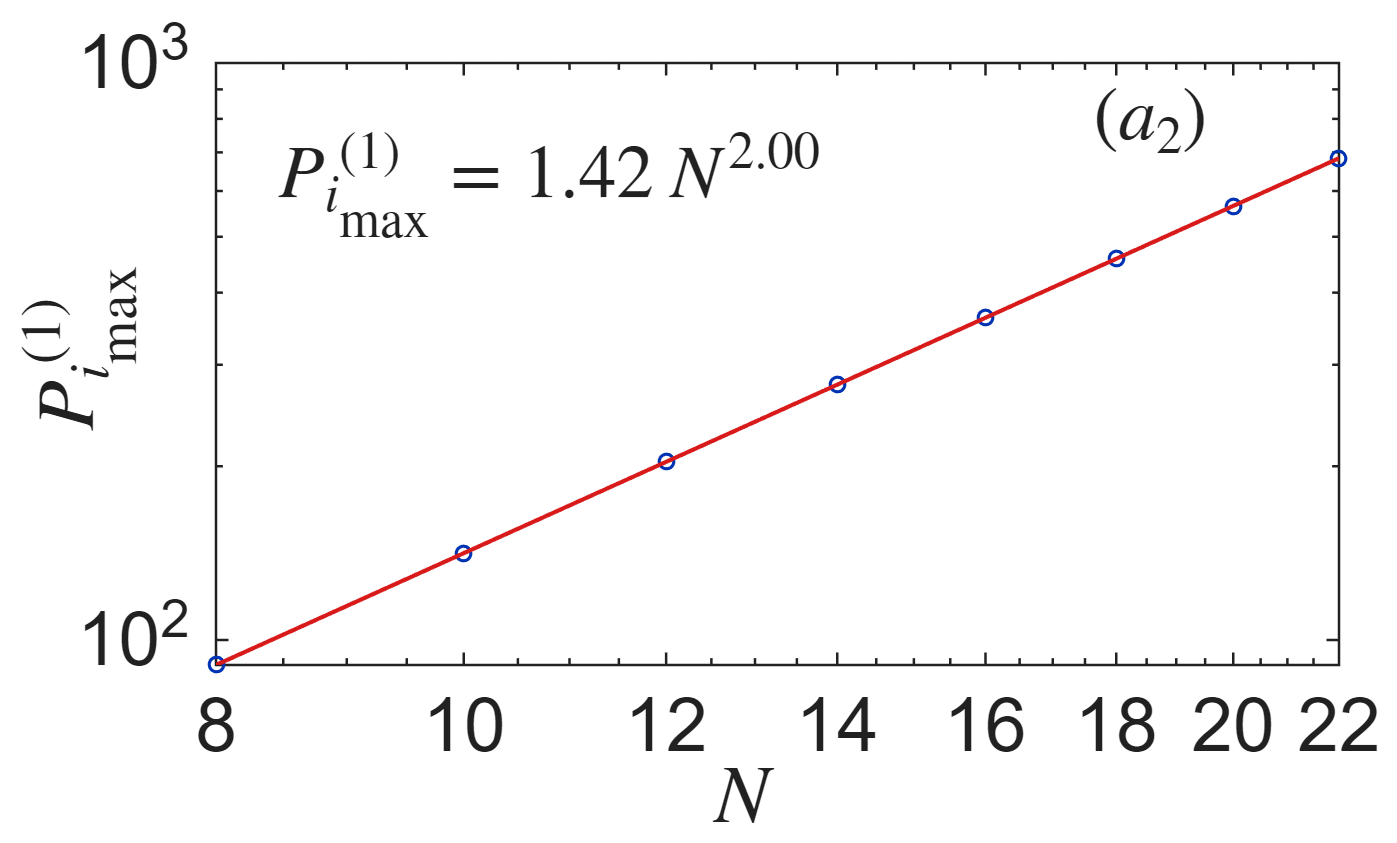}
 \includegraphics[width=.32\linewidth, height=.22\linewidth]{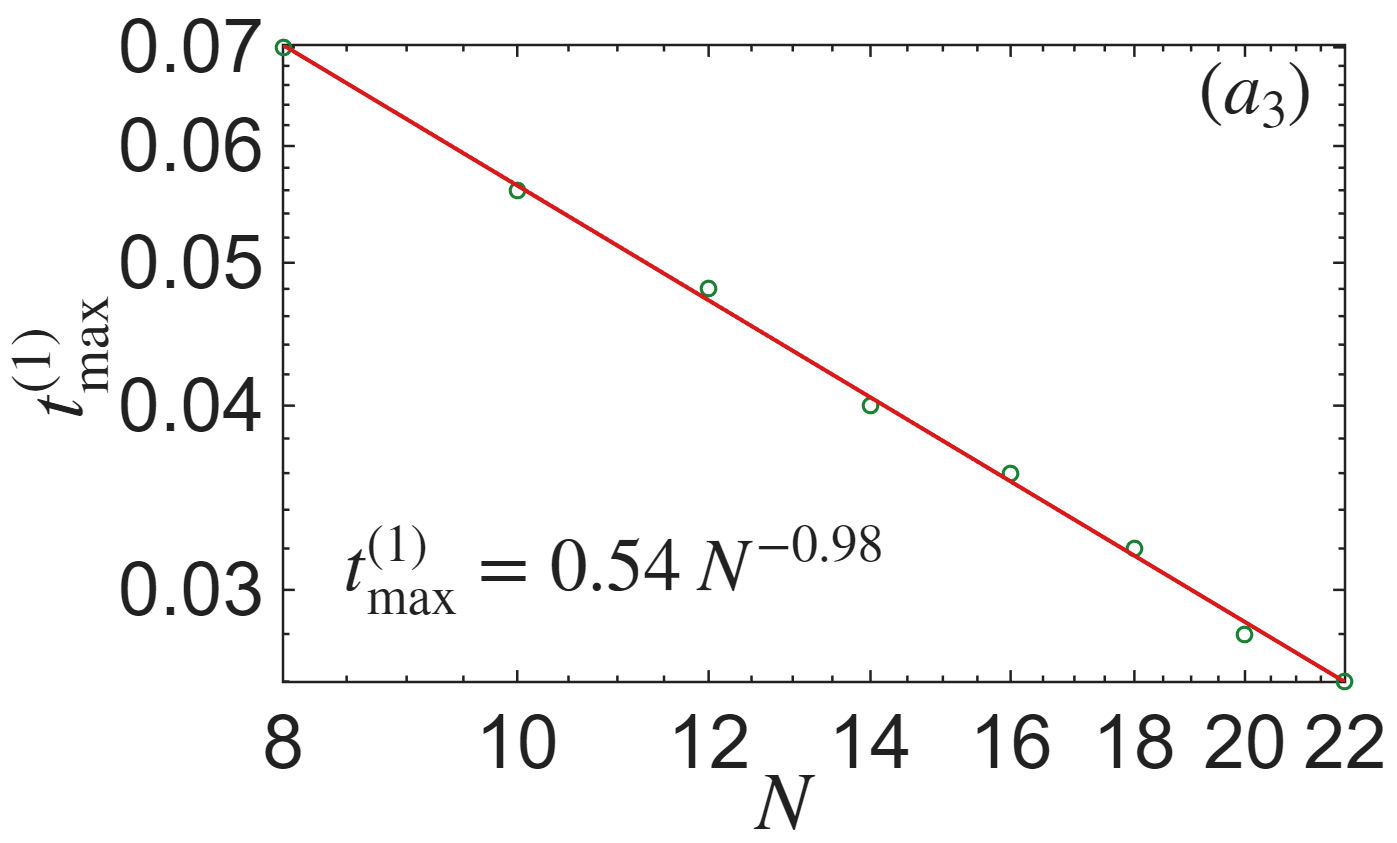}
 \includegraphics[width=.32\linewidth, height=.22\linewidth]{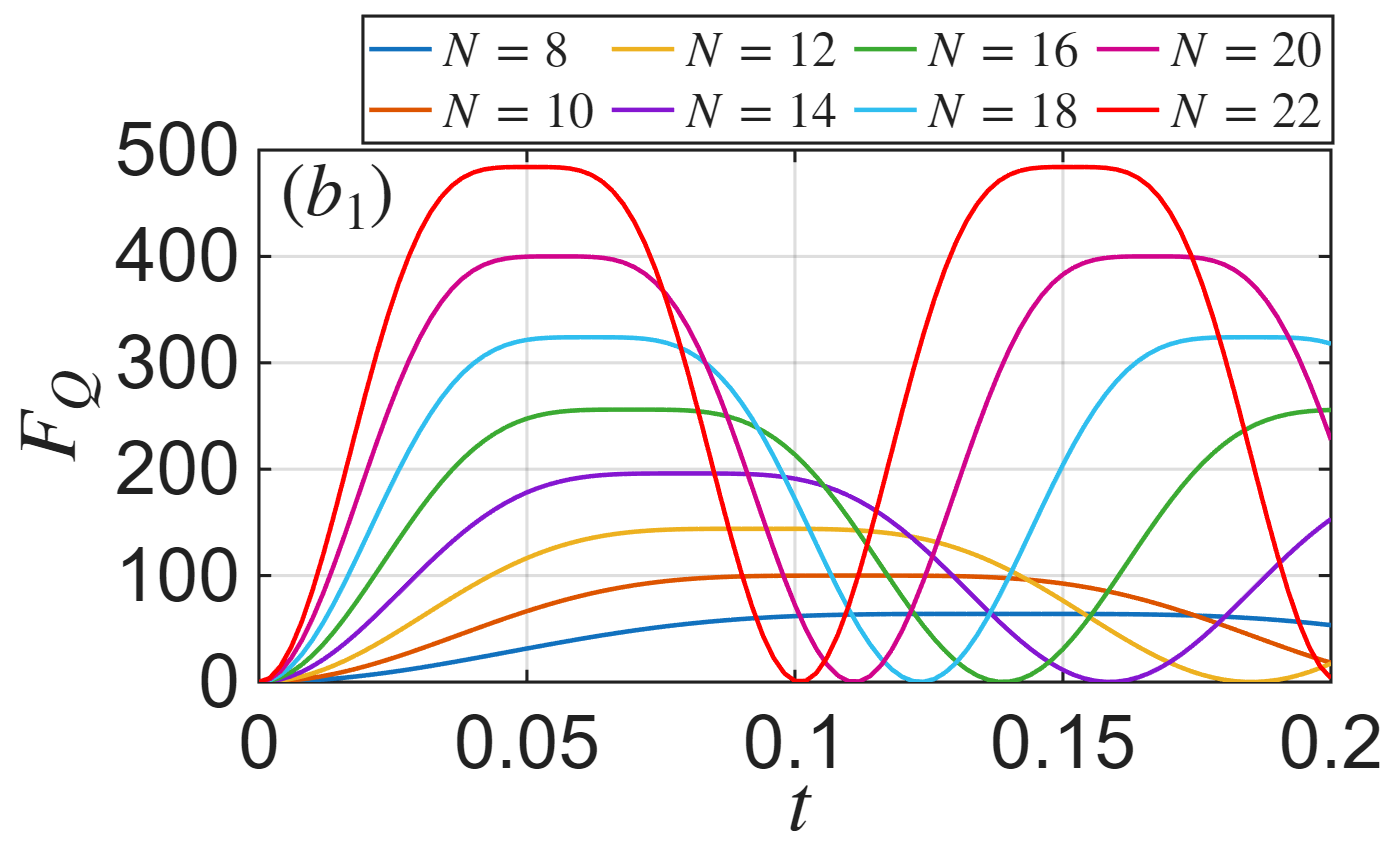}
\includegraphics[width=.32\linewidth, height=.22\linewidth]{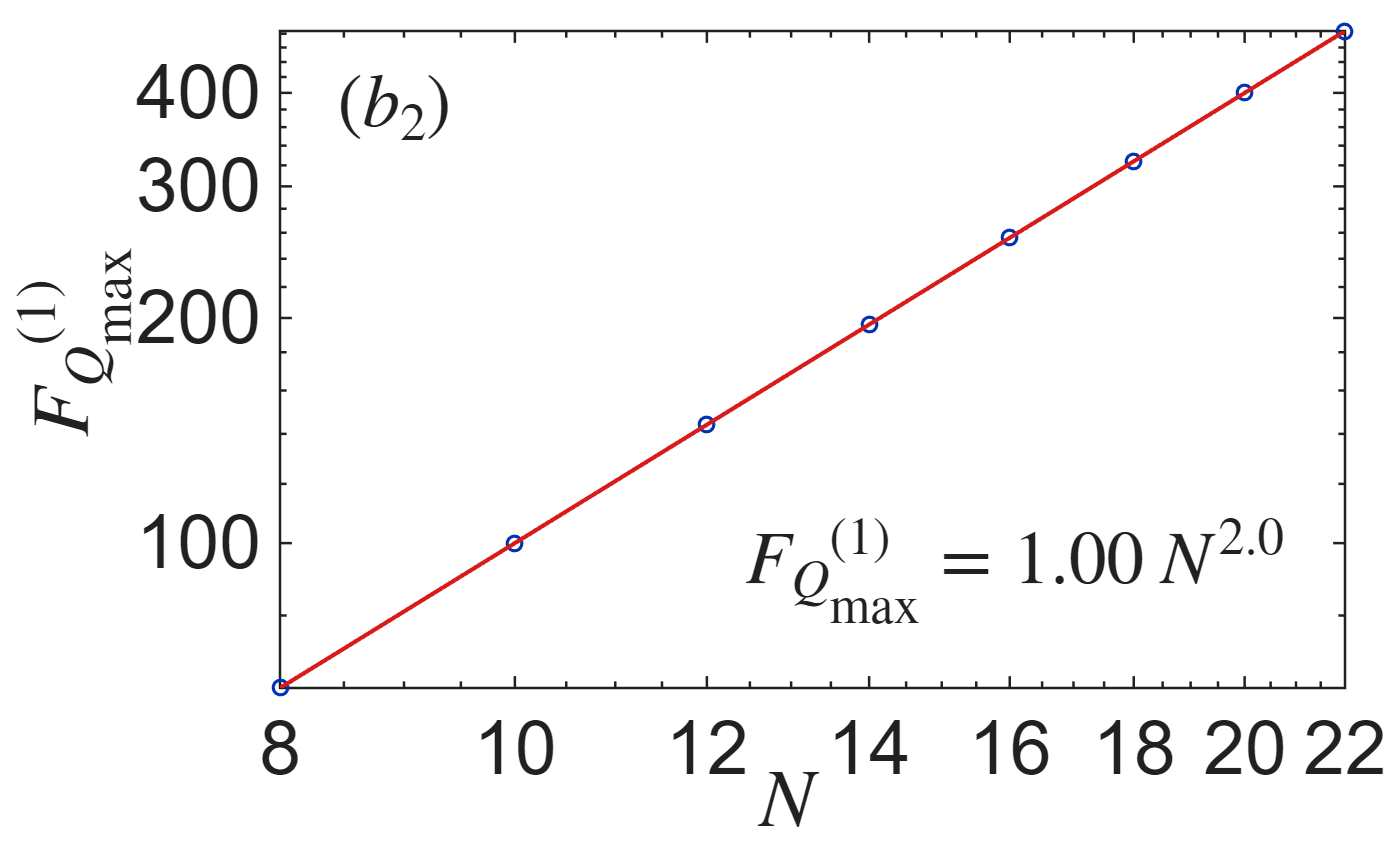}
\includegraphics[width=.32\linewidth, height=.22\linewidth]{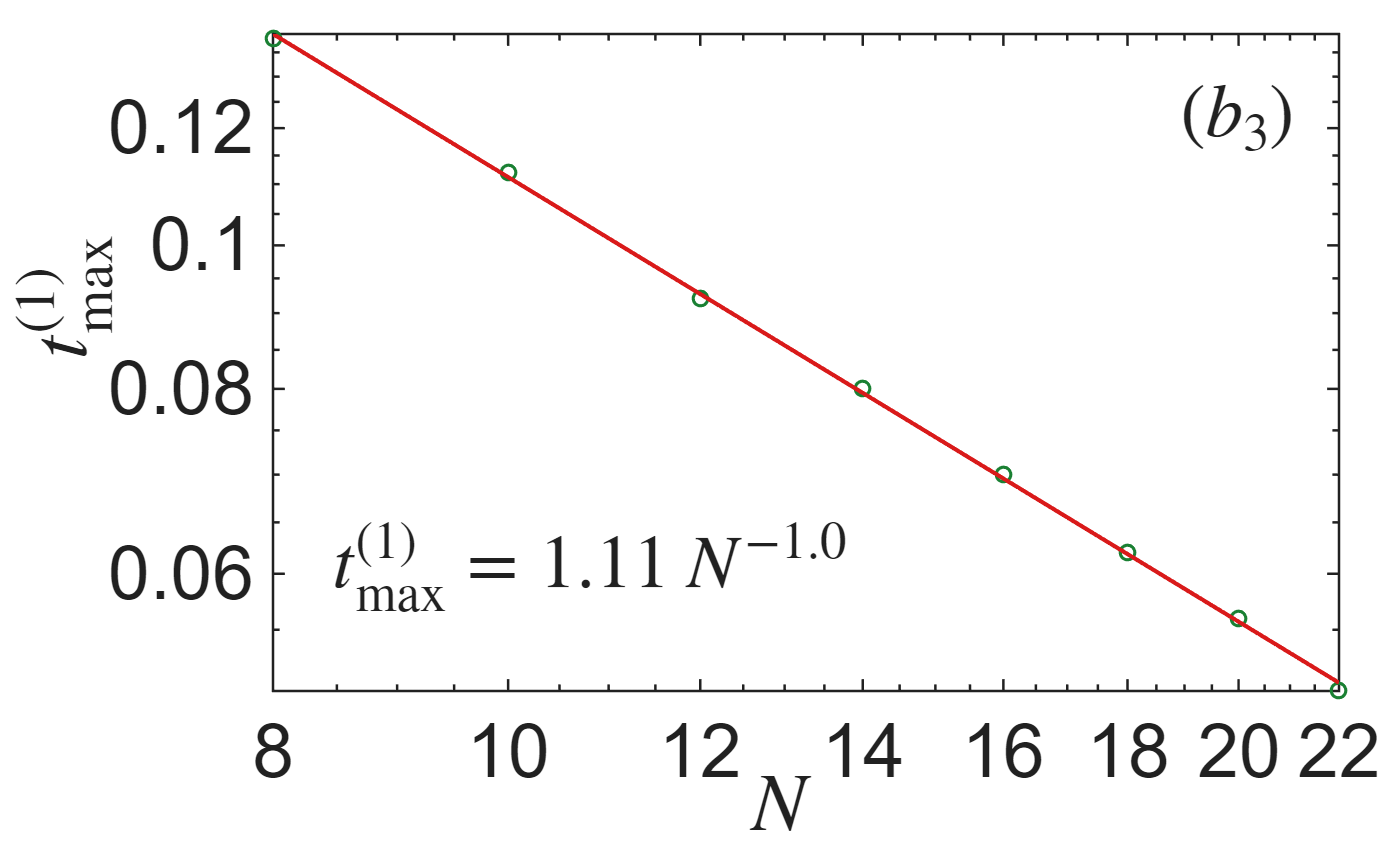}
 \includegraphics[width=.32\linewidth, height=.22\linewidth]{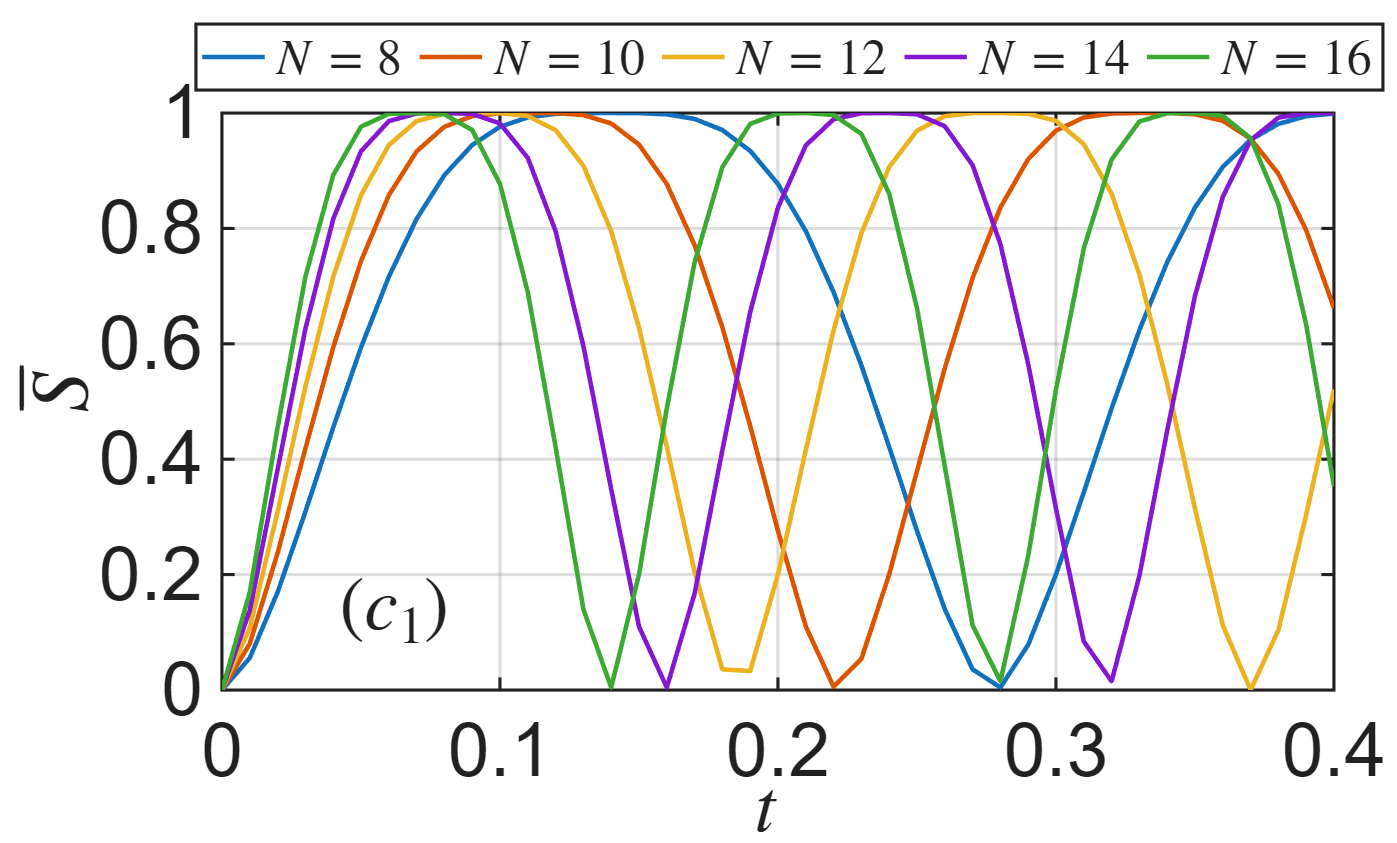}
\includegraphics[width=.32\linewidth, height=.22\linewidth]{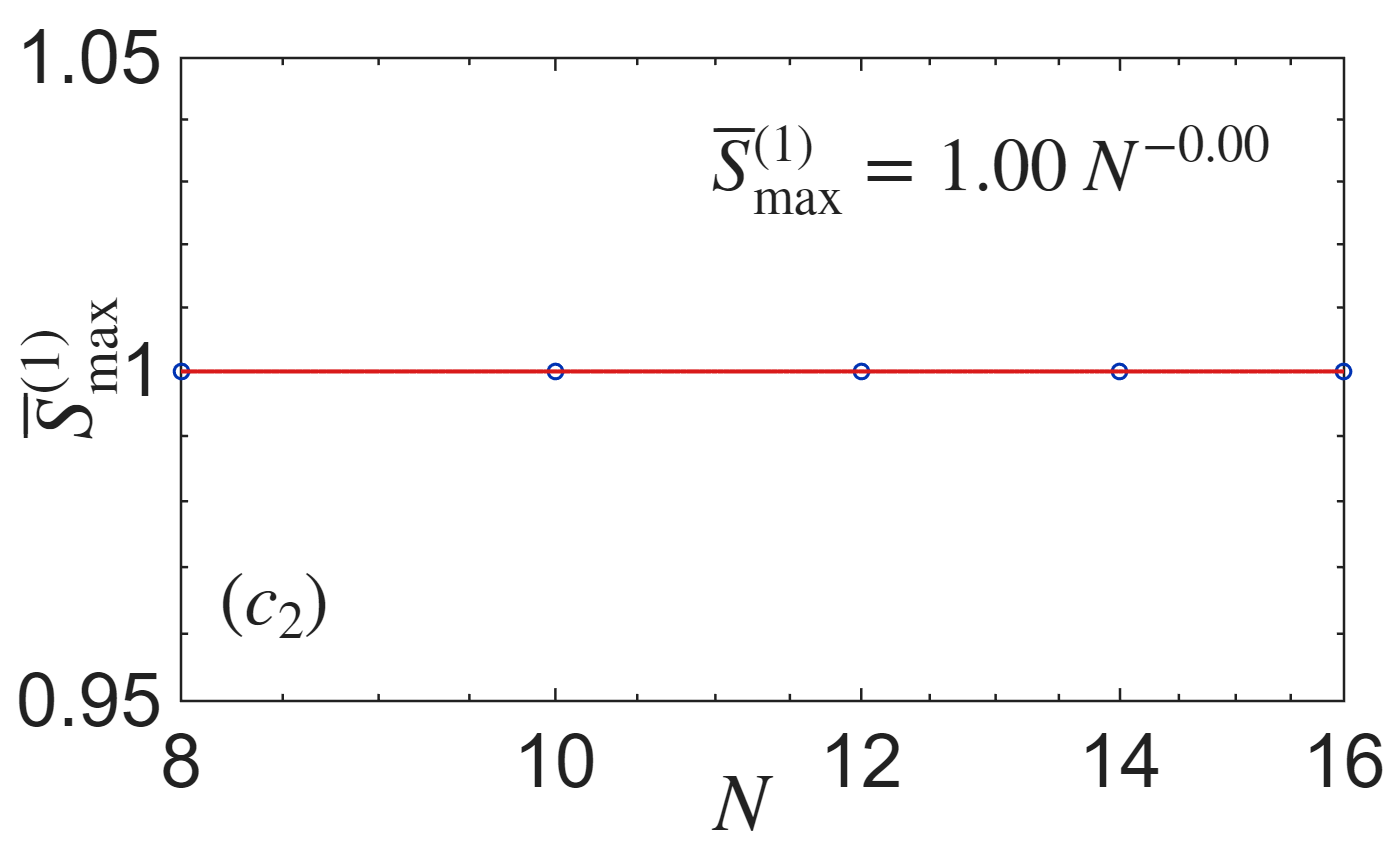}
\includegraphics[width=.32\linewidth, height=.22\linewidth]{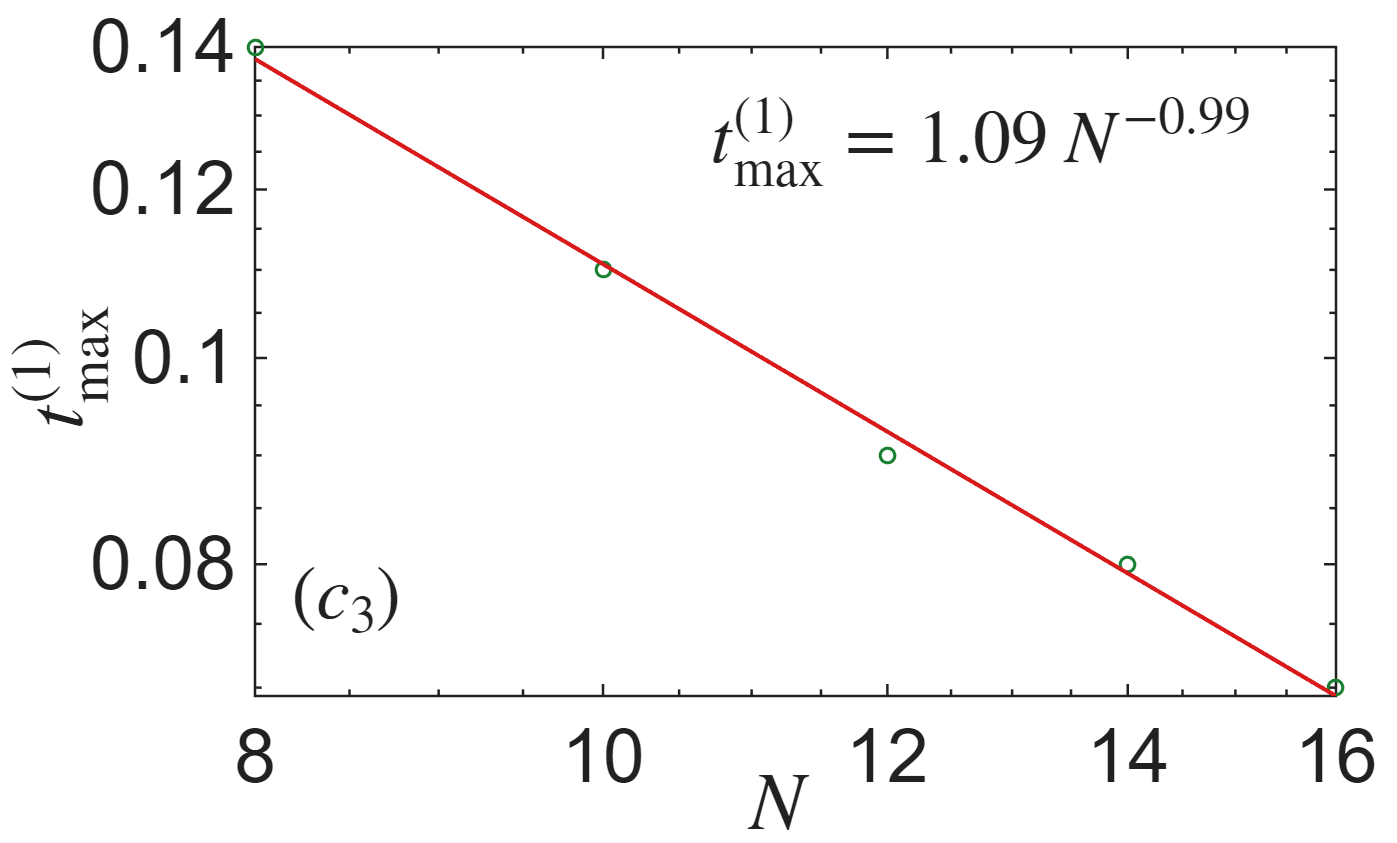}
\caption{
\textbf{System-size scaling for the $\boldsymbol{\kappa = N}$ protocol.}
A noninteracting battery with $h_z = 1$ is driven by a fully collective ($\kappa = N$) interacting chargoid in the absence of a transverse field ($h_x = 0$). The charging Hamiltonian is normalized such that $\lVert \hat{H}_C^{\rm (I)} \rVert = N$ by rescaling the interaction strength $J_x$. The system size is varied as $N = 8$--$22$. Panels ($a_1$)--($c_1$) show the time evolution of the instantaneous charging power $P_i$, QFI $F_Q$, and ABEE $\overline{S}$ for different system sizes. Panels ($a_2$)--($c_2$) display the corresponding first maxima of these quantities as functions of system size, while panels ($a_3$)--($c_3$) show the corresponding times required to reach these maxima.}
 \label{Pi_QFI_SA_N}
  \end{figure*}

\paragraph*{Finite-size scaling of the fully collective charging protocol ($\kappa=N$):} To assess the robustness of our conclusions beyond the $N=8$ case, we perform a finite-size analysis for the fully collective charging protocol ($\kappa=N$), considering system sizes in the range $N=8\text{--}22$, depending on computational feasibility. To ensure a fair comparison across different system sizes, we impose the fair-charging condition $|\hat{H}_C^{\rm (I)}|=N$ by appropriately rescaling the interaction strength $J_x$. The corresponding results are presented in Fig.~\ref{Pi_QFI_SA_N}.

As shown in Fig.~\ref{Pi_QFI_SA_N}($a_1$), the instantaneous charging power exhibits qualitatively similar dynamics for all system sizes, while its peak value increases systematically with increasing $N$. This demonstrates that the charging advantage associated with the fully collective ($\kappa=N$) protocol remains robust in larger systems. Furthermore, the time required to reach the maximum charging power decreases as the system size increases, indicating increasingly rapid charging dynamics [Fig.~\ref{Pi_QFI_SA_N}($a_1$)].

\begin{figure}
    \centering
\includegraphics[width=0.95\linewidth,height=0.65\linewidth]{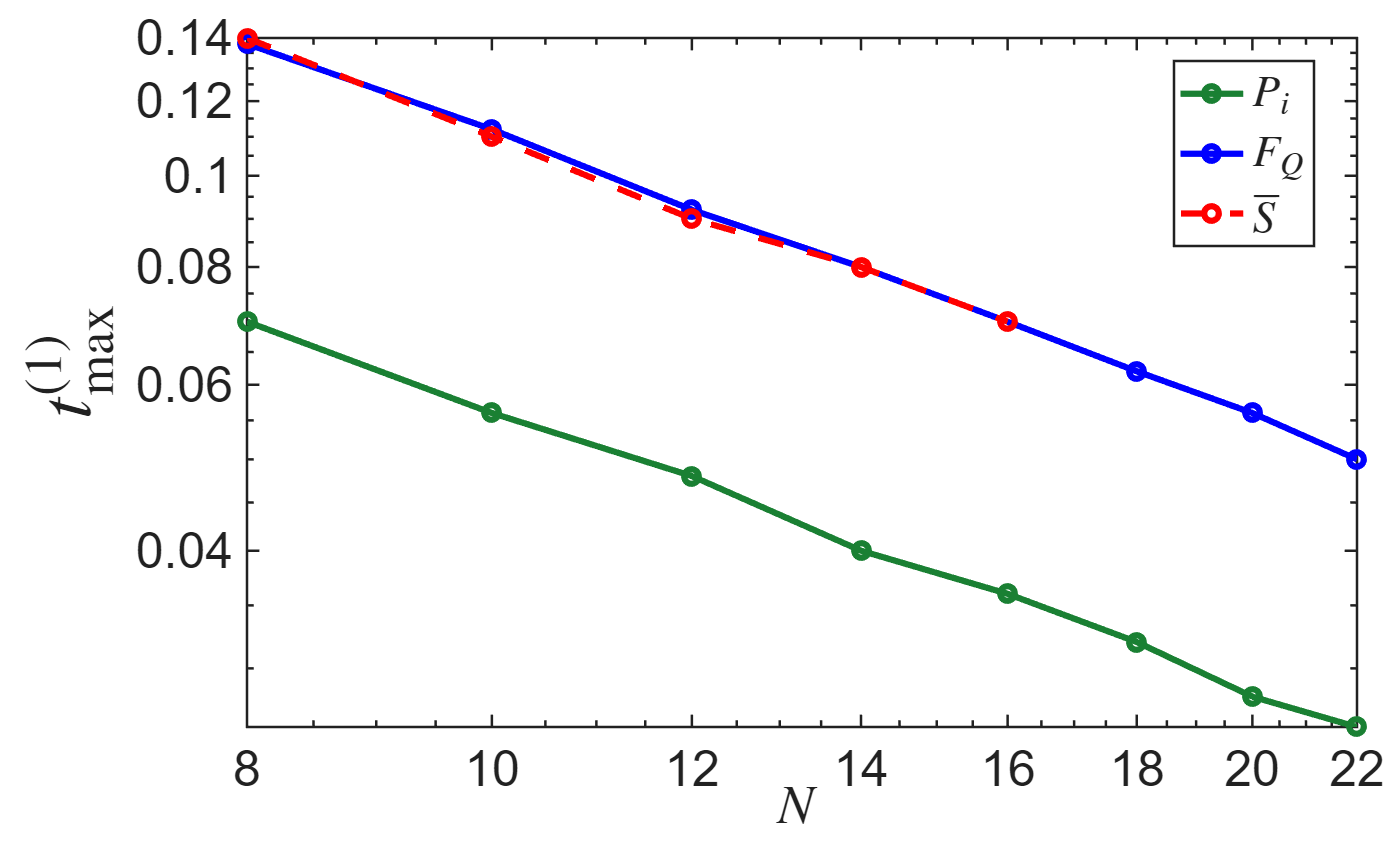}
    \caption{Time required to reach the first maxima of $P_i$, $F_Q$, and $\overline{S}$ as a function of system size $N$, using the same parameter set as in Fig.~\ref{Pi_QFI_SA_N}.}
    \label{t_max}
\end{figure}

To quantify these trends, we analyze the scaling of both the maximum instantaneous power and the time required to reach it. Since the power dynamics exhibit several local maxima of comparable amplitude, we focus on the first local maximum, denoted by $P_{\rm max}^{(1)}$, shown in Fig.~\ref{Pi_QFI_SA_N}($a_2$). The corresponding time at which this maximum occurs is denoted by $t_{\rm max}^{(1)}$ and is presented in Fig.~\ref{Pi_QFI_SA_N}($a_3$). Here, the superscript ``$(1)$'' denotes the first local maximum of the corresponding quantity. Our results reveal that the first maximum charging power grows quadratically with system size, following the scaling relation $P_{\rm max}^{(1)} \approx 1.42 \times N^{2.0}$ [ Fig.~\ref{Pi_QFI_SA_N}($a_2$)]. At the same time, the time required to reach this maximum decreases approximately inversely with system size and is well described by $t_{\rm max}^{(1)} \approx 0.54 \times N^{-0.98}$  [Fig.~\ref{Pi_QFI_SA_N}($a_3$)]. These results confirm that the fully collective charging protocol not only enhances the achievable charging power but also accelerates the charging process as the system size increases.
\par
The above scaling behavior of the charging power is also reflected in the dynamics of multipartite quantum correlations, as quantified by the QFI. The QFI dynamics, shown in Fig.~\ref{Pi_QFI_SA_N}($b_1$), indicate that multipartite quantum correlations are continuously generated for all system sizes considered. Moreover, the magnitude of the QFI increases significantly with increasing system size, suggesting that larger systems support stronger multipartite quantum correlations under the fully collective charging protocol. In addition, the time required to reach the maximum of the QFI decreases as the number of spins increases.

To further characterize this behavior, we perform a finite-size scaling analysis of both the maximum QFI and the time required to reach it. Building on the observed growth and time-shift of the QFI peak with system size, we again focus on the first local maximum of the QFI, denoted by $F_{Q,\rm max}^{(1)}$, where the superscript ``$(1)$'' indicates the first local maximum of the corresponding dynamical quantity. As shown in Fig.~\ref{Pi_QFI_SA_N}($b_2$), the maximum QFI increases approximately quadratically with the system size, demonstrating a substantial enhancement of multipartite quantum correlations in larger systems. At the same time, the corresponding time $t_{\rm max}^{(1)}$ required to attain this maximum decreases approximately inversely with system size, as illustrated in Fig.~\ref{Pi_QFI_SA_N}($b_3$). These results reveal that the generation of multipartite quantum correlations becomes both stronger and faster as the system size increases, mirroring the scaling behavior observed for the charging power.
\par
A similar analysis is also performed for ABEE, as shown in Fig.~\ref{Pi_QFI_SA_N}($c_1$--$c_3$). The ABEE exhibits a qualitatively different scaling behavior compared to the charging power and the QFI. In particular, while it does not show an increase in its maximal attainable value with system size, it is bounded and reaches the same maximum value of $1$ for all system sizes considered. This behavior arises from the fact that the ABEE, being an entanglement entropy for a bipartition, is upper bounded by one ebit in our normalization, and therefore cannot grow in magnitude beyond this limit. Instead, the system-size dependence appears in the dynamics of entanglement generation. As shown in Fig.~\ref{Pi_QFI_SA_N}($c_1$), the time required for the ABEE to reach its maximal value decreases systematically with increasing system size.

To further characterize this behavior, we analyze the maximum value and the corresponding time to reach it of first local maxima of ABEE.  As shown in Fig.~\ref{Pi_QFI_SA_N}($c_2$), the maximal ABEE remains constant for all system sizes. However, the time required to attain this maximum decreases approximately inversely with system size, as illustrated in Fig.~\ref{Pi_QFI_SA_N}($c_3$). This indicates that the rate at which maximal entanglement is generated becomes faster for larger systems, consistent with the overall acceleration of the dynamical processes observed for the charging power and QFI.

Notably, the peak of the ABEE remains delayed with respect to the peak of the instantaneous charging power. This persistent temporal ordering across all investigated system sizes indicates that the separation between the timescales of maximal energy transfer and maximal correlation buildup is not restricted to the $N=8$ case. Within the accessible system sizes, these results therefore support the robustness of both the temporal hierarchy and the enhanced charging performance associated with the fully collective ($\kappa=N$) charging protocol. 
\par
We compare the time required to reach the first maxima of the instantaneous power, QFI, and ABEE, as shown in Fig.~\ref{t_max}. We find that the instantaneous power reaches its first maximum in a shorter time compared to both the QFI and ABEE. In contrast, the QFI and ABEE exhibit nearly identical time scales for attaining their first maxima, as illustrated in Fig.~\ref{t_max}.

\subsubsection*{Fair Charging with Increasing Participation Number}
\begin{figure}
 \includegraphics[width=.95\linewidth, height=.65\linewidth]{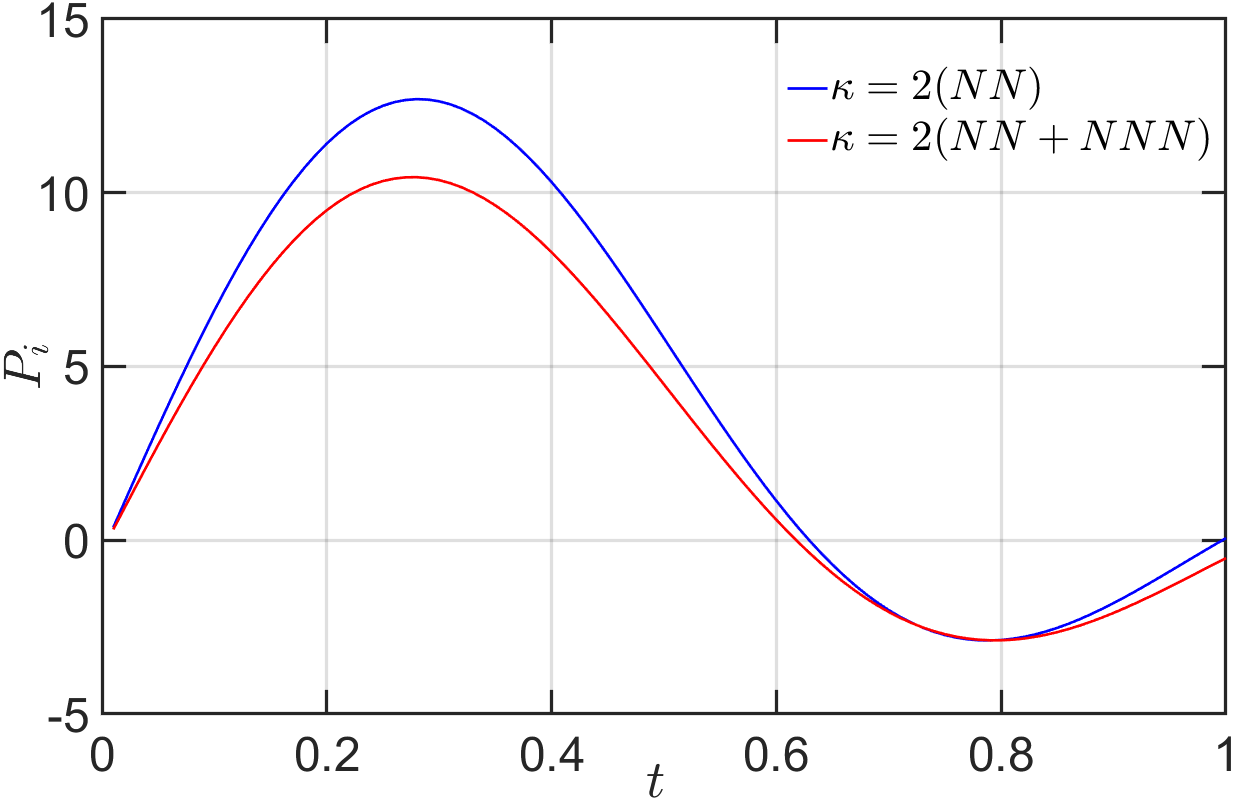}
\caption{
{\bf Fair charging with increasing participation number:} 
Noninteracting battery with $h_z=1$, charged by a $2$-local interacting chargoid in the absence of a transverse field ($h_x=0$). 
The chargoid Hamiltonian is normalized as $\lVert \hat H_C^{\rm (I)} \rVert = N (=8)$ by tuning the interaction strengths $J_{x_1}$ and $J_{x_2}$, corresponding to NN and NNN couplings, respectively, with $J_{x_1} > J_{x_2}$.
}
 \label{Pi_NN_NNN}
  \end{figure}
We now examine how extending the interaction range (pariticipation number) in the chargoid affects the charging dynamics under fair charging conditions. This is achieved by incorporating next-nearest-neighbor (NNN) couplings in the NN interaction, which increase the number of spins participating in the interaction without introducing fully global connectivity. The interaction part of the charging Hamiltonian is modified as
\begin{equation}
\hat H_{\rm C}^{\rm (I)}
=
J_{x_1}\sum_{j=1}^{N-1}\hat\sigma_j^x\hat\sigma_{j+1}^x
+
J_{x_2}\sum_{j=1}^{N-2}\hat\sigma_j^x\hat\sigma_{j+2}^x,
\label{Charging_Ham_NNN}
\end{equation}
where $J_{x_1}$ and $J_{x_2}$ denote the strengths of NN and NNN interactions, respectively. To ensure a meaningful comparison between different interaction structures, we impose the fair charging condition by tuning the interaction strength, $J_{x_1} > J_{x_2}$, such that the operator norm of the charging Hamiltonian satisfies $\lVert \hat H_C^{\rm (I)} \rVert = N$. This normalization eliminates any classical advantage arising from increased energy scales and isolates the effect of interaction geometry.

We compare the instantaneous power of chargoids with NN interactions alone to those including both NN and NNN couplings, maintaining a fixed operator norm, $\lVert H_C^{\rm (I)} \rVert = N$.  
The results indicate that adding NNN interactions decreases the maxima of instantaneous power $P_i$ compared to the NN-only protocol [Fig.~\ref{Pi_NN_NNN}]. This demonstrates that increasing the participation number through short-range interaction extensions does not automatically translate into enhanced charging power.

The observed suppression of power coincides with the introduction of additional interaction pathways through next-nearest-neighbor couplings. In this regime, the presence of extended interactions is accompanied by a redistribution of dynamical contributions across different parts of the system, which is reflected in the reduced sharpness of the instantaneous power peak. These results indicate that charging performance depends sensitively on both the number of interacting spins and the structure of the interaction network. In particular, partially extended interaction schemes are associated with a redistribution of dynamical correlations, in contrast to the stronger collective enhancement observed in fully connected charging protocols.

\section{Conclusion and Outlook}
\label{conclusion}

In this work, we investigated the dynamical interplay between energy transfer and correlation generation in many-body quantum battery charging protocols. By systematically comparing the time evolution of the instantaneous charging power with a hierarchy of correlation measures, we examined how energetic and correlation-based observables develop under different interaction structures, charging protocols, and system sizes.

A central result of our study is the emergence of a robust temporal hierarchy between energy transfer and correlation buildup. Across all battery--charger configurations considered, the instantaneous charging power consistently reaches its maximum before the corresponding peaks of bipartite, tripartite, and multipartite correlation measures. This ordering remains stable for different interaction types and system sizes, revealing a clear separation of timescales between charging dynamics and correlation growth. 

To further elucidate the role of interaction structure, we analyzed $\kappa$-local charging protocols in both unconstrained and norm-constrained charger Hamiltonian settings. In the unconstrained regime, the enhancement of instantaneous charging power with increasing interaction order is found to originate predominantly from the corresponding growth of the charging Hamiltonian norm. By imposing the fixed charging Hamiltonian ( $\lVert \hat H_C^{{\rm (I)}} \rVert = N$), we disentangled this trivial energy-scale effect from the genuine influence of interaction geometry and many-body participation.

Within this constrained framework, increasing the interaction order alone does not generically lead to enhanced charging performance when the local spin-flip strength is equal to the interaction strength ($J_x=h_x$). This demonstrates that multipartite connectivity, by itself, is insufficient to guarantee a charging advantage. A significant enhancement emerges only when interaction terms dominate the dynamics ($J_x>h_x$), emphasizing the crucial role of cooperative many-body processes. In particular, interaction-only charging protocols with fully collective interactions ($\kappa=N$) coherently involve all spins and exhibit enhanced charging performance that remains remarkably stable across the system sizes investigated. By contrast, partially extended interactions ($\kappa<N$) fail to efficiently engage the entire system and therefore do not provide a comparable advantage.

We further showed that extending the interaction range through next-nearest-neighbor couplings can reduce the charging power within the parameter regimes considered. This behavior indicates the presence of competing dynamical pathways that can hinder efficient energy transfer, emphasizing that charging performance is governed not only by the interaction range but also by the detailed structure of the underlying many-body dynamics.

Finally, we emphasize that the present work focuses on the dynamical aspects of quantum battery charging, namely energy injection and correlation generation, rather than a complete thermodynamic resource-theoretic characterization. Important quantities such as ergotropy, extractable work, switching costs, and control overhead have not been addressed here and remain important directions for future research. Moreover, extending the present analysis to open-system dynamics and experimentally realistic noisy environments would provide valuable insight into the robustness of the mechanisms identified in this work. We therefore hope that our framework will serve as a foundation for future studies connecting charging dynamics, many-body correlations, and thermodynamic resources in quantum batteries.

\section*{Acknowledgments}
RKS gratefully acknowledges Anupam and Sibasish Ghosh for their insightful and stimulating discussions.

\bibliography{QB}
\appendix
\section{System-size analysis: non-interacting battery and interacting chargoid (Protocol I)}
\label{Finit_size_ana}
\begin{figure*}
    \centering
\includegraphics[width=0.32\linewidth,height=0.20\linewidth]{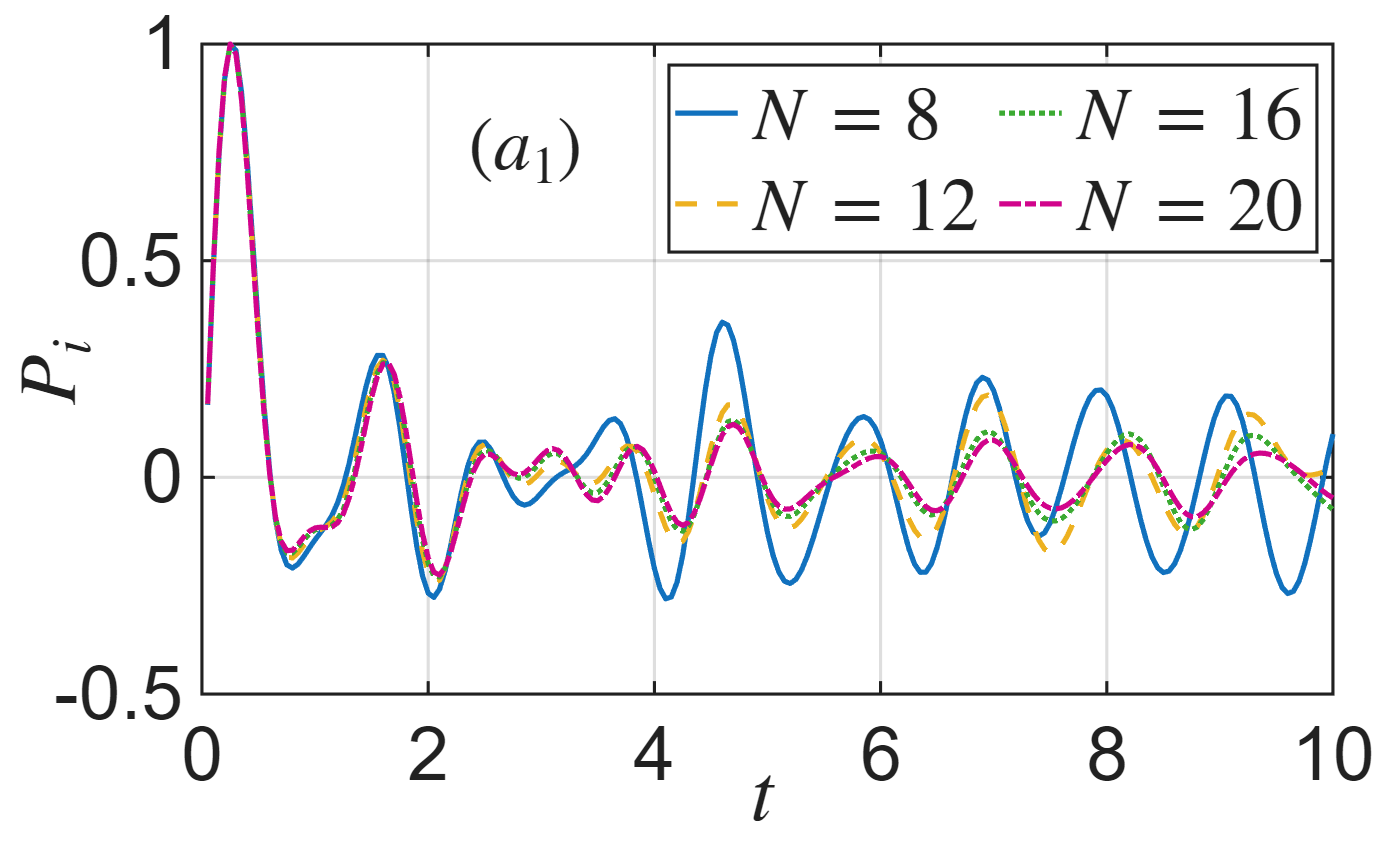}
\includegraphics[width=0.32\linewidth,height=0.20\linewidth]{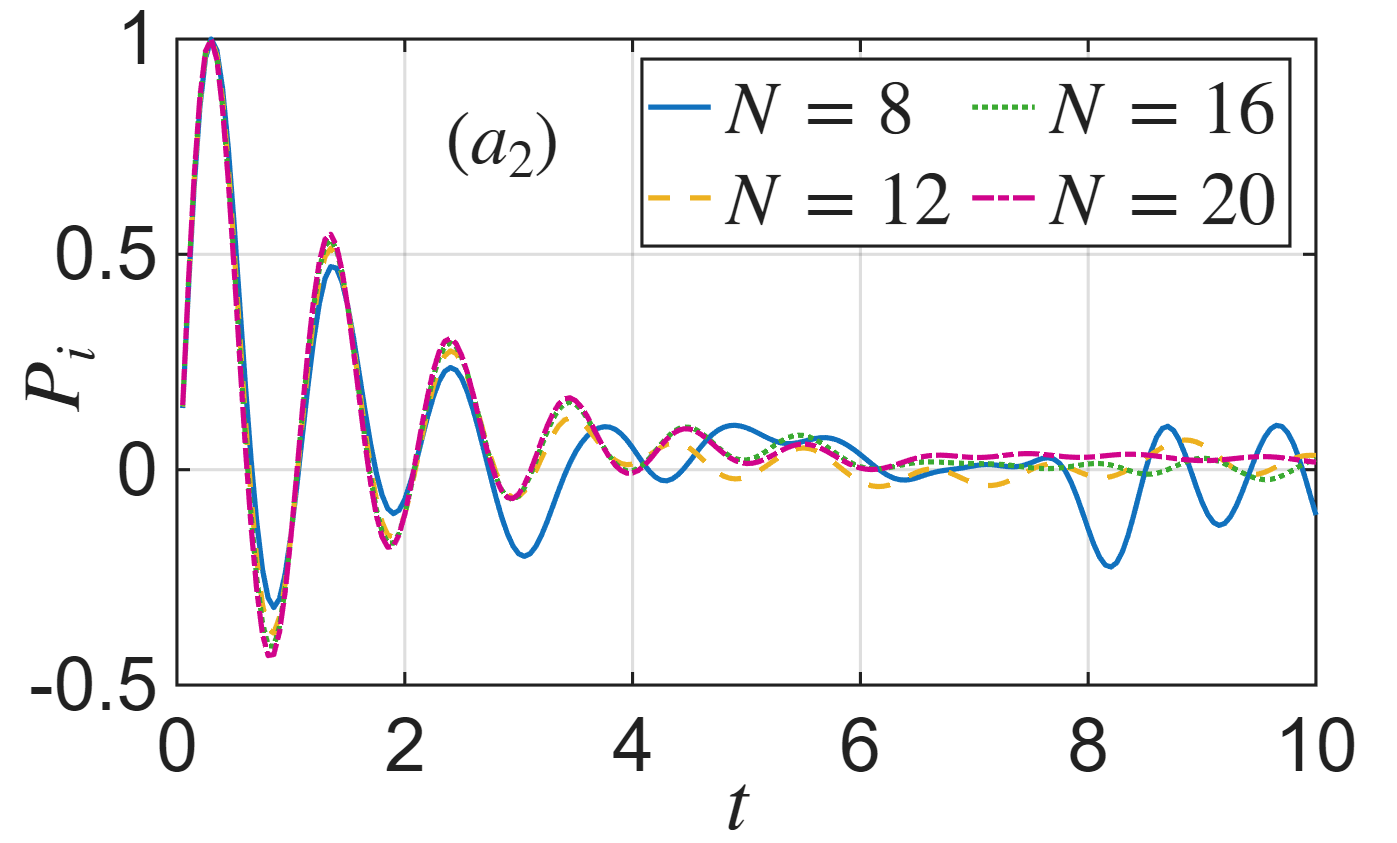}
\includegraphics[width=0.32\linewidth,height=0.20\linewidth]{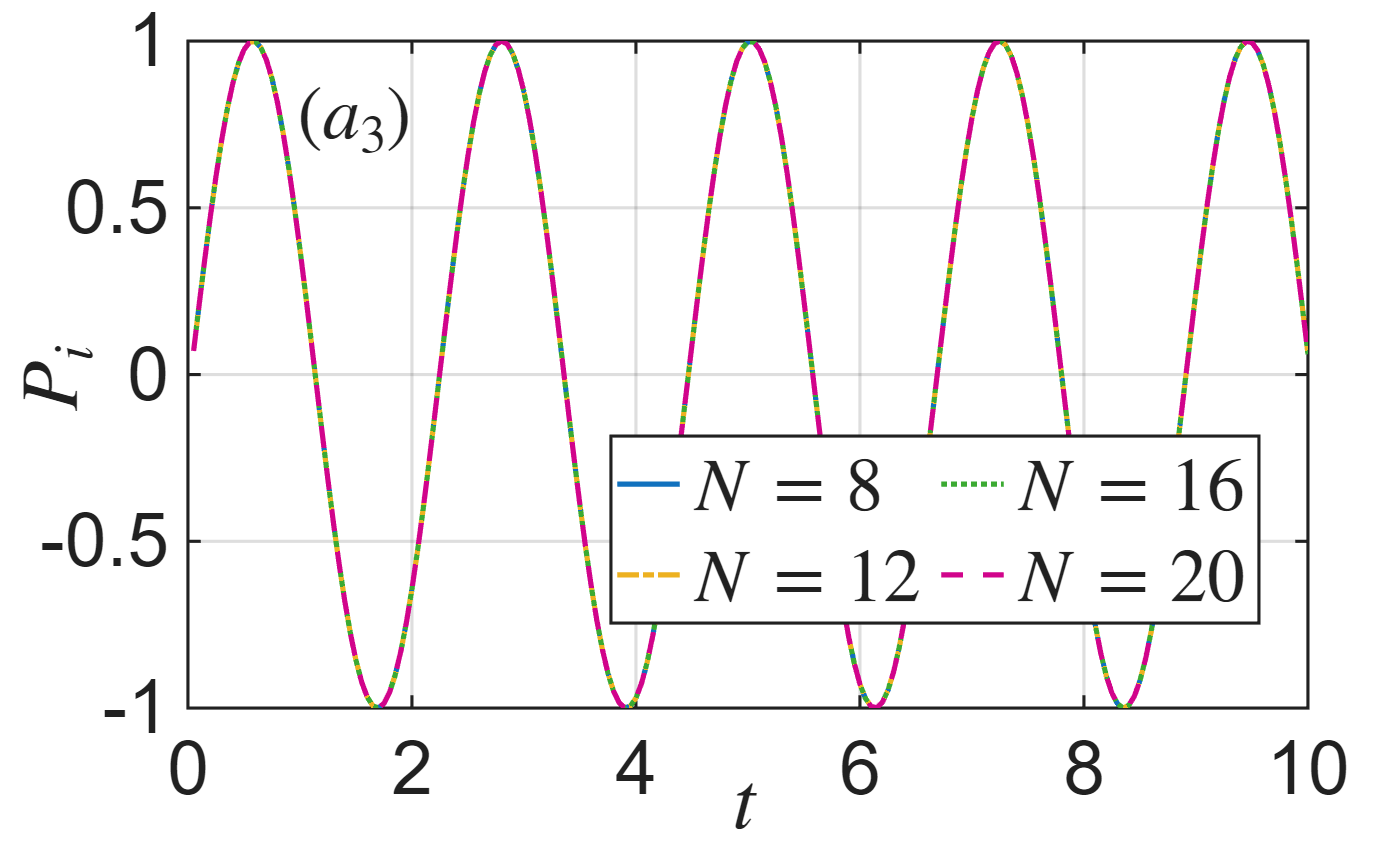}
\includegraphics[width=0.32\linewidth,height=0.20\linewidth]{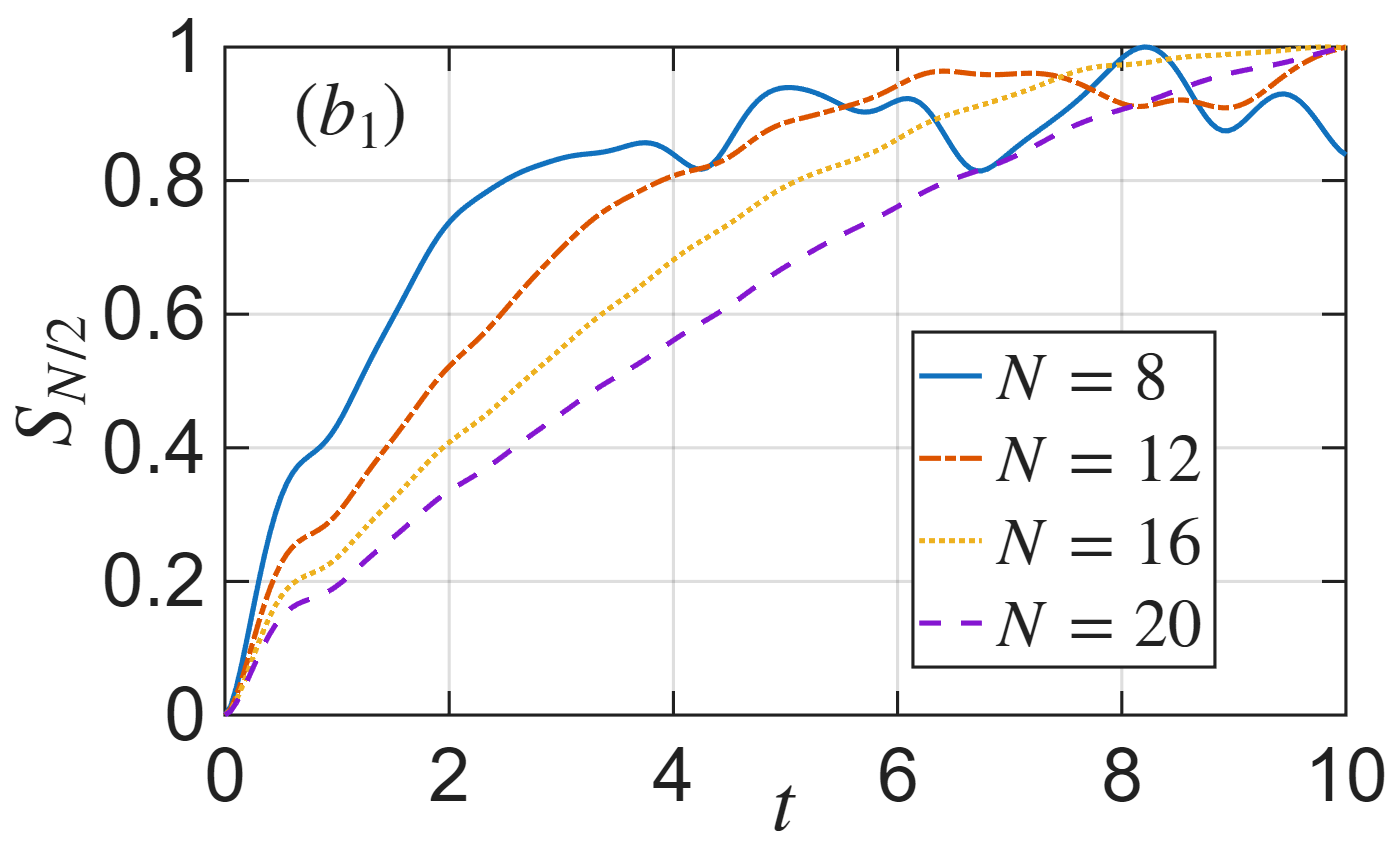}
\includegraphics[width=0.32\linewidth,height=0.20\linewidth]{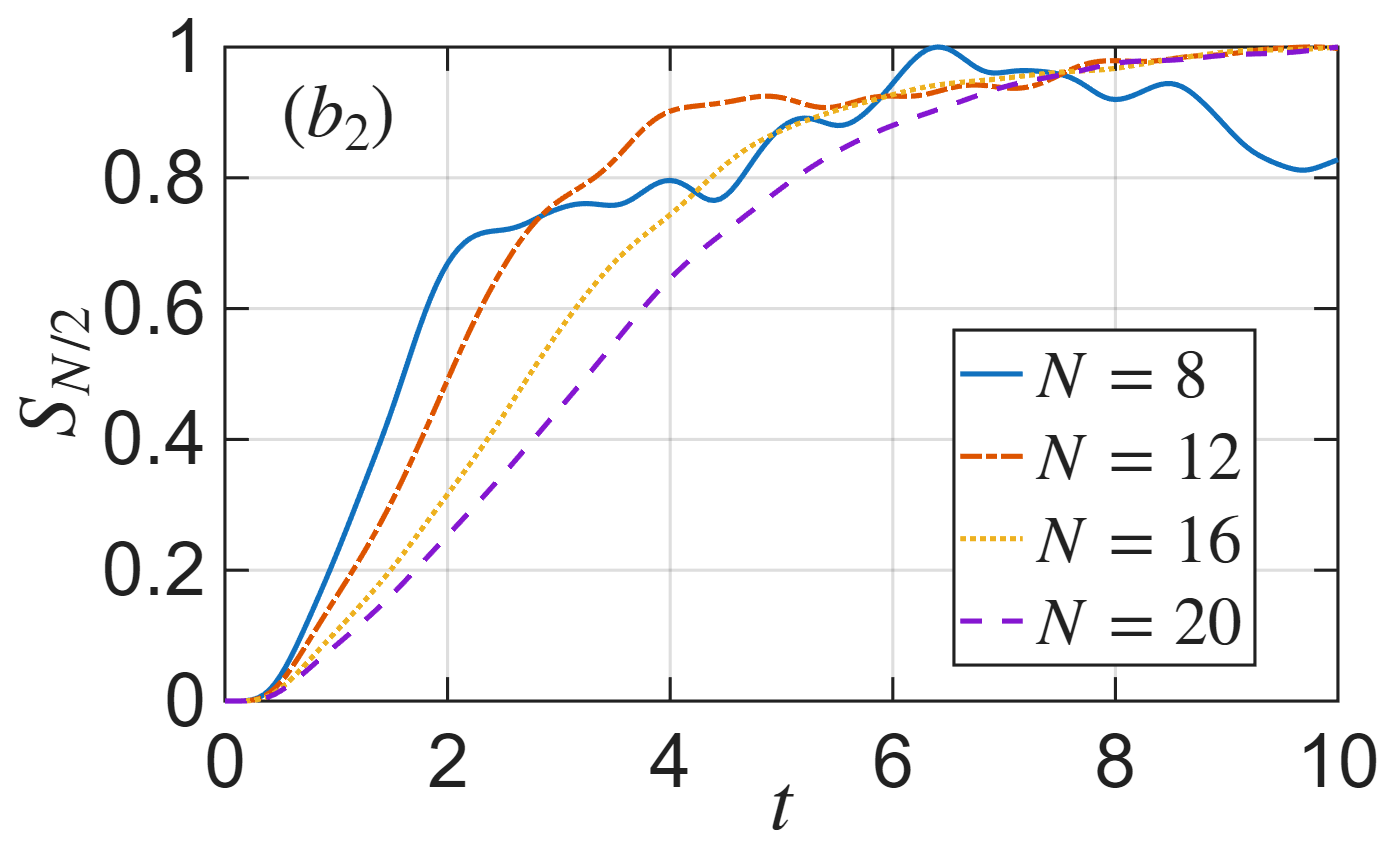}
\includegraphics[width=0.32\linewidth,height=0.20\linewidth]{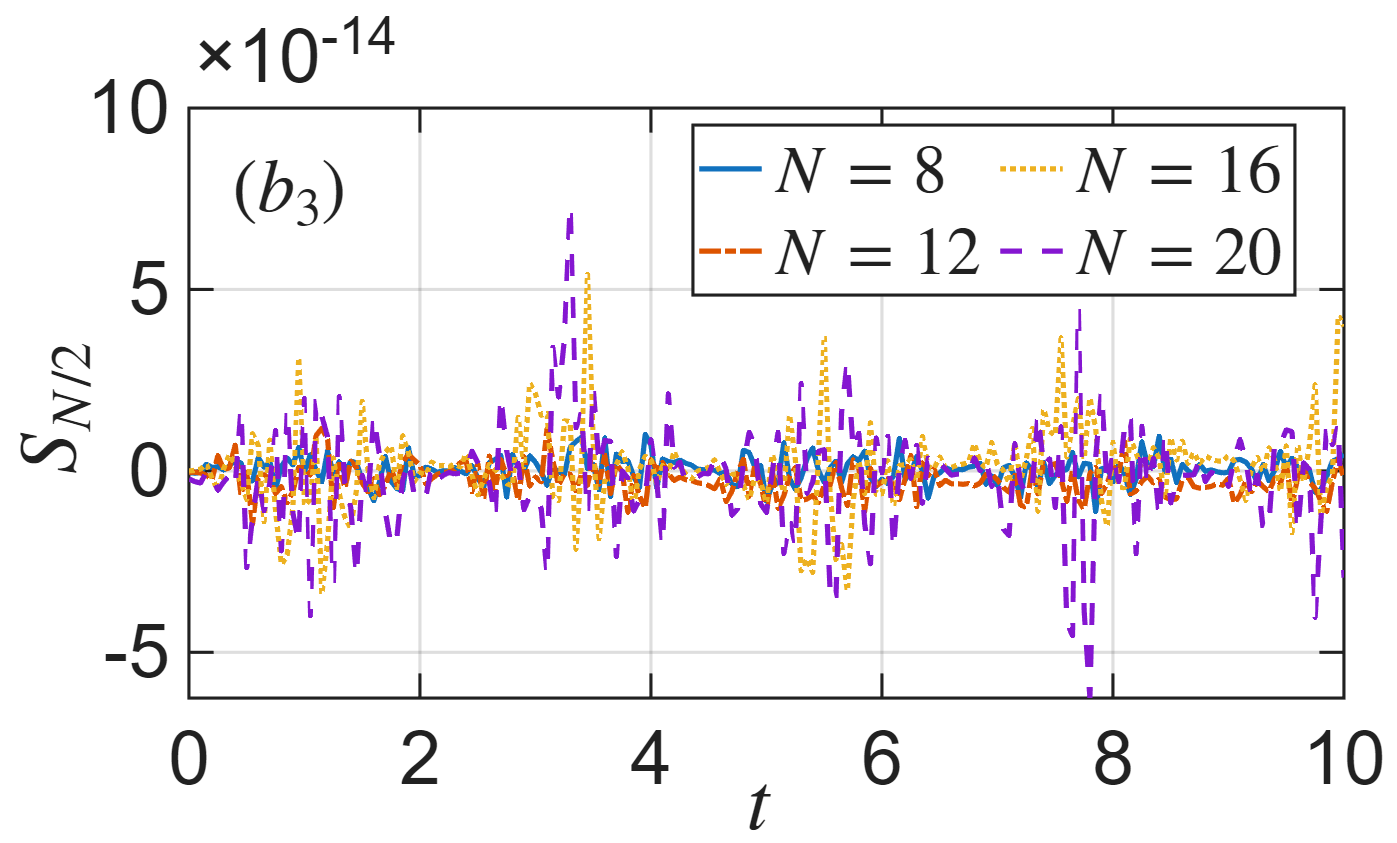}
\includegraphics[width=0.32\linewidth,height=0.20\linewidth]{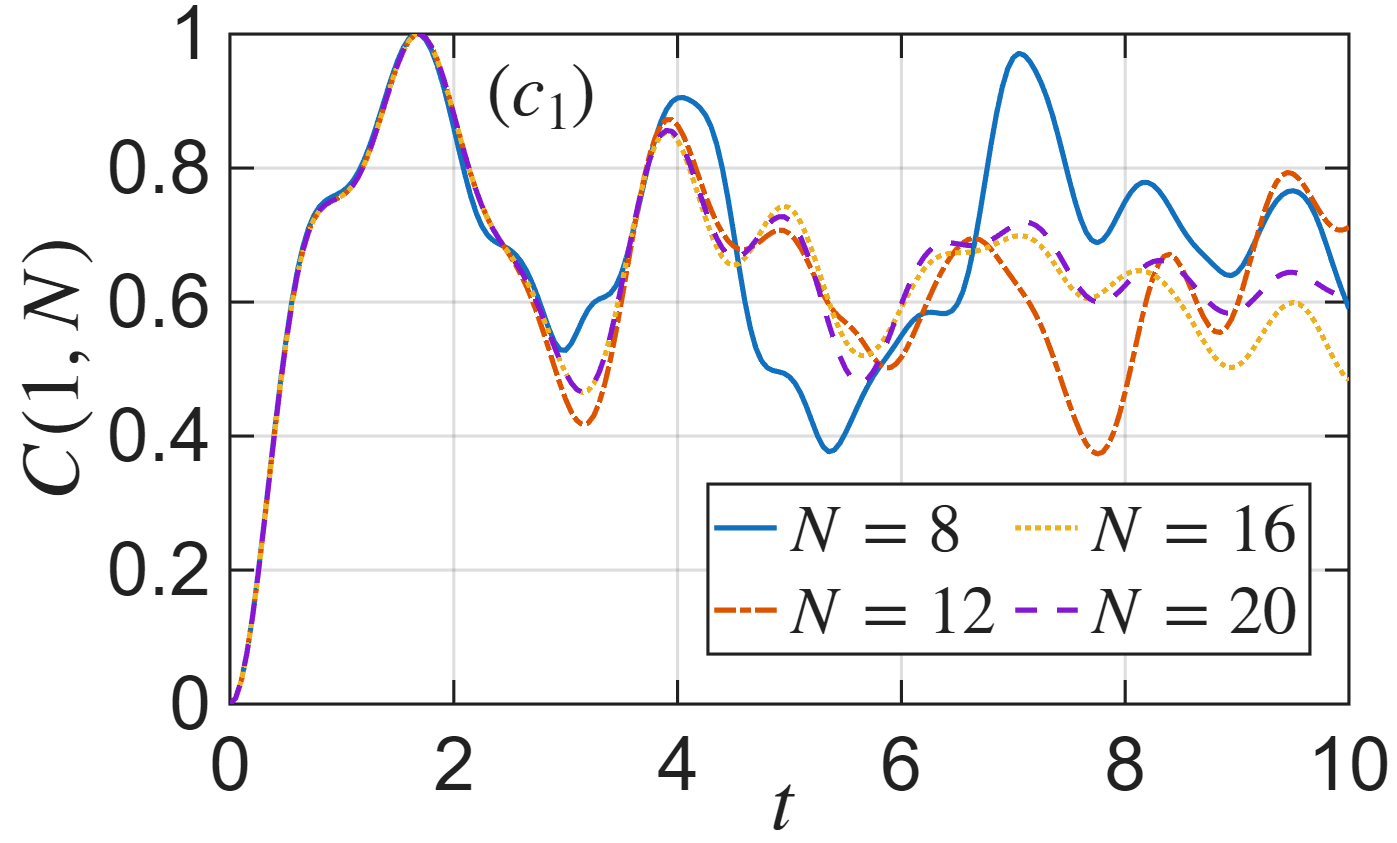}
\includegraphics[width=0.32\linewidth,height=0.20\linewidth]{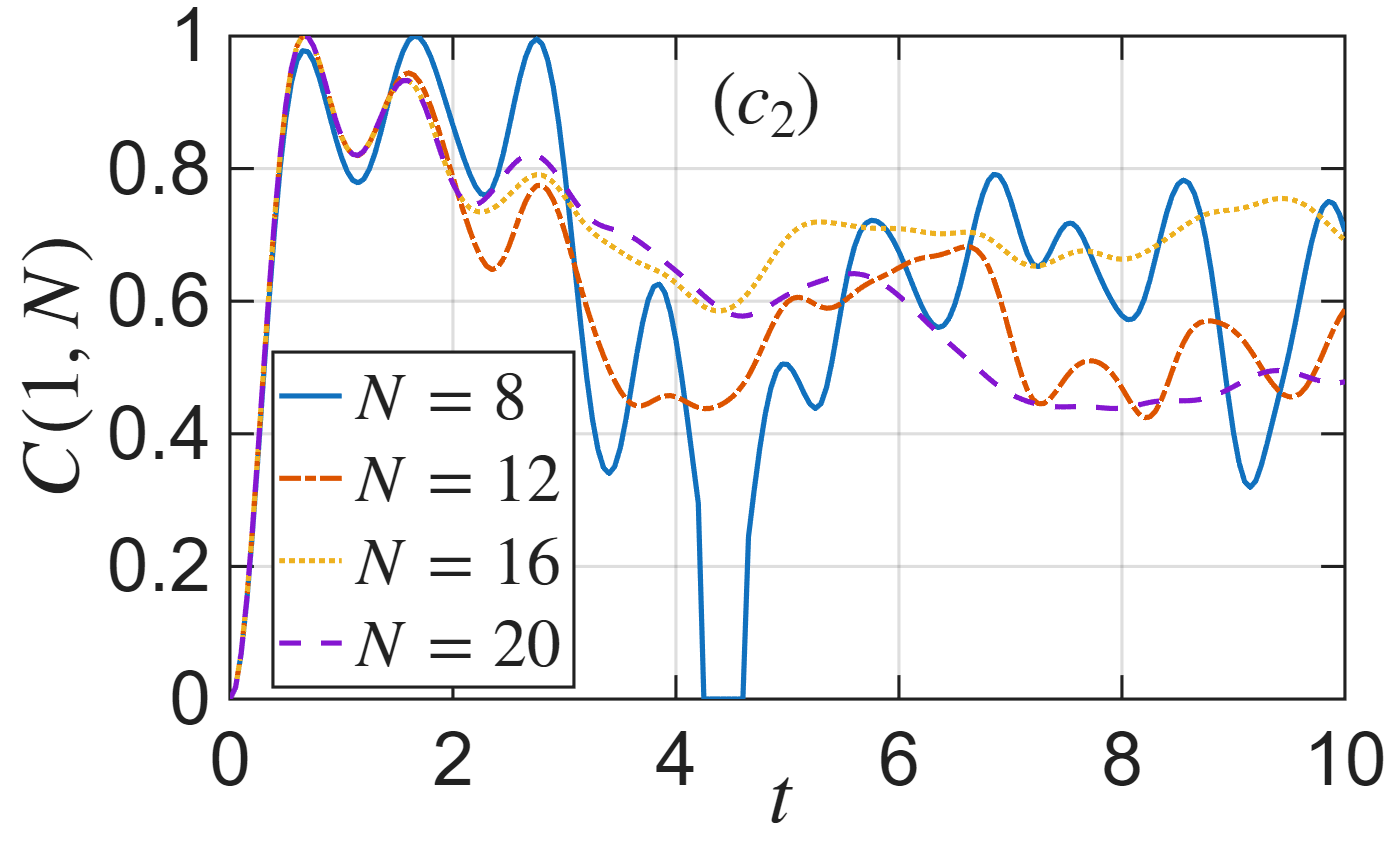}
\includegraphics[width=0.32\linewidth,height=0.20\linewidth]{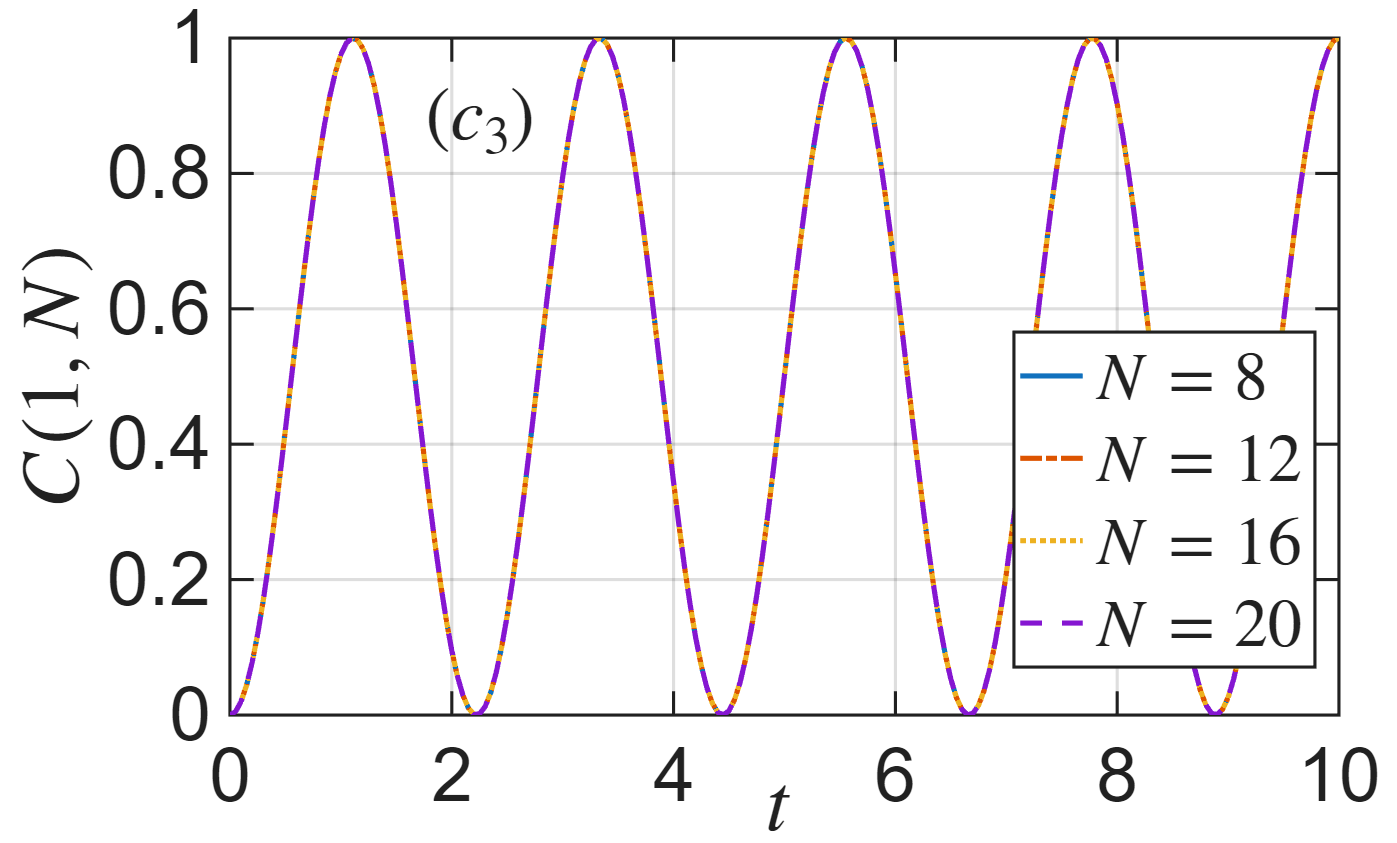}
\includegraphics[width=0.32\linewidth,height=0.20\linewidth]{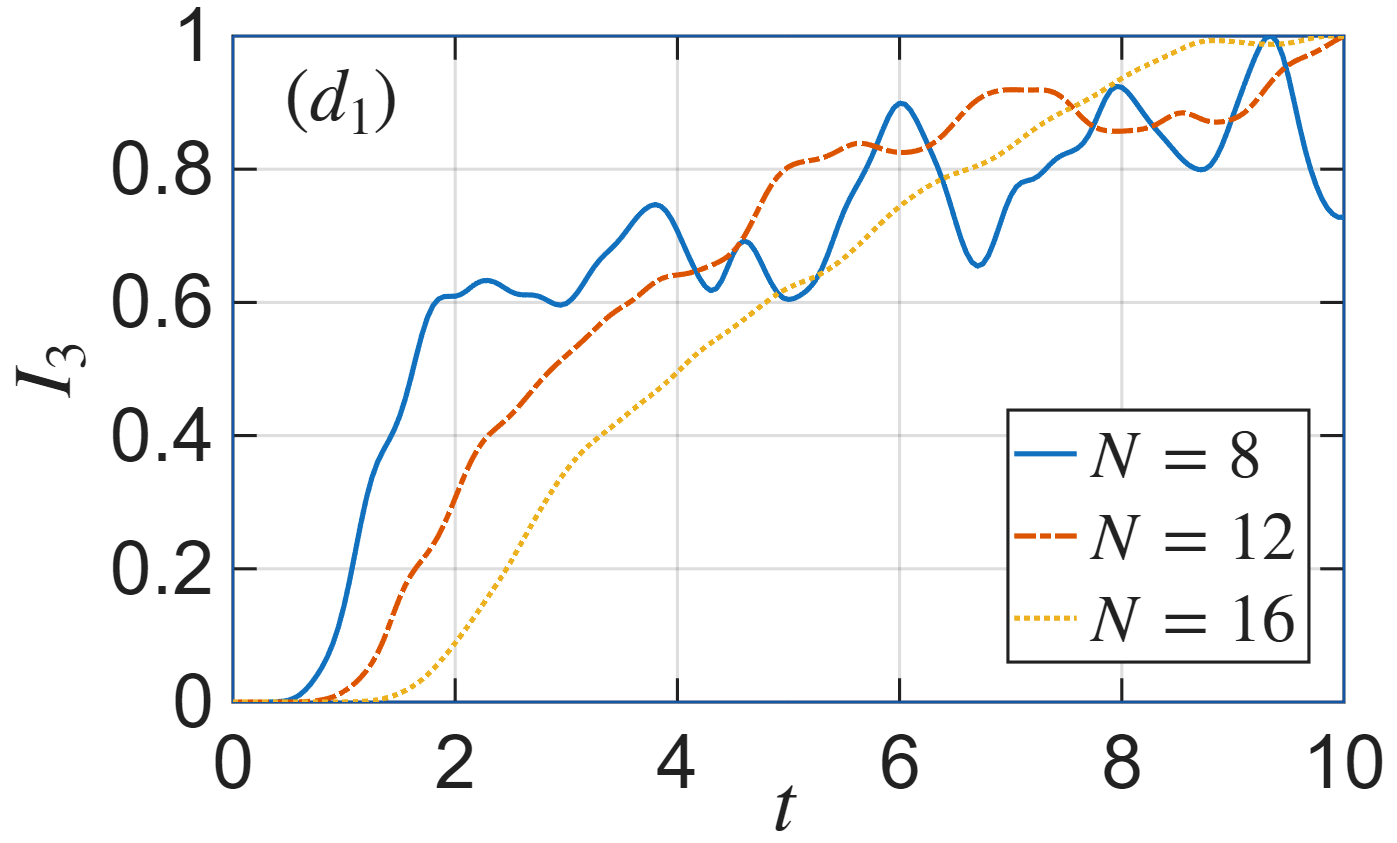}
\includegraphics[width=0.32\linewidth,height=0.20\linewidth]{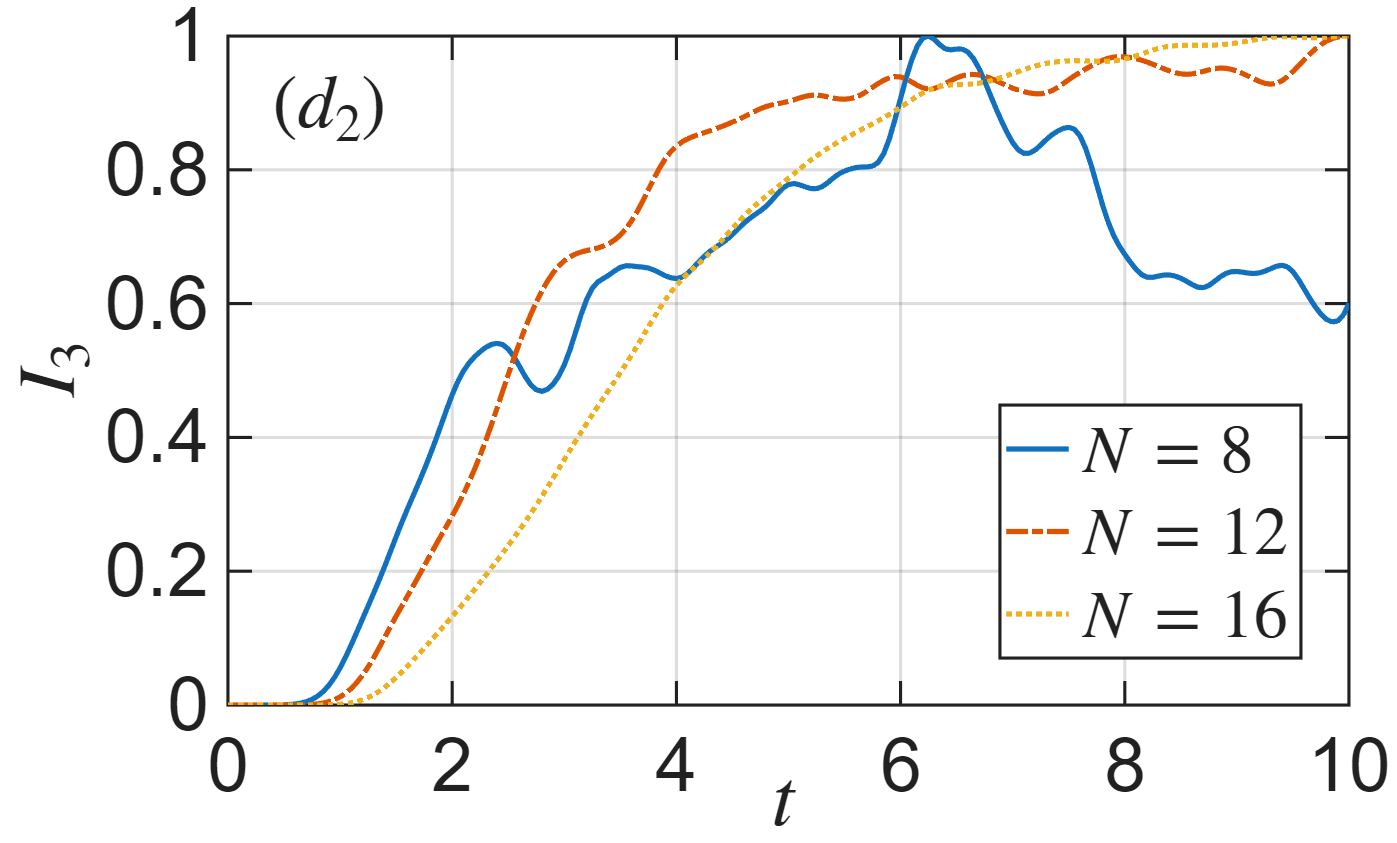}
\includegraphics[width=0.32\linewidth,height=0.20\linewidth]{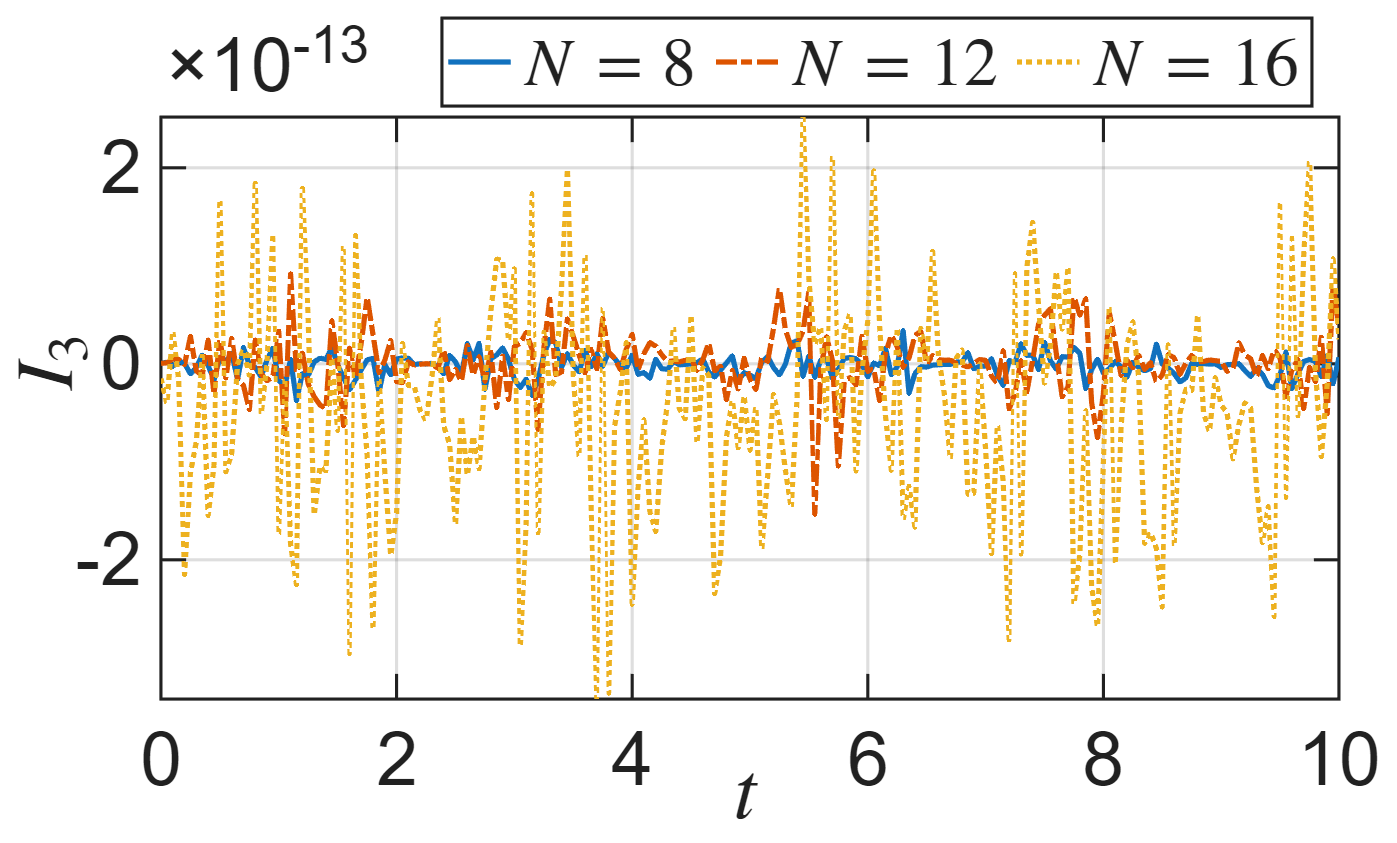}
\includegraphics[width=0.32\linewidth,height=0.20\linewidth]{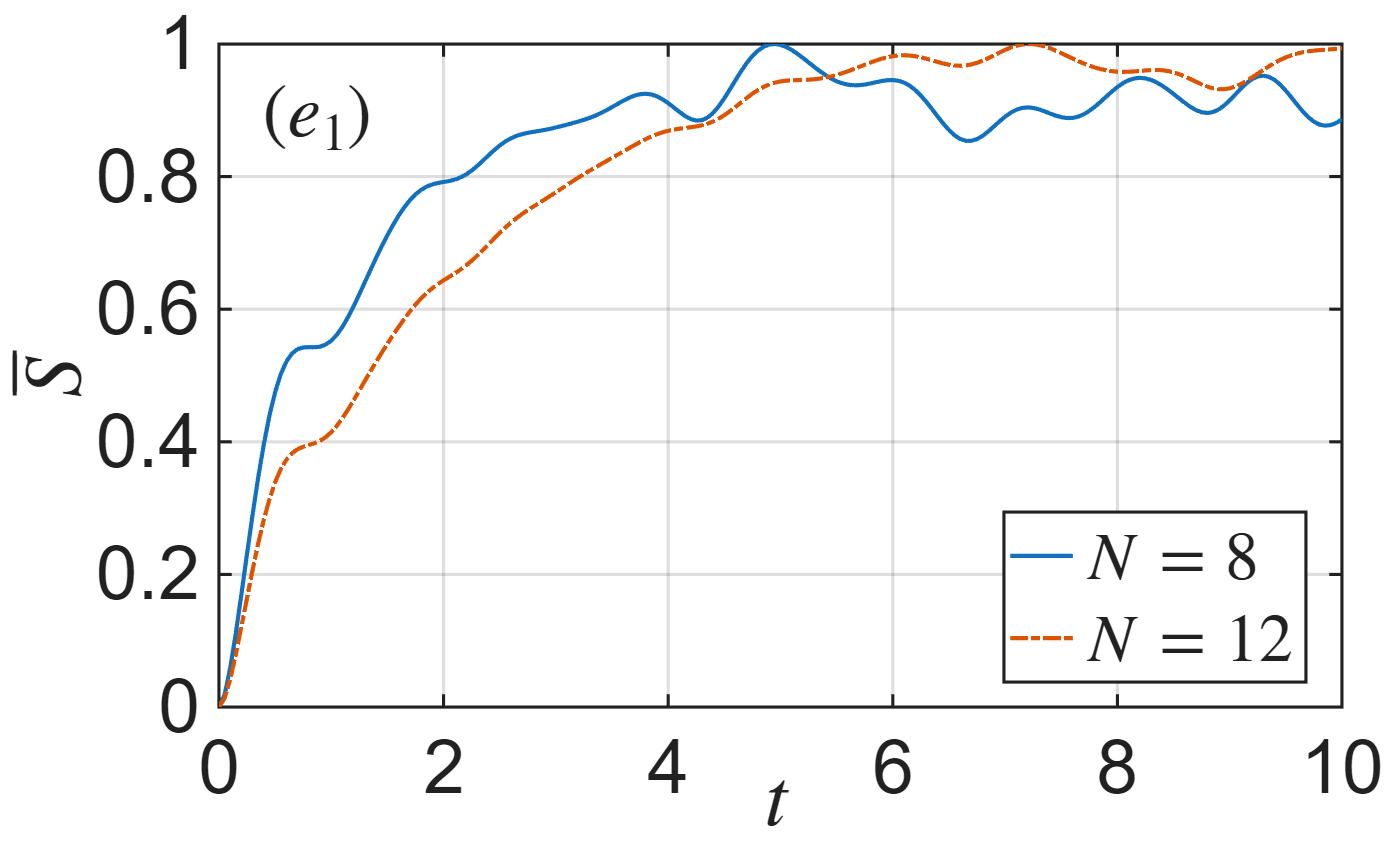}
\includegraphics[width=0.32\linewidth,height=0.20\linewidth]{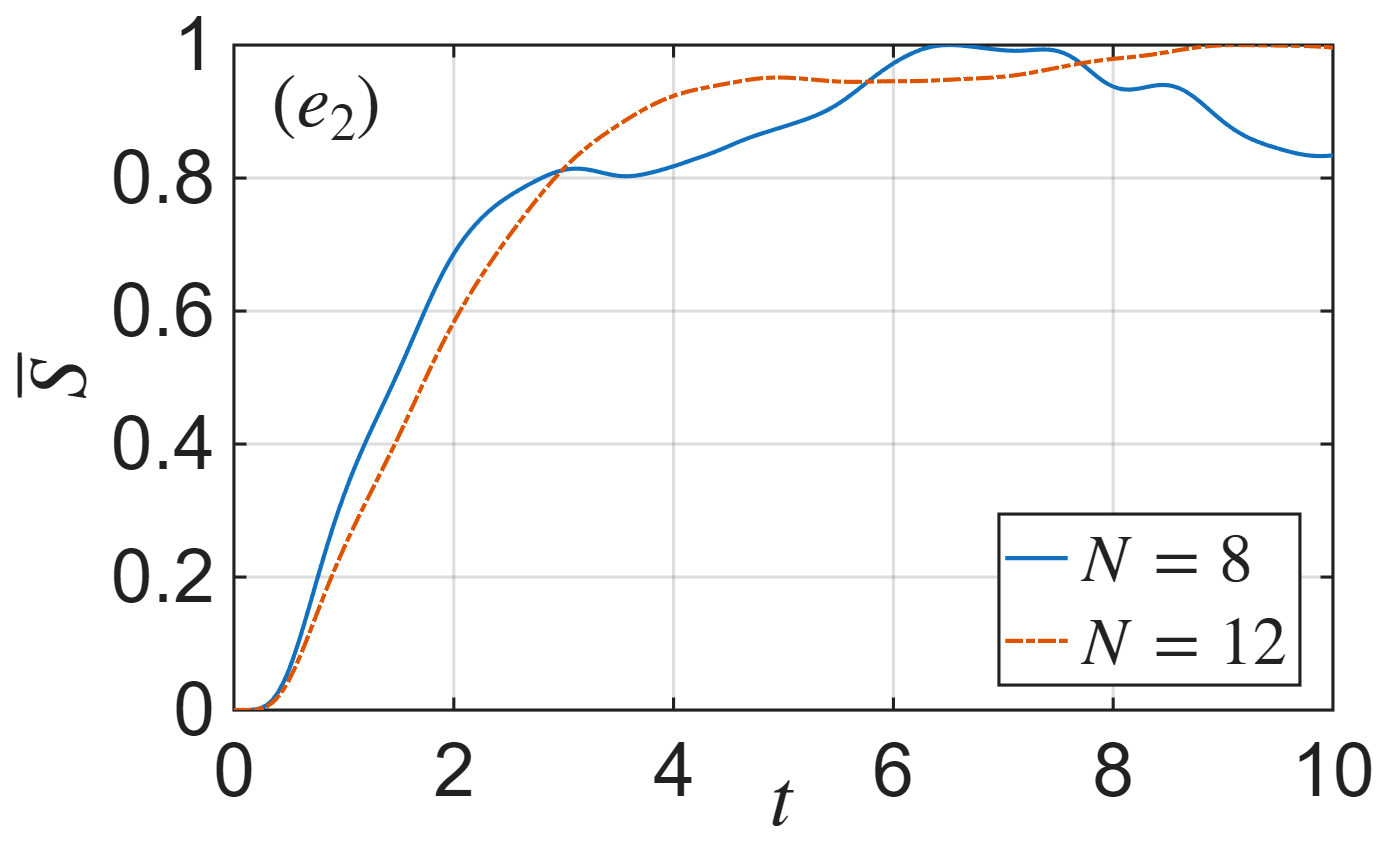}
\includegraphics[width=0.32\linewidth,height=0.20\linewidth]{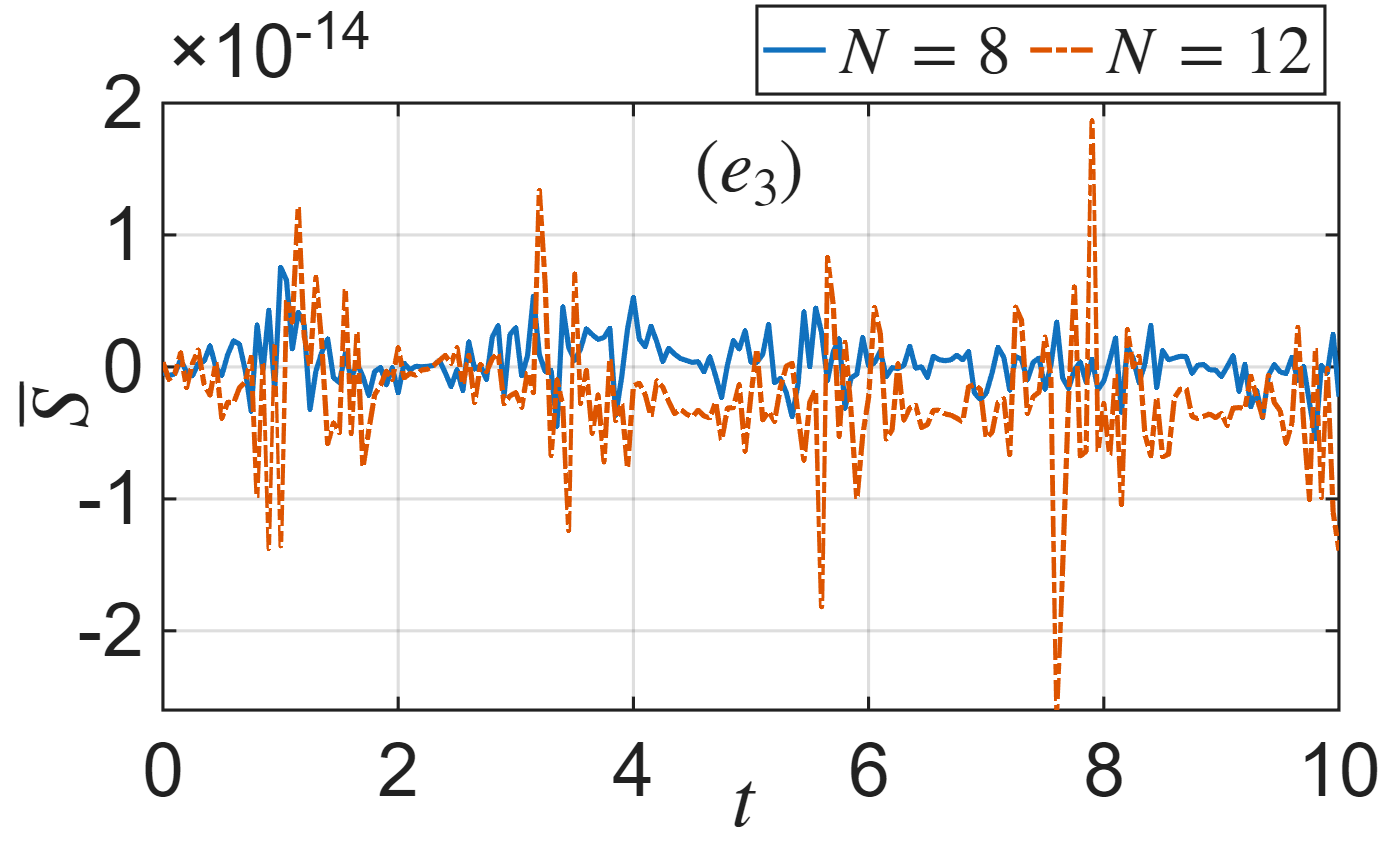}
\includegraphics[width=0.32\linewidth,height=0.20\linewidth]{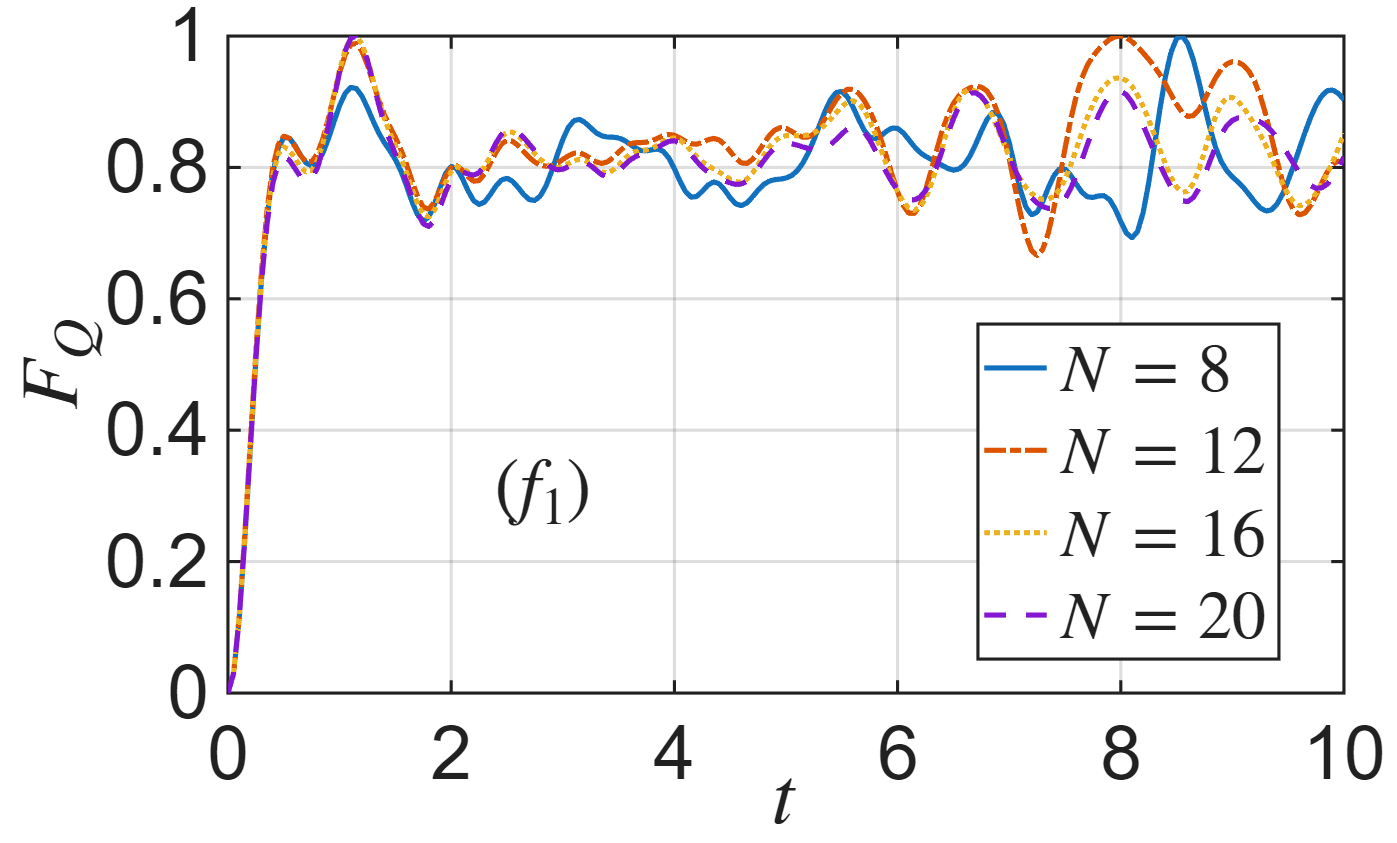}
\includegraphics[width=0.32\linewidth,height=0.20\linewidth]{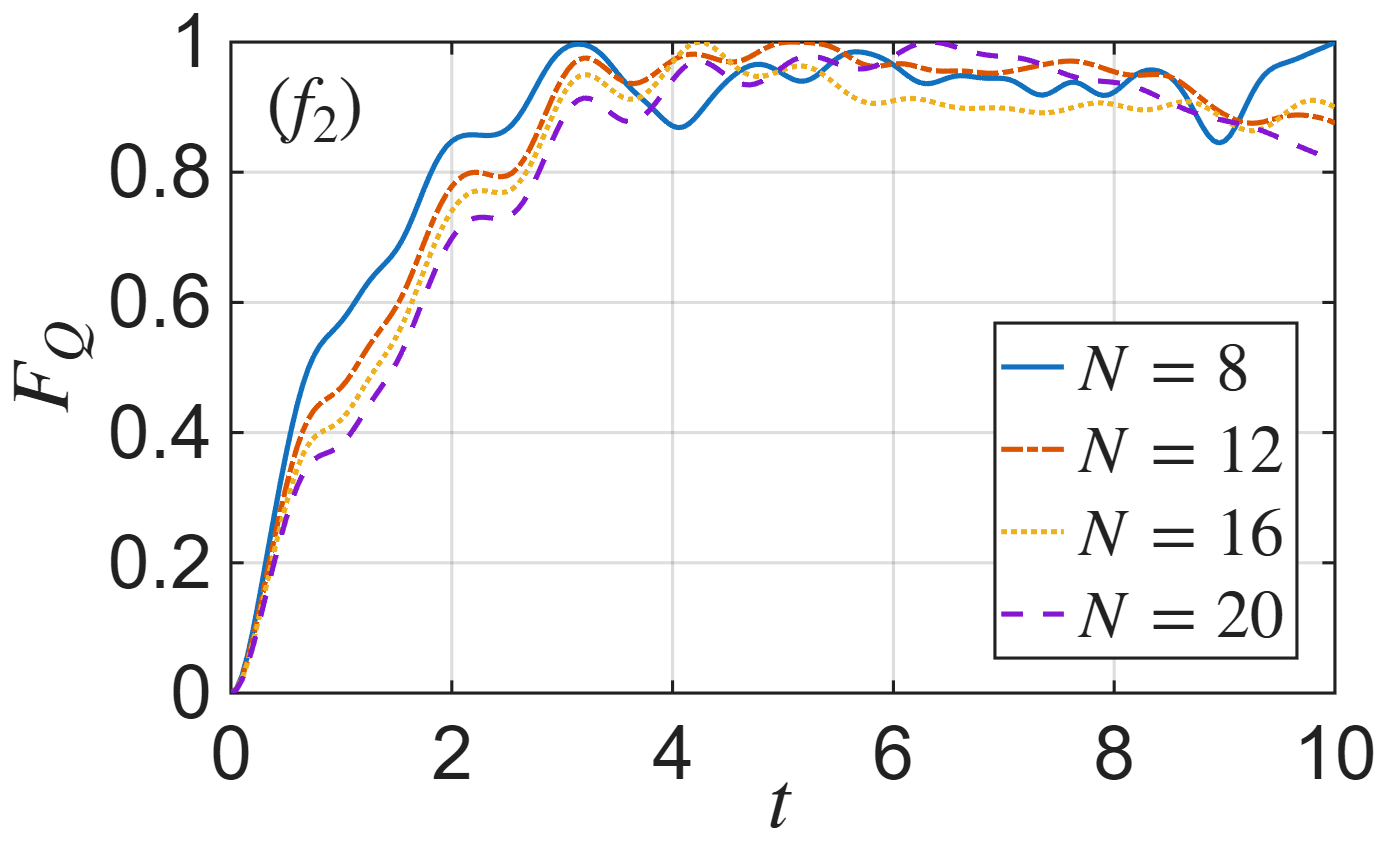}
\includegraphics[width=0.32\linewidth,height=0.20\linewidth]{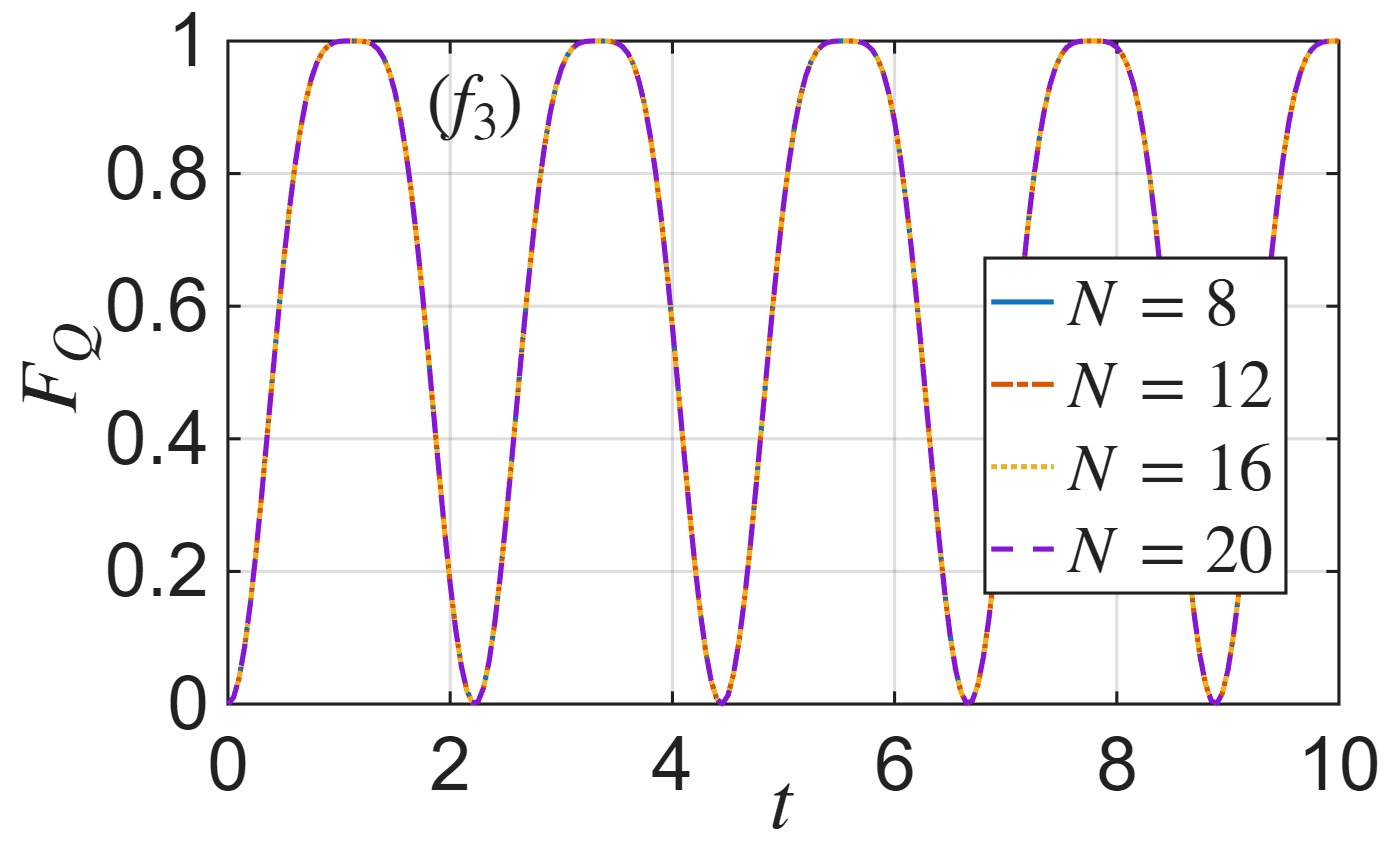}
    \caption{ Noninteracting battery with $h_z = 1$ driven by an interacting charger with fixed $\kappa = 2$ and $h_x = 1$, as described in Protocol ${\rm I}$. Normalized instantaneous charging power $P_i$, BEE $S_{N/2}$, concurrence $C(1,N)$, TMI $I_3$, ABEE $\overline{S}$, and QFI $F_Q$ are shown as functions of time for different system sizes $N$, as indicated in the legends. Panels ($a_1–a_3$), ($b_1–b_3$), ($c_1–c_3$), ($d_1–d_3$), ($e_1–e_3$), and ($f_1–f_3$) correspond to $P_i$, $S_{N/2}$, $C(1,N)$, $I_3$, $\overline{S}$, and $F_Q$, respectively. The left, middle, and right columns correspond to the Ising ($J_x=1$, $J_y=J_z=0$), XY ($J_x=J_y=1$, $J_z=0$), and Heisenberg ($J_x=J_y=J_z=1$) interactions, respectively.}
\label{P_N}
\end{figure*}

The temporal ordering of the maxima of the instantaneous charging power and entanglement measures remains robust for larger system sizes. In the main text, we presented results for $N=12$; here, we extend the analysis to system sizes in the range $N=8$--$20$. For this study, we consider the battery–chargeroid configuration introduced in Protocol~${\rm I}$ with fixed interaction range $\kappa=2$. We analyze the normalized instantaneous charging power, $P_i/P_i^{\mathrm{max}}$, as a function of time for different system sizes. In addition, we examine several correlation measures in a hierarchical manner, including bipartite, tripartite, and multipartite quantities defined in the main text. For bipartite correlations, we consider the BEE, $S_{N/2}/S_{N/2}^{\mathrm{max}}$, and concurrence, $C(1,N)/C(1,N)^{\mathrm{max}}$. For tripartite correlations, we analyze the TMI, $I_3/I_3^{\mathrm{max}}$. For multipartite correlations, we consider  ABEE, $\overline{S}/\overline{S}^{\mathrm{max}}$, and the QFI, $F_Q/F_Q^{\mathrm{max}}$, for increasing system sizes.

For both the Ising interaction ($J_x=1$, $J_y=J_z=0$) [Fig.~\ref{P_N}($a_1$)] and the XY interaction ($J_x=J_y=1$, $J_z=0$) [Fig.~\ref{P_N}($a_2$)], the instantaneous power exhibits nearly identical short-time behavior across all system sizes. At later times, however, noticeable deviations emerge, indicating a clear system-size dependence [Fig.~\ref{P_N}($a_1,a_2$)]. In contrast, for the isotropic Heisenberg interaction ($J_x=J_y=J_z=1$) [Fig.~\ref{P_N}($a_3$)], the dynamics of $P_i$ collapse onto a single curve for all system sizes, demonstrating an almost complete independence of the charging dynamics on system size. Importantly, the time required to reach the maximum charging power remains unchanged with increasing system size, as illustrated in Fig.~\ref{P_N}($a_3$).

Next, we analyze the BEE, shown in Fig.~\ref{P_N}($b_1$--$b_3$). For all  Ising [Fig.~\ref{P_N}($b_1$)], and XY interaction [Fig.~\ref{P_N}($b_2$)], the BEE exhibits qualitatively similar dynamical behavior across system sizes. However, for Ising and XY interactions, the time required to reach the maximum entanglement is consistently larger than the time at which the instantaneous charging power attains its peak. For the isotropic Heisenberg interaction [Fig.~\ref{P_N}($b_2$)], the BEE remains identically zero throughout the entire time evolution.

We then analyze the concurrence, as shown in Fig.~\ref{P_N}($c_1$--$c_3$). For the Ising interaction [Fig.~\ref{P_N}($c_1$)] and the XY interaction [Fig.~\ref{P_N}($c_2$)], the initial growth of concurrence is similar across system sizes, and the time required to reach its maximum is approximately the same, while deviations appear at later times. In contrast, for the isotropic Heisenberg interaction [Fig.~\ref{P_N}($c_3$)], the concurrence dynamics are identical for all system sizes considered, indicating a complete collapse of the dynamics with respect to system size. The time required to attain the maximum concurrence is also consistent across all system sizes. Importantly, for all system sizes and interaction types considered, the time at which the instantaneous charging power reaches its maximum is always shorter than the time required for concurrence to reach its maximum, establishing a clear temporal ordering in which energy charging precedes bipartite entanglement buildup [Fig.~\ref{P_N}($c_1$--$c_3$)].

Next, we analyze the TMI, as shown in Fig.~\ref{P_N}($d_1$--$d_3$). For the Ising [Fig.~\ref{P_N}($d_1$)] and XY [Fig.~\ref{P_N}($d_2$)] interactions, the TMI increases with time and reaches its maximum value. In both cases, the time required to attain this maximum is larger than that of the instantaneous charging power. In contrast, for the isotropic Heisenberg interaction [Fig.~\ref{P_N}($d_3$)], the TMI remains identically zero for all times, indicating the absence of tripartite correlations throughout the dynamics.

We next consider the ABEE, shown in Fig.~\ref{P_N}($e_1$--$e_3$). For the Ising [Fig.~\ref{P_N}($e_1$)] and XY interactions [Fig.~\ref{P_N}($e_2$)], the ABEE increases in time and reaches its maximum value. In both cases, the time required to reach the maximum is larger than that of the instantaneous charging power. In contrast, for the isotropic Heisenberg interaction [Fig.~\ref{P_N}($e_3$)], it remains identically zero throughout the dynamics.

Finally, we analyze the QFI, as shown in Fig.~\ref{P_N}($f_1$--$f_3$). For the Ising [Fig.~\ref{P_N}($f_1$)] and XY interactions [Fig.~\ref{P_N}($f_2$)], the QFI exhibits similar short-time growth across system sizes, while noticeable size dependent deviations appear at longer times. In contrast, for the isotropic Heisenberg interaction [Fig.~\ref{P_N}($f_3$)], the QFI dynamics collapse onto a single curve for all system sizes, indicating scale-independent behavior. In all cases, the peak of the QFI occurs after the peak of the instantaneous charging power.

Overall, our results reveal a robust hierarchical temporal structure across all observables and interaction types. The instantaneous charging power consistently reaches its maximum earlier than all correlation measures, independently of the system size.

\end{document}